\documentclass[fleqn,usenatbib]{mnras}

\usepackage{newtxtext,newtxmath}

\usepackage[T1]{fontenc}
\usepackage{lipsum}
\DeclareRobustCommand{\VAN}[3]{#2}
\let\VANthebibliography\thebibliography
\def\thebibliography{\DeclareRobustCommand{\VAN}[3]{##3}\VANthebibliography}

\usepackage{graphicx}	
\usepackage{amsmath}	
\usepackage{float}
\usepackage[figuresright]{rotating}
\usepackage[utf8]{inputenc}
\usepackage{afterpackage}
\usepackage[english]{babel}

\usepackage{dblfloatfix}
\usepackage{rotating}

\title[{\it AstroSat} observations of GRS~1915$+$105]{Unraveling the foretime of GRS~1915$+$105 using \textit{AstroSat} observations: Wide-band spectral and temporal characteristics}

\author[Athulya et al.]{Athulya M. P.,$^{1}$\thanks{E-mail: athulyamp9425@gmail.com, mpathulya.res-physics@dsu.edu.in
}
Radhika D.,$^{1}$
V. K. Agrawal$^{2}$,
Ravishankar B. T.$^{2}$,
Sachindra Naik$^{3}$,
 Samir Mandal$^{4}$,
 \newauthor
Anuj Nandi$^{2}$
\\
$^{1}$Department of Physics, Dayananda Sagar University, Bengaluru, 560068, India\\
$^{2}$Space Astronomy Group, ISITE Campus, U R Rao Satellite Centre, Bengaluru, 560037, India \\
$^{3}$Astronomy and Astrophysics Division, Physical Research Laboratory, Ahmedabad, 380009, India\\
$^{4}$Department of Earth and Space Sciences, Indian Institute of Space Science and Technology, Thiruvananthapuram, 695547, India \\
}

\date{Accepted XXX. Received YYY; in original form ZZZ}

\pubyear{2021}

\begin{document}
\label{firstpage}
\pagerange{\pageref{firstpage}--\pageref{lastpage}}
\maketitle

\begin{abstract}
We present a comprehensive study of GRS~1915$+$105 in wide energy band ($0.5-60$ keV) using {\it AstroSat} - \textit{SXT} and \textit{LAXPC} observations during the period of 2016$-$2019. 
The \textit{MAXI} X-ray light curve of the source shows rise and decay profiles similar to canonical outbursting black holes. However, the source does not follow the exemplary `q'-diagram in the Hardness-Intensity Diagram (HID). Model independent analysis of light curves {suggests} that GRS~1915$+$105 displays various types of variability classes ($\delta$, $\chi$, $\rho$, $\kappa$, $\omega$ and { $\gamma$}). We also report possible transitions from one class to another ($\chi \rightarrow \rho$, $\rho \rightarrow \kappa$ via an `unknown' class and $\omega \rightarrow \gamma \rightarrow \omega+\gamma$) within a few hours duration. The rising phase of the light curve witnesses the evolution of the source through the classes $\chi \rightarrow \rho \rightarrow \kappa \rightarrow \omega \rightarrow \gamma \rightarrow \delta$ and the descent follows the transition from $\delta$ to $\chi$. Broadband energy spectra are well modeled with multi-coloured disc blackbody and Comptonised components. We explore the `spectro-temporal' features of the source in the different variability classes, transitions between classes, and evolution during the { three year period considered}. Detailed analysis indicates a gradual increase in the photon index ($\Gamma$) from 1.83 to 3.8, disc temperature ({${kT_\textrm{in}}$}) from 1.33 to 2.67 keV, and Quasi-periodic Oscillation (QPO) frequency ($\nu$) from 4 to 5.64 Hz during the rise, while the parameters decrease to $\Gamma$ $\sim$1.18, {${kT_\textrm{in}}$} $\sim$1.18 keV, and $\nu$ $\sim$1.38 Hz {respectively in} the decline phase. Our findings confirm that QPO rms amplitude increases with frequency ($\nu$) till $\sim$2.4 Hz beyond which an anti-correlation exists between the two. The source shows maximum bolometric luminosity { (L$_\textrm{bol}$)} during the peak at $\sim$36\% of Eddington luminosity (L{\textsubscript{\textrm{EDD}}}), and a minimum of $\sim$2.4\% of L$_\textrm{EDD}$ {during the decay phase.  {Further evolution of the source towards an obscured low-luminosity (L$_\textrm{bol}$ of $\sim$ 1\% L$_\textrm{EDD}$) phase,  with a decrease in the intrinsic bolometric luminosity of the source due to obscuration, has also been indicated from our analysis.}} The implication of our results are discussed in the context of accretion disc dynamics around the black hole.

\end{abstract}




\section{Introduction}
\label{sec1}

Accretion powered X-ray binaries (XRBs) are binary systems consisting of a compact object $-$ a neutron star or a black hole (BH), with a disc around it formed by accreting matter from a companion star. The XRBs provide ideal opportunity to probe a black hole and to understand the implications of general relativity by studying the strong gravitational influence of the source on the matter at the inner edge of the accretion disc. The XRBs are revealed during an X-ray activity (persistent or transient), which is triggered and fueled by accretion of matter. Depending on the mass of the companion, the accretion of matter from the companion can happen via Roche-lobe over flow for a Low Mass X-ray Binary (LMXB) or through stellar winds in a High Mass X-ray Binary (HMXB) system \citep{1978ApJ...224..625P, 1991ApJ...371..684B}. Most of the LMXBs discovered are transient in nature \citep{1996ApJ...464L.127K, 1997ApJ...477L..41C, 2015A&A...579A..33V} i.e., they spend most of their time in quiescent  {phase} where there is a steady mass transfer from the companion onto the compact object. Eventually, as the temperature in the disc increases, a thermal-viscous instability is triggered within the disc causing a rapid in-fall of matter onto the compact object leading to an X-ray outburst \citep{1976MNRAS.175..613S, 1990ApJ...353..585H}. 

The outburst duration in a BH-XRB can range from several months to a few years. The energy spectrum of the BH-XRBs is usually characterised by a multi-colour disc component \citep{1973A&A....24..337S} resulting in soft X-ray photons, and Comptonisation of the seed disc photons by the hot corona {producing hard X-rays} (\citealt{1980A&A....86..121S, 1987ApJ...319..643L, 1995xrbi.nasa..126T, 1995ApJ...455..623C, 2010LNP...794...17G};  {\citealt{2015ApJ...807..108I}}). The sources exhibit variability in their light curves which result in broadband noise features in the Power Density Spectrum (PDS) with or without the presence of QPOs \citep{1995xrbi.nasa..252V}. 

Based on the evolution of the energy spectra and temporal characteristics during an outburst, the BH source progresses sequentially through `canonical' spectral states (\citealt{1992ApJ...391L..21M, 2001ApJS..132..377H, 2005A&A...440..207B, 2006ARA&A..44...49R, 2010MNRAS.403...61D, 2012A&A...542A..56N}) in the Hardness-Intensity Diagram (HID). The HID exhibits a `q'-shaped plot ({such as for GX 339--4 as in Figure \ref{fig4} plotted in purple  {colour}}) trailed in an anti-clockwise direction (\citealt{2003MNRAS.338..189M, 2005ApJ...629..403C, 2012A&A...542A..56N, 2014AdSpR..54.1678R, 2019MNRAS.487..928S};  {\citealt{2021MNRAS.502.1334S} and references therein}). The vertical branch emerging at the right edge of the HID represents the initial phase of the outburst where the source is in the Low/Hard State (LHS) with the spectrum dominated by the Comptonised component. The PDS is associated with strong fractional rms variability ($\sim$30 -- 40\%). The source then pivots to {a} horizontal track in the HID representing the transition to the Hard Intermediate State (HIMS) and then to the Soft Intermediate State (SIMS). The energy spectrum in both states are softer in nature with substantial contribution from the disc photons. The PDS displays remarkable differences during the HIMS and the SIMS. Prominent features like Type-C QPOs (see \citealt{1999ApJ...526L..33W, 2005ApJ...629..403C, 2011MNRAS.418.2292M, 2012A&A...542A..56N, 2014AdSpR..54.1678R} for the definitions and characteristics of QPOs) along with a flat top noise are observed during the HIMS. A powerlaw noise component with a drop in the fractional rms variability along with Type-A/B QPOs {are} seen in the PDS during the SIMS. {  The source} diverts to the vertical branch in the left corner of the HID thus making a transition to the High/Soft State (HSS). The energy spectrum is dominated by a thermal disc component with weak rms variability in the PDS ($\sim$1\%). The source reverts back to SIMS, HIMS {,} and LHS in the final stage of the `q'-shape. However, the persistent BH-XRBs like Cyg X$-$1 (\citealt{2020A&A...637A..66M};  {\citealt{2021MNRAS.507.2602K}}), the aperiodically variable GRS~1915$+$105 (see Figure \ref{fig4} in this paper), and a few black hole transients like 4U 1630$-$472 \citep{2020MNRAS.497.1197B} and MAXI J0637$-$430 (\citealt{2021MNRAS.tmp.2464B}) do not follow the exact `q'-profile in the HID, and hence are an exception to this generalisation.

The superluminal X-ray binary GRS~1915$+$105 was discovered by \textit{WATCH}, an All Sky Monitor, on-board \textit{GRANAT} during its outburst in August 1992 \citep{1992IAUC.5590....2C}. It had been exhibiting persistent brightness since its discovery till mid 2018, after which an exponential decrease in the X-ray flux was observed. This makes the source one of the most remarkable LMXB sources unlike others (e.g. XTE J1859$+$226, GRO J1655$-$40, 4U~1630$-$472, GX~339{--}4) which generally have long quiescence periods followed by an outburst lasting for months to a few years. \cite{2001A&A...373L..37G} estimated the mass of the black hole in the binary system to be $14\pm4$ M$_\odot$ with an orbital period of 33.5 days. The long orbital period indicated the large accretion disc that fueled its long-lasting outburst. However, \cite{2014ApJ...796....2R}, using \textit{VLT}, identified the companion as a K-giant donor star. Through the trigonometric parallax, they estimated the distance to the source as 8.6$_{-1.6}^{+2.0}$ kpc and the mass of the BH was revised to 12.4$_{-1.8}^{+2.0}$ M$_\odot$. Recent studies by \citet{2020MNRAS.499.5891S} also {reveal} the mass of the BH to be around  12.44 -- 13.09 M$_\odot$ with a spin value of 0.99 -- 0.997. GRS~1915$+$105 was  {one of the} first Galactic source to display apparent superluminal radio ejections with the characteristics of a micro-quasar \citep{1994Natur.371...46M}. 

GRS~1915$+$105, throughout its 28 years long extreme X-ray activity, has displayed diverse variability in its light curve. It has, so far, exhibited 15 classes of variability \citep{2000A&A...355..271B, 2002MNRAS.331..745K, 2005A&A...435..995H} namely $\alpha, \beta, \gamma, \delta, \theta, \kappa, \lambda, \mu, \nu, \rho, \phi, \chi, \omega, \eta, \xi$. These classes could be structured down to transition between three states; \textit{quiescent state C} $-$ where the flux is relatively low, \textit{outburst state B} $-$ where the count rate is high in the light curve with red-noise like variability in it, and \textit{flare state A} $-$ where flux is seen to be rapidly varying between the two flux levels \citep{1997ApJ...479L.145B, 2000A&A...355..271B}. 

Based on short and long term observations of the source, the `spectro-temporal' characteristics associated with each class have been  {widely} studied hitherto, thus giving an idea about the  {accretion dynamics around} the source within a particular class and/or during a class transition. GRS~1915$+$105 has so far exhibited Low Frequency QPOs (LFQPOs) in 0.1 $-$ 10 Hz and High Frequency QPOs (HFQPOs) at 34 Hz, 41 Hz and 67 Hz \citep{2001ApJ...554L.169S, 2013MNRAS.432...19B, 2019MNRAS.489.1037B, 2020MNRAS.499.5891S}. 
Type-C QPOs in the frequency range 2 -- 5.4 Hz, during $\chi$ variability class, have been reported by \cite{2015AJ....149...82Z} and \cite{2020ApJ...889L..36M}. \cite{2001ApJ...548..401T} investigated the properties of strong $0.8 - 3$ Hz QPOs in 60 -- 124 keV energy band during the LHS and observed the decrease in QPO amplitude at high energies. During $\rho$ variability class, QPOs at frequencies 6 -- 8 Hz were observed. \cite{2002MNRAS.330..487N} {report} a strong correlation between the QPO frequency and the source flux which was also assessed by noticing the presence of QPO at different frequencies during burst and quiescent phase in class $\alpha$. The overall change in the broadband noise and the strength of the QPO during a class transition from class $\chi$ to the `heart-beat' state was studied by \cite{2019ApJ...870....4R}. The correlation between the QPO frequency in the $1 - 10$ Hz range and the duration of the quiescent phases { are} emphasized by \cite{2000ApJ...531L..41C}. The energy dependence of LFQPOs in GRS~1915$+$105 was investigated by \cite{2002A&A...386..271R} and it was concluded that the frequency variations are better correlated with the soft X-ray {($2 - 5$ keV) flux}. 

In the spectral domain, \cite{2001ESASP.459..437R} deduced a specific combination of models to describe the spectrum of the source during each  {class}. The spectral characteristics during HSS ($\delta$ class) is described using a multi-colour disc blackbody with a powerlaw{,} while that for LHS ($\chi$ class) consists of {a} Comptonisation {component} due to hot plasma along with the disc component. The LHS with steady radio emission requires an additional powerlaw component. The spectra of the source during states B and C were studied by \cite{2001ApJ...554L..45Z}{,} where Comptonisation of disc photons from a hybrid plasma, a  {plasma distribution} consisting of thermal and non-thermal electrons, is discussed (see also \citealt{2005MNRAS.360..825Z}). 
For the $\chi$ and $\theta$ { classes}, spectral variabilities in 2 -- 120 keV energy band were studied by \cite{2010ApJ...713..257U} and a dominant Comptonising component at low electron temperature was observed. The investigation of spectral variabilities during the $\rho$ and $\alpha$ classes by \cite{2002MNRAS.330..487N} revealed a steady disc temperature and optical depth of the Compton cloud identical for both classes and inferred that the source spectrum is identical during the burst and quiescent phases. The nature of the accretion disc of the source during soft and steady conditions were studied by \cite{2009ApJ...695..888U} for the classes $\chi, \gamma$ and $\phi$. The spectrum during $\phi$ variability showed a dominating Comptonisation component, in disparity with a typical soft-state spectrum where minimal Comptonised component and a maximum disc component are observed. Similarly, an increased Comptonised flux contribution was observed in the source by \cite{2010MNRAS.409..903P} during the $\omega$ class.

After a year of steady low X-ray activity since June 2018, the source exhibited a few unexpected sequence of events like an unusual low luminosity phase (\citealt{2019ATel12742....1H}) subsequently followed by several X-ray rebrightenings with associated radio flares (\citealt{2019ATel12855....1T, 2019ATel12787....1I}) throughout the year 2019. Recent studies explore the possibility of internal obscuration in the system since May 2019 (\citealt{ 2020ApJ...904...30M, 2021ApJ...909...41B, 2021MNRAS.503..152M, 2021A&A...647A.173K}). \cite{2020ApJ...904...30M} detected absorption lines, that reflected the signature of obscuration. Study of \textit{Swift} data by \cite{2021ApJ...909...41B}, reveals the similarities in the spectra of GRS~1915$+$105 with that observed in highly obscured Seyfert-2 AGN. \cite{2021A&A...647A.173K} {discuss} the possible fate of the outburst; that is, the eventual return of GRS~1915$+$105 to quiescence or continued activity {, while }the source  {is} in a heavily obscured but intrinsically bright accretion state.

In this paper, we perform a systematic and detailed study of the source,  {GRS 1915+105}, for the period of $2016 - 2019$ using \textit{AstroSat} observations and characterise { {the source} behaviour}.  We study the long term \textit{MAXI} \citep{2009PASJ...61..999M} light curve of the source and investigate its possible similarity with the rise and decay pattern of the general outburst profile of a canonical outbursting BH source. The HID of the source is also examined to track the evolution of the source through different spectral states and a detailed comparison with canonical BHs is performed. We explore in depth  `spectro-temporal' properties during the different class variabilities exhibited by the source. The evolution of the broadband `spectro-temporal' features, nature of the light curve, and colour-colour diagram (CCD) of the source as it evolves through different classes are investigated. 
Evolution of the characteristics during the { three} years of observation period are studied to understand the variation of spectral and temporal parameters, and thereby compare with canonical BH transients. 

This paper is structured as follows: In Section~\ref{sec2}, we present the details of {\it AstroSat} observations of the source and describe the process of data reduction using {\it SXT} and {\it LAXPC} pipeline software. In Section~\ref{sec3}, we briefly describe the different  {analyses} and modeling procedures applied to study the temporal and spectral properties of the source. We present the results obtained from our  {analyses} in Section~\ref{sec4}  and interpret the same by considering the scenario of accretion dynamics, and arrive at our conclusions in Section~\ref{sec5}. 

\vspace{-0.5cm}
\section{Observations and Data Reduction}
\label{sec2}
\vspace{-0.1cm}
\textit{AstroSat} \citep{2006AdSpR..38.2989A} is India’s first satellite launched exclusively for astronomical studies and is capable of studying various celestial objects in near and far UV, soft ($0.3 - 8$ keV) and hard X-rays ($3 - 100$ keV) by means of four co-aligned instruments: Ultra-Violet Imaging Telescope  (\textit{UVIT}) \citep{2017AJ....154..128T}, Soft X-ray Telescope (\textit{SXT}) \citep{2016SPIE.9905E..1ES, 2017JApA...38...29S}, Large Area X-ray Proportional Counter (\textit{LAXPC}) \citep{2016ApJ...833...27Y, 2017ApJS..231...10A}, and the Cadmium Zinc Telluride Imager (\textit{CZTI}) \citep{2017JApA...38...31B}. We exclusively consider all the available \textit{SXT} and \textit{LAXPC} observations of GRS~1915$+$105 for our study during the { period of 2016 $-$ 2019}. The spectral coverage in $0.3 - 8$ keV by \textit{SXT} along with the advantage of large effective area of \textit{LAXPC} in $3 - 80$ keV, jointly provide a wide-band spectral view. The temporal resolution of \textit{LAXPC} of 10 ${\mu}s$ also provides excellent opportunity for detecting QPOs in a wide range of frequencies \citep{2016ApJ...833...27Y, 2017ApJ...841...41V, 2020MNRAS.499.5891S}.
 
\textit{AstroSat} observations of GRS~1915$+$105 between November 2016 to June 2019 were carried out as a part of Announcement of Opportunity (AO), Guaranteed Time (GT), and  Target of Opportunity (ToO) cycles. Table~\ref{tab1} represents the list of all the \textit{AstroSat} observations of GRS~1915$+$105 during this period, along with the effective exposure time and average count rate for each Epoch of observation. We also mention the segments i.e., the time interval in light curve having continuous observation which also excludes the orbit constraints like South Atlantic Anomaly (SAA) passage of the spacecraft. The level-2 data of \textit{SXT} and level-1 data of \textit{LAXPC} for all the observations at various Epochs were obtained from the ISSDC data dissemination archive\footnote{\url{https://astrobrowse.issdc.gov.in/astro_archive/archive/Home.jsp}}. We also extract the {\it MAXI} light curve within this period to study the long term evolution of the source.

 \begin{table*}
     \centering
      \vspace*{-0.39cm}  
     	\hspace*{+1cm}
	\caption{Log of X-ray observations of GRS~1915$+$105 from November 2016 to June 2019 using   \textit{SXT} and \textit{LAXPC} onboard \textit{AstroSat}. The source count rate and variability classes (see Sections~{\ref{4.2}} and ~{\ref{sec4.3}}) are also included along with exposure times. Also quoted is the segment number corresponding to the time interval in light curve with continuous observation. Here, IM corresponds to an intermediate state, while HB ($\rho^\prime$) represents a `heart-beat' state.   {The first observation, MJD 57705.50 corresponds to 13 November 2016.}}
     	\label{tab1}
\addtolength{\tabcolsep}{-3pt}    
		\begin{tabular}{ccccccccccccccccc} 
		\hline
				    Observation ID & &MJD &Epoch && Segment& &\multicolumn{2}{c}{Exposure Time (ks)} &&\multicolumn{2}{c} {Average Count Rate} (cts s$^{-1}$) & { Variability Class}\\
		     \cline{8-12}
               &&Start Time&&&& &\textit{SXT} &\textit{LAXPC}  &&\textit{SXT} & \textit{LAXPC} &	\\	         
		\hline
		G06\_033T01\_9000000792 &&57705.50  &1 &  &3 &    &0.89    &2.68  & & {78}  & {3230}       &$\delta$\\
		G06\_033T01\_9000001116 &&57840.87  &2 &&1 &   &1.22    &3      && {21}  & {860}      &$\chi$ $\longrightarrow$ IM $\longrightarrow$ $\chi$ $\longrightarrow$ HB ($\rho^\prime$)\\
		G07\_046T01\_9000001124 &&57844.07 &3${^a}$  &  &1&&1.02    &2.04  & & {24}   & {810}      &$\chi$\\
		                          &&57844.13  &3${^b}$ &  &2& &1.15    &2.52  & & {26 } & {845}     &IM\\
		                            && 57844.21 &3${^c}$ &  &3  &&1.14    &2.15  & & {27} & {815}     &$\chi$\\
		                            && 57844.53 &3${^d}$ &  &8 &&1.28    &2.58  & & {26} & {980}     &HB ($\rho^\prime$)\\
    	G07\_046T01\_9000001162	& &57857.59 &4    &  &6& &1.45    &3     & & {24}  & {775}     &HB ($\rho^\prime$)\\
		G07\_028T01\_9000001166 & &57858.82 &5      &  &3 &&1.36    &3     & & {23 } & {770}     &HB ($\rho^\prime$)\\
		G07\_028T01\_9000001232  &&57891.82 &6${^a}$ &  &2 &&2.07    &3     & & {20}  & {635}     &HB ($\rho^\prime$)\\
		                           & & 57891.89  &6${^b}$ &  &3 &&1.03    &3     & & {19}  & {670}     &HB ($\rho^\prime$)\\
		G07\_046T01\_9000001236 & &57892.50 &7${^a}$ &  &3 &&1.22    &3    &  & {19}  & {660}     &$\rho$\\
		                            & & 57892.70 &7${^b}$ &  &6 &&2.09    &3     & & {17}  & {660}     &$\rho$\\
		G07\_028T01\_9000001272  &&57909.76 &8       &  &1& &1.19    &2.43  & & {26}  & {1060 }     &{ `Unknown'} \\
		G07\_046T01\_9000001274  &&57910.11 &9${^a}$ &  &13 &&2.26    &2.86  & & {27}  & {1055}      &{ `Unknown'}\\
		                            &&  57910.63 &9${^b}$ &  &14 &&1.74    &0.78  & & {27}  & {1145}      &{ `Unknown'} \\
		G07\_028T01\_9000001370 & &57943.71 &10     &  &2& &2.26    &3   &   & {23}  & {1250}      &$\kappa$\\
		G07\_046T01\_9000001374  &&57946.42 &11      &  &14& &1.99    &2.73  & & {34}   & {1500}      &$\kappa$\\
		G07\_028T01\_9000001406  &&57961.39&12      &  &3& &1.67    &1.97  & & {31}  & {1750}      &$\kappa$\\
		G07\_046T01\_9000001408  &&57961.64 &13      &  &2& &2.28    &3    && {42}  & {2230}      &$\kappa$\\
		G07\_028T01\_9000001500 & &57995.33 &14      &  &5& &1.67    &2.71  & & {56}  & {1080}      &$\omega$\\
		G07\_046T01\_9000001506 & &57996.55 &15      &  &2& &2.18    &3     & & {38}  & {2970}      &$\omega$\\
			G07\_046T01\_9000001534  &&58007.66 &16${^a}$ &  &2 &&1.95  &2.78  &  & {69}  & {3085}     &$\omega$\\
		                        & &58008.05 &16${^b}$ &  &11   &&0.68   &   1.47 &  &   {73}      &   { 3450}      &$\gamma$
		 \\
		                        & &58008.19 &16${^c}$ &  &14     & &  0.93    &   2.32   &  &    {62}    &   {3240}        &$\omega + \gamma$\\
		G07\_028T01\_9000001618  &&58041.42 &17    &  &1     & &1.32  &2.63 &   & {111}  & {4340}      &$\delta$\\
		A04\_180T01\_9000001622  &&58045.94 &18     &  &2    & &2.6   &2.41  &  & {54}  & {4015}      &$\delta$\\
		A04\_042T01\_9000001630 & &58047.36 &19     &  &2     & &1.23  &3     &  & {101}  & {3830 }    &$\delta$ \\
		G08\_028T01\_9000001656 & &58058.18 &20    &  &2     & &1.10  &3     &  & {54}  & {1655}      &$\delta$ \\
	    A04\_180T01\_9000002000 & &58209.13 &21    &  &1     & &1.25  &2.8   &  & {45}  & {1700}      &$\chi$\\
	    G08\_028T01\_9000002006 & &58211.76 &22     &  &1    & &2.19  &3.0    & & {42}  & {1480 }     &$\chi$\\
	    G08\_028T01\_9000002080  &&58246.66 &23     &  &8     & &1.84  &3.0    & & {17}  & {720}     &$\chi$ \\
	    G08\_028T01\_9000002110  &&58260.60 &24     &  &3     & &2.11  &2.98   & & {16}  & {630}    &$\chi$ \\
	    G08\_028T01\_9000002112 & &58262.77 &25     &  &3    & &2.16  &2.82   & & {17}  & {610}    &$\chi$\\
	    G08\_028T01\_9000002220 & &58313.57 &26     &  &3    & &2.26  &3.0    & & {7}   & {360}     &$\chi$\\
	    G08\_028T01\_9000002306 & &58344.70 &27    &  &2   & &2.36  &2.95   & & {5}   & {305}    &$\chi$\\
	    G08\_028T01\_9000002334 & &58358.10&28    &  &1    & &-     &2.74  &  &-      & {295 }    &$\chi$\\
	    A05\_173T01\_9000002812  &&58565.98 &29     &  &3    & &1.11  &3     &  & {6}   & {300   }  &$\chi$\\
	    T03\_116T01\_9000002916 & &58618.09 &30     &  &19   & &1.84  &2.54  &  & {2}   & {100 }    &$\chi$(with flares)\\
	    T03\_117T01\_9000002988 & &58648.51 &31     &  &18   & &2.19  &2.78  &  & {1}   & {95}     &$\chi$\\
	\hline
\end{tabular}
\addtolength{\tabcolsep}{3pt}
\end{table*} 

 \vspace{-0.4cm}
\subsection{{\it MAXI} Observations}
\begin{figure*}
\hspace*{-0.5cm}
  \includegraphics[width=18.5cm,height=8cm]{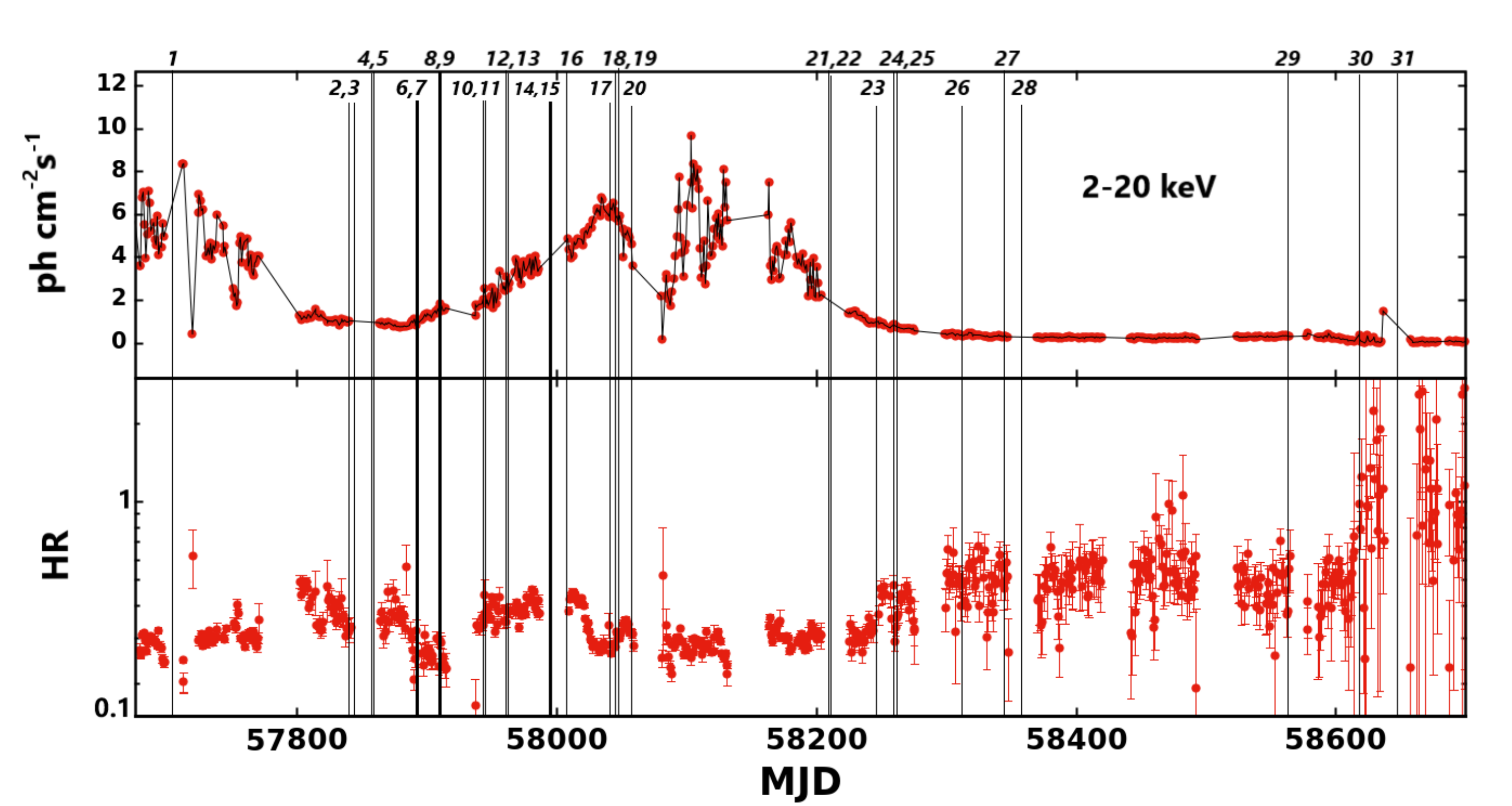}
  \caption{Light curve of GRS~1915$+$105 as observed with the \textit{MAXI/GSC} from July 2016 to June 2019 in $2 - 20$ keV range (\textit{top panel}). The hardness ratio, defined as the ratio between the X-ray count rate in $6-20$ keV and $2-6$ keV, is plotted in the \textit{bottom panel}. The \textit{AstroSat} observations of GRS~1915$+$105 during the Epochs mentioned in Table~\ref{tab1} are marked with vertical black lines along with their Epoch numbers.}
  \label{fig1}
\end{figure*}

\vspace{-0.1cm}
GRS~1915$+$105 was continuously monitored by the \textit{Gas Slit Camera (GSC)} onboard \textit{Monitor of All sky X-ray Imaging (MAXI)} mission \citep{2011PASJ...63S.623M} during the period of {\it AstroSat} observations. We, therefore, use \textit{MAXI} mission data obtained from \textit{MAXI/GSC} on-demand web interface\footnote{\url{http://maxi.riken.jp/mxondem/}} to study the light curve in $2-20$ keV, and Hardness Ratio (HR) of the source as ratio of count rates in $6-20$ keV and $2-6$ keV, as shown in the top and bottom panels of Figure~\ref{fig1}, respectively. The figure also highlights the \textit{AstroSat} observations of the source in black vertical lines corresponding to the different Epochs. With the intention of comparing the evolution of the source with the canonical evolution of an outbursting BH, we obtain \textit{MAXI} observations of GX~339{--}4 during its outburst in $2010-2011$ for the same energy ranges. 

 \vspace{-0.4cm}
\subsection{{\it SXT} Data Reduction}
\begin{figure}
\centering
\includegraphics[width=\columnwidth]{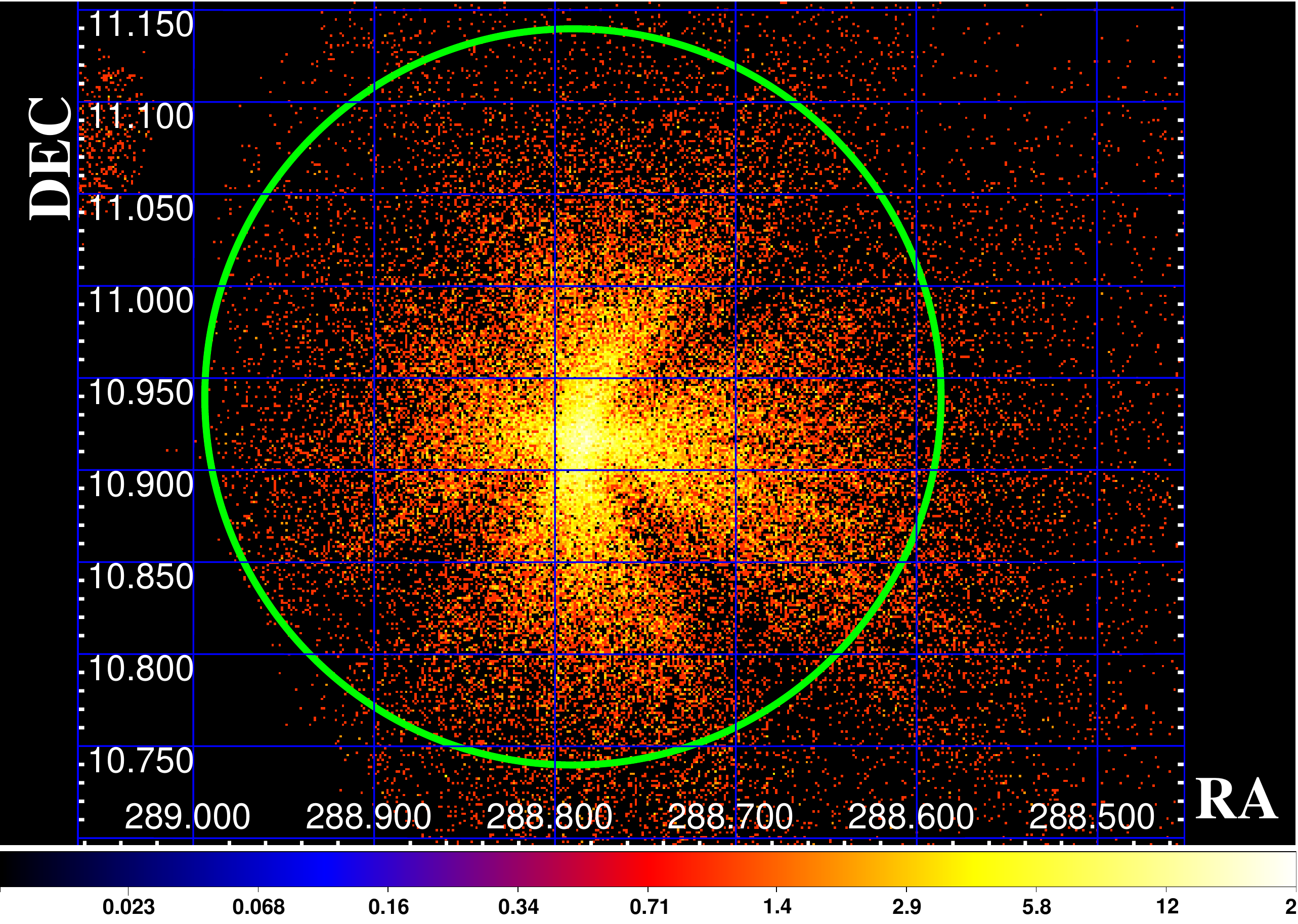}
\caption{\textit{AstroSat/SXT} image of GRS~1915$+$105 corresponding to segment 3 of Epoch~1, obtained using PC mode, is shown. The green circle of 12$\arcmin$ radius encloses the extracted region of the source.}
\label{fig2}
\end{figure}

\vspace{-0.1cm}
We obtained {\it SXT} Level-2 data generated by the latest Level-2 \texttt{SXTPIPELINE version 1.4b} released on January 2019. The Photon Counting (PC) mode data were chosen for the analysis during all the Epochs. The extraction of \textit{ SXT} data was performed following the guidelines provided by the {\it SXT} team\footnote{\url{https://www.tifr.res.in/~astrosat_sxt/dataanalysis.html}} and following \cite{2019MNRAS.487..928S,2020MNRAS.497.1197B}. The tools for \textit{SXT} data extraction along with the background, spectral response, and effective area files are provided by the {\textit{SXT}} instrument team\footnotemark[3]. All the events from individual orbits belonging to each observation were merged into a single clean merged event file using Julia based \texttt{SXTevtmerger} script. \texttt{XSELECT v2.4g} was used to generate the light curve, spectrum, and image from the merged event file. In order to implement simultaneous analysis of \textit{SXT} and \textit{LAXPC} data, we chose one segment from each observation with a minimum of 1 ks exposure time. We also made sure that simultaneous {\it LAXPC} observation was available for the selected Good Time Interval (GTI). We, therefore, filtered the GTI accordingly before generating the light curve, spectrum and the image. We initially attempted to include the source data points within a circular region of 13$\arcmin$ -- 16$\arcmin$  radius for extraction which contained 88 -- 93\% of event-list. However, during certain Epochs none of these regions could be accommodated within the Field Of View (FOV) due to off-axis pointing. We, therefore, used a circular region of 12$\arcmin$ (enclosing $\sim$85\% of event-list) for extraction during all the observations. None of the Epochs showed pile up (>  40 cts s$^{-1}$ as indicated in the \textit{AstroSat Handbook)\footnote{\url{http://www.iucaa.in/~astrosat/AstroSat_handbook.pdf}}}. \texttt{sxtARFModule} provided by \textit{SXT} instrument team\footnotemark[3] is used to generate the suitable {ARF} file for the selected region by applying correction for any offset pointing. Figure~\ref{fig2} shows \textit{SXT} image generated for GRS~1915$+$105 during segment 3 of Epoch~1 (see Table\ref{tab1}). The green circle of 12$\arcmin$ radius encloses the extracted region.

\vspace{-0.4cm}
\subsection{{\it LAXPC} Data Reduction}
\vspace{-0.1cm}
Level-1 {\it LAXPC} data are converted to level-2 using {\tt LaxpcSoft v 3.1.2}\footnote{\url{http://astrosat-ssc.iucaa.in/?q=laxpcData}} released on 15 Feb 2020, following procedures described in \cite{2017ApJS..231...10A}. Amongst the three LAXPC detectors (\textit{LAXPC} 10, \textit{LAXPC} 20, \textit{LAXPC} 30) onboard \textit{AstroSat}, we analysed data from \textit{LAXPC} 20, since \textit{LAXPC} 30 and \textit{LAXPC} 10 exhibited some abnormal gain changes during 2018\footnote{\url{http://astrosat-ssc.iucaa.in/}}. 
{\it LAXPC} data extraction and processing was performed following \cite{2018MNRAS.477.5437A, 2019MNRAS.487..928S} to obtain the source spectra and light curves. The GTI was considered, thereby the background and spectral response were generated as mentioned in \cite{2017ApJS..231...10A}. In order to align the  background and source spectra from \textit{LAXPC} instrument, the code \texttt{backshiftv2.e} was used. The background model chosen for an observation of the source during a particular time corresponds to the closest background model made available by {\tt LaxpcSoft v 3.1.2} for that particular observation period. In order to retain minimal residue in the spectrum beyond 30 keV, we opt for single event, top layer data from \textit{LAXPC} 20 to generate spectra and light  {curves }in $3-60$ keV range \citep{2019MNRAS.487..928S,2020MNRAS.499.5891S}. 

 \vspace{-0.5cm}
 \section{Analysis and Modeling}
 \label{sec3}
 \subsection{\textbf{Temporal Analysis}}
 \vspace{-0.1cm}
 \begin{figure*}
\vspace*{-0.4cm}
\hspace*{-0.3cm}
  \includegraphics[width=19cm,height=7.4cm]{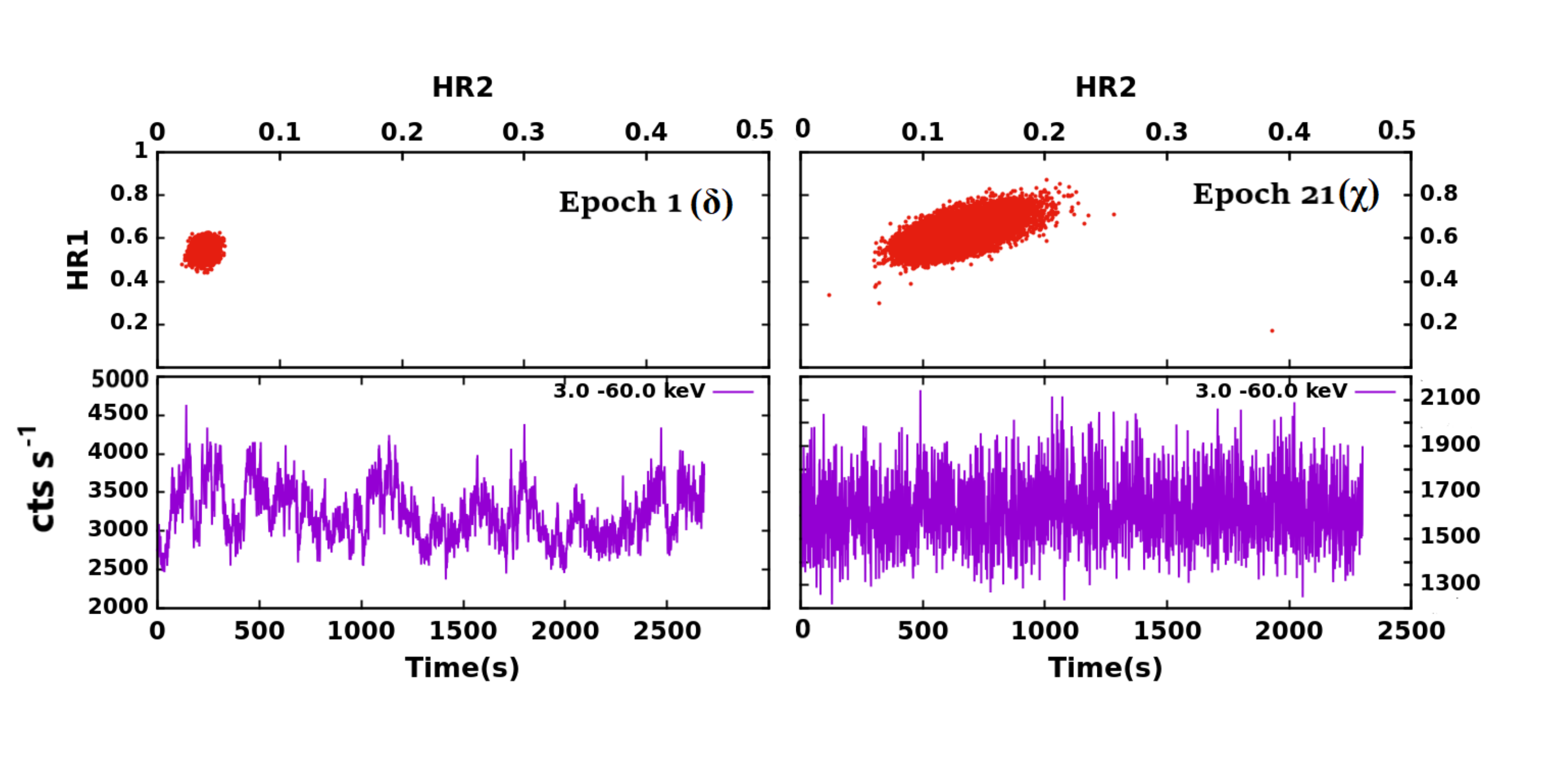}
  \vspace{-1cm}
  \caption{Background subtracted light curve (3 -- 60 keV) and CCD of GRS~1915$+$105 corresponding to Epoch~1 (left panel) and Epoch~21 (right panel), both binned at 1~s. The CCD is plotted as HR1 vs HR2, where HR1 is the ratio between the count rate in 6 -- 15 keV and 3 -- 6 keV ranges and HR2 is the ratio between the count rate in 15 -- 60 keV and 3 -- 6 keV ranges. See text for details.}
  \label{fig3}
\end{figure*}

Initially, we generated a background corrected \textit{LAXPC} light curve with a time bin of $1$~s in the energy range of $3-60$ keV. The 1~s binned background corrected light curves in the energy ranges $3-6$ keV, $6-15$ keV and $15-60$ keV were also generated in order to plot the CCD. In comparison, the corresponding energy ranges are $2 - 5$ keV, $5 - 13$ keV and $13 - 60$ keV in \cite{2000A&A...355..271B} for generating CCDs. The \textit{bottom-left} and \textit{bottom-right} panels of Figure~\ref{fig3} show the light curves corresponding to { Epochs}~1 and 21 in $3 - 60$ keV respectively. The \textit{top-left} and \textit{top-right} panels of Figure~\ref{fig3} represent the CCDs corresponding to the same Epochs, where hard colour, HR2 (ratio of count rates in $15 - 60$ keV and $3 - 6$ keV ranges), is plotted against the soft colour, HR1 (ratio of count rates in $6 - 15$ keV and $3 - 6$ keV ranges). We then generated $0.01$~s time resolution light curve in the energy band $3-60$ keV to obtain PDS with 65536 bins resulting in a Nyquist frequency of $50$ Hz (1/(2$\times$10 ms)) and a lowest frequency of 0.001 Hz (1/(65536$\times$10 ms)). 

The PDS was geometrically re-binned by a factor of 1.05 and also normalised to the rms space \citep{1990A&A...230..103B,1999ApJ...510..874N}. The total rms variability of the PDS was computed for the frequency range $0.01 - 50$ Hz using the rectangle rule integration method \citep{riemann_1867} { as \textit{rms} = ${\sqrt{(P(\nu)\times\delta\nu)}}\times100$ (in \%) (see also \citealt{2018Ap&SS.363..189D} and references therein), where $P(\nu)$ is units of \emph{$rms^2 Hz^{-1}$} and $\delta\nu$ is the frequency interval width in {\it Hz}. The error for rms was estimated by propagating the error on power and $\delta\nu$}. The broadband noise components associated with the PDS were fitted using multiple \textit{Lorentzians} and \textit{powerlaw} \citep{2002ApJ...572..392B, 2012MNRAS.427..595M}. Any narrow features observed in the PDS were modeled with a \textit{Lorentzian} and those features with Q-factor \big(Q = $\frac{\textit{v}}{\Delta\textit{v}}$\big) $\geq 3$ \citep{2000ARA&A..38..717V} and Significance { \big($\sigma$ = $\frac{Norm}{ {Err}_\textrm{norm}}$\big)} $\geq 3$ \citep{2012MNRAS.426.1701B, 2014MNRAS.445.4259A, 2019MNRAS.487..928S} were classified as QPOs. Here, $\nu$ is the QPO frequency, ${\Delta\textit{v}}$ is its full width at half maximum (FWHM), $Norm$ is the normalisation of the {\it Lorentzian} and $ {Err}_\textrm{norm}$ is the negative error in norm value (see also \cite{1989ARA&A..27..517V, 1995xrbi.nasa..252V, 2006ARA&A..44...49R, 2014AdSpR..54.1678R} and references therein). The QPO-rms amplitude in percentage was obtained by finding the area under the {\it Lorentzian} corresponding to the QPO {feature}. 

 \vspace{-0.2cm}
\subsection{\textbf{Spectral Analysis}}
\label{sec 3.2}
\vspace{-0.1cm}

Broadband spectral analysis and { modeling across} $0.7 - 60$ keV (\textit{SXT}: $0.7 - 7$ keV and \textit{LAXPC}: $3 - 60$ keV) range was performed using \texttt{XSpec v12.10.1f} \citep{1996ASPC..101...17A} and \texttt{HEASOFT v6.26}.  
While performing the simultaneous fit, a normalisation constant was taken into account in order to address the difference in calibration between the two instruments. A systematic error of 2\% was incorporated to both {\it SXT} and \textit{LAXPC} data in the spectral fitting as suggested in {\it SXT} user manual\footnote{\url{https://www.tifr.res.in/~astrosat\_sxt/dataanalysis.html}} and \cite{2017ApJS..231...10A}. A gain fit command with the slope fixed to 1 was used to alter the \textit{SXT} response to fit the edges at 1.8 and 2.4 keV produced due to Si and Au edges (see \citealt{2017JApA...38...29S} and {\textit{Neil Gehrels Swift-XRT}} website\footnote{\url{https://www.swift.ac.uk/analysis/xrt/digest\_cal.php}}). \textit{SXT} data were grouped with 25 counts in each bin before fitting. The interstellar absorption was taken into account with the {\it TBabs} model implementing the Galactic elemental abundance \citep{2000ApJ...542..914W}. The hydrogen column density ($n_{\textrm{H}}$) was kept free for most of the observations and found to vary from $\sim$5 -- 6.5 $\times10^{22}$ atoms cm$^{-2}$, which is comparable to the previous estimates \citep{2016ApJ...833...27Y, 2020MNRAS.499.5891S}.

The best spectral fit for all the Epochs was obtained with different combinations of two model components: (1) multi-colour disc blackbody \textit{diskbb} to address the soft X-ray emission from the disc (\citealt{1986ApJ...308..635M, 1984PASJ...36..741M}), (2) thermal Comptonisation \textit{nthcomp} (\citealt{1996MNRAS.283..193Z, 1999MNRAS.309..561Z}). A non-thermal hard component \textit{powerlaw} was also occasionally used (see \citealt{2020MNRAS.499.5891S}). The sporadic appearances of fluorescent Fe line emission in 6 -- 7 keV and smeared edge profile at $\sim$8 keV were addressed using \textit{gaussian} and \textit{smedge} components respectively. An additional \textit{edge} was considered for the instrumental Xenon edge at $\sim$33 keV \citep{2017ApJS..231...10A,2019MNRAS.487..928S}. 

We initially modeled the \textit{SXT} data using the model combination - \textit{TBabs(diskbb)}, which resulted in a good fit for all the Epochs with physically reasonable {parameter }values. We then extended the above combination to fit the broadband {\it SXT} and {\it LAXPC} spectrum. The combination of models, \textit{TBabs*constant(diskbb+nthcomp)}, hereafter referred to as Model-1, was used to fit all the Epochs. Here, we set the seed temperature in \textit{nthcomp} model to inner disc temperature in \textit{diskbb} model so as to constrain the value of disc temperature ({$kT_\textrm{in}$}). The fit obtained for Epoch 2 using Model-1 yielded a {$\chi^{2}$/dof of 562.12/446}. Further, \textit{gaussian} and \textit{edge} components were added to Model-1 in order to fit the Fe line emission at 6.9 keV and the Xenon edge at 34.6 keV. The combination of models \textit{TBabs*constant(diskbb+gaussian+edge*nthcomp)} gave the best fit parameters for Epoch~2 with a {$\chi^{2}$/dof of 559.91/513}. However, for  {Epoch 12}, \textit{nthcomp} model alone could not produce a satisfactory fit in the high energies. For example, the spectral fit of Epoch~12 data using Model-1 produced {a $\chi^{2}$/dof of 667.46/439}. We, therefore, incorporated an additional \textit{powerlaw} model component to fit { at} the high energy range. {This significantly improved the {$\chi^{2}$/dof } to 629.41/527 resulting in a F-test probability of 7$\times$10$^{-15}$.  { In order to ensure that the additional component is not a result of any statistical fluctuation \citep{2002ApJ...571..545P}, we performed simulations using the \textit{simftest}\footnote{\url{https://heasarc.gsfc.nasa.gov/xanadu/xspec/manual/node125.html}} script from \texttt{XSPEC} (\cite{2015MNRAS.447.2274B,2016ApJ...822...57L} and references therein) to generate 1000 fake spectra. The maximum change in fit-statistic ($\Delta \chi^{2}$) obtained from our simulations was 10, whereas the actual $\Delta \chi^{2}$ for the data was 161, which is way higher than those obtained from simulations and hence reconfirms the significance of the additional {\it powerlaw} component.} This combination of models, \textit{TBabs*constant(diskbb+nthcomp+powerlaw)}, will hereafter be referred to as Model-2.}

We also computed the total unabsorbed flux along with the individual contributions from the disc and Comptonised components by implementing the \textit{cflux} model{, in the energy range 0.7 -- 40 keV}.  {The bolometric luminosity of the source is calculated for the energy range 0.3 -- 100 keV, using the formula,} $\textit{L} = flux\times(4\pi D^2)$ \citep{1985apa..book.....F},  where D is the distance to the source in kpc (8.6 kpc; see Section \ref{sec1}).{ However, in this paper, the luminosity of the source is described in terms of Eddington luminosity (L$_\textrm{EDD}$), where L$_\textrm{EDD}$ $\approx$ 1.3$\times$10$^{38}$(M/M$_\odot$) erg~s$^{-1}$ \citep{1986rpa..book.....R}.}

In the following section, we present the results based on our in-depth analysis of the light curve, HID, CCD, PDS and energy spectra.

\vspace{-0.7cm}
\section{Results}
\label{sec4}
\vspace{-0.1cm}
\subsection{\textbf{Light curve and Hardness-Intensity Diagram (HID)}}
\label{sec4.1}
\vspace{-0.1cm}

The long term variation of the source intensity observed by \textit{MAXI} from MJD 57700 (8 November 2016) to 58700 (5 August 2019) within the energy band of { 2 -- 20} keV is shown in the top panel of Figure~\ref{fig1}, while the bottom panel shows the variation of HR with time. The source is initially spotted to be in { a} high X-ray activity phase with flux value greater than 8 ph cm$^{-2}$ s$^{-1}$ { and} HR value between $0.28 - 0.32$. The flux decreased to $\sim$1.5 ph cm$^{-2}$ s$^{-1}$ as it reached Epoch~2, $\sim$100 days after the first \textit{MAXI} observation. The flux remained relatively low for $\sim$100 days, until it reached Epoch~7, which was also associated with an increase in HR to $0.4 - 0.5$. 
The source, however, showed an unusual dip in the HR during Epochs~6 and 7. Post Epoch~7, there was a slow rise in the intensity (over $\sim$140 days) towards the peak, till Epoch~17. The HR values varied between $0.35 - 0.5$.{ The source then exhibited a highly variable flux between MJD 58100 to 58180 (Epochs 20 and 21 for $\sim$180 days).} There was a corresponding drop in the HR to $\sim$0.3. Subsequent to Epoch~21, the source showed decreasing X-ray activity and beyond MJD 58300 (1 July 2018), the flux remained steadily low. The HR value increased from 0.5 to 1 during the Epochs~21 to 29 corresponding to the hard state of the source, except Epoch~30 and 31, where the HR value increased suddenly from 1 to 6. { Further increase in the HR, beyond MJD~58600, corresponds to the `harder' spectral states exhibited by the source. Similar high HR values (> 2) have been observed earlier in a few other BH binaries \citep{2021MNRAS.502.1334S} including MAXI J1535$-$571 \citep{2018ApJ...867..114B, 2019ApJ...875....4S, 2019MNRAS.487..928S},  XTE J1752$-$223 \citep{2020MNRAS.493.2452C} etc., during their hard  {state}.}

\begin{figure}
	\includegraphics[width=\columnwidth, height=6cm]{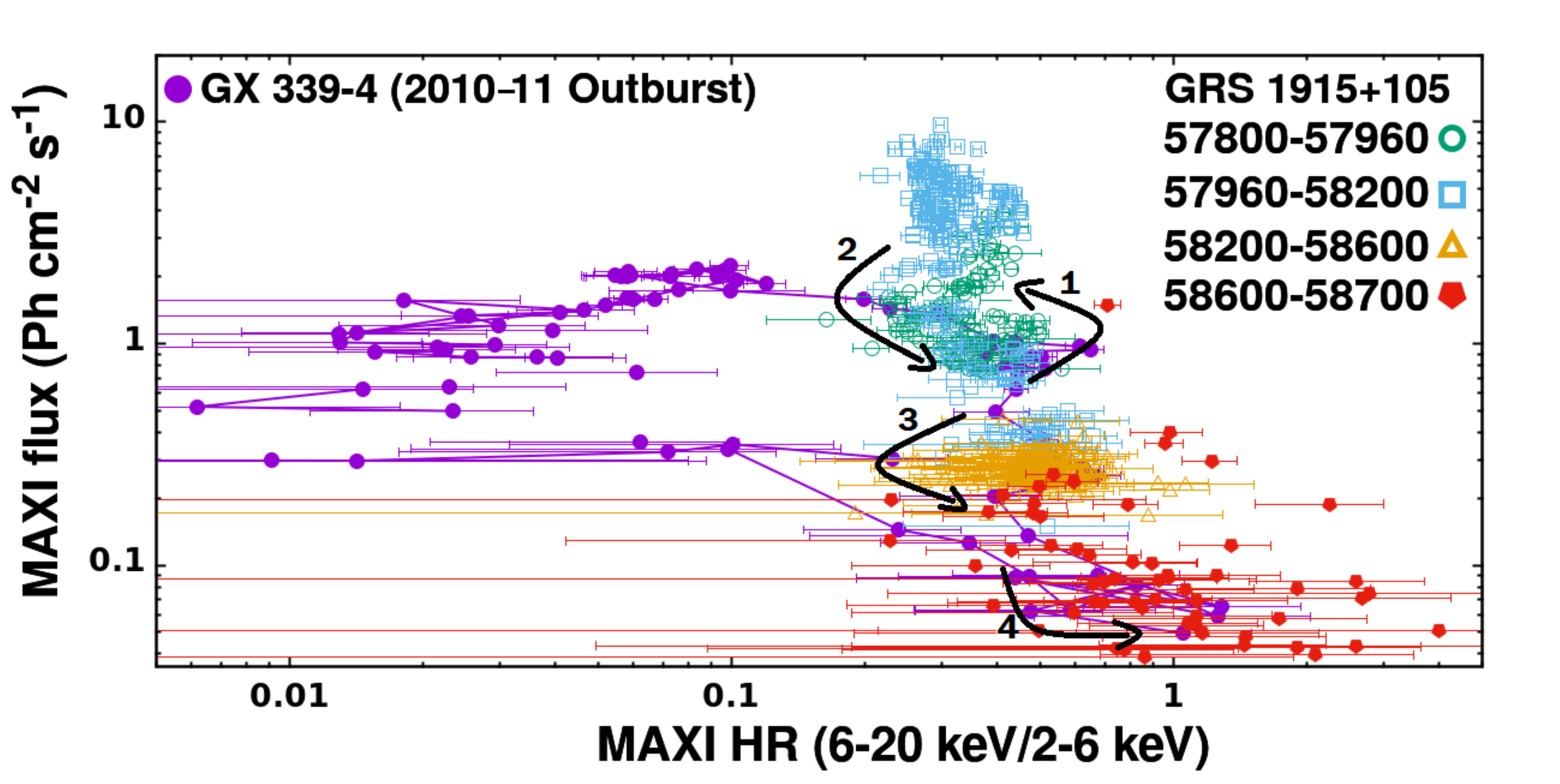}  
    \caption{Hardness-Intensity Diagram (HID) of GRS~1915$+$105 showing evolution of hardness ratio (6 -- 20 keV/ 2 -- 6 keV) with MAXI 2 -- 20 keV flux from {MJD 57800 (16 February 2017) to MJD 58700 (5 August 2019). The numbered black arrows in the figure denote the direction of evolution of GRS~1915$+$105 along the HID.} The green circles {(Arrow-1)} represent points from MJD 57800 (16 February 2017) to MJD 57960 (26 July 2017); the blue squares {(Arrow-2)} correspond to data points starting from MJD 57971 (27 July 2017) to MJD 58200 (23 March 2018). The yellow triangles {(Arrow-3)} represent the data points starting from MJD 58200 (23 March 2018) to MJD 58600 (27 April 2019), and the red points { (Arrow-4)} represent the data points from {MJD 58600 (27 April 2019) to MJD 58700 (5 August 2019)}; the purple points represent the HID of GX~339{--}4 during its 2010 -- 11 { outburst}. See text for details.} 
    \label{fig4}
\end{figure}

{The HID of GRS~1915$+$105, starting from MJD 57800 to 58700 (16 February 2017 to {5 August 2019}), is shown in Figure~\ref{fig4}, where the {\it MAXI} flux in 2 -- 20 keV range is plotted against the X-ray colour (HR). The chronic variation of HR (Figure \ref{fig4}) is explained in comparison to the flux variation with time seen in the light curve (Figure~\ref{fig1}). The black arrows in Figure \ref{fig4} represent the direction of evolution of the source during the entire observation period. The first arrow (Arrow - 1) tracks the HR variation from MJD 57800 to MJD 57960 (green), corresponding to the initial low luminosity phase of the light curve. Arrow - 2 shows the variation in HR during the rising phase between MJD 57960 -- 58200 (blue). Arrow - 3 tracks the HR variation during the low intensity phase of the light curve from MJD 58200 to MJD 58600 (blue to yellow), while Arrow - 4 represents for observations beyond MJD 58600 (red).}

{In order to compare the evolution of HID with that of canonical LMXB BH transients, we have over-plotted in Figure~\ref{fig4} the HID of GX~339{--}4 during its 2010$-$11 outburst using data obtained from \textit{MAXI} (represented by the purple points). The energy range chosen to plot the HID for GX~339{--}4 is the same as that of GRS~1915$+$105. This comparison between the HIDs is done by considering a few differences and similarities between the sources, and  {their} possible effects on the HID. GRS~1915$+$105 is located at a distance of $\sim$8.6 kpc with an absorption column density of $\sim$5$\times$10$^{22}$ atoms cm$^{-2}$, while GX~339{--}4 is located at a distance of 8 -- 12 kpc \citep{2019MNRAS.488.1026Z} with an absorption column density of $\sim$0.5$\times$10$^{22}$ atoms cm$^{-2}$ along the line of sight \citep{2012A&A...542A..56N}. The relatively high $n_{\textrm{H}}$ value of GRS~1915$+$105 results in a higher absorption of low energy photons from the source. This causes an apparent increase in the HR of GRS~1915$+$105 with respect to that of GX~339{--}4 for the same flux, thus making it difficult to compare/quantify the HR of both sources. Nevertheless, we choose to compare the evolution of GRS~1915$+$105 with that of GX~339{--}4, in view of the similarity in the light curves of both sources, since GRS~1915$+$105 shows a rise and decay profile (Figure \ref{fig1}) similar to canonical outbursting black holes.}

The HID of GX~339{--}4 during the outburst shows a typical `q'-shaped track,  {with the HR varying from 1.6 to 1.2 during the hard state branch, 1.2 to 0.25 while in the HIMS, 0.25 to 0.06 during the SIMS, and 0.06 to 0.01 in the soft branch.  The HIMS and  {the} SIMS are seen as elongated horizontal branches in the HID. The left vertical and right vertical branches indicate the LHS and the HSS, respectively.  During each spectral state, GX 339--4 exhibits a definite pivot in its path in the HID, thus exhibiting the `q'-shaped track. However, in case of GRS~1915$+$105, no particular structure could be assigned to its HID. The HR varies from 0.2 to 0.5 during the rising phase, 0.2 to 0.7 during the peak phase, and 0.3 to 0.8 during decay phase of the light curve. Beyond MJD 58600, the HR is seen to vary from 0.5 to 4. Every phase of the light curve of GRS~1915$+$105 occupies common range of HR values, whereas GX 339--4 has a distinct HR range corresponding to each phase of its light curve. A broad variation is seen in the HR of GX~339--4 during the intermediate states corresponding to the rise/decay of the flux in the light curve. This broad variation in the HR is not observed in   GRS~1915$+$105.  GX 339--4 also shows a significant drop in the luminosity as the source makes a transition from the SIMS to the Soft state \citep{2011A&A...534A.119C,2012A&A...542A..56N}. This drop in the luminosity (hysteresis) is also not very clearly noticed in GRS~1915$+$105.} GRS~1915$+$105 shows an evolution in the intermediate state similar to another persistent source GRS 1758--258 \citep{2011AcPol..51f..49O} {eventually returning to the LHS.}

In the following sections, we have presented the broadband `spectro-temporal' features exhibited by the source during the {\it AstroSat} observations based on multiple criteria: (1) when the source is seen to be exhibiting features corresponding to a {particular variability class}, (2) when the source makes transitions from a `defined' class to another class, and (3) the overall evolution of the spectral and timing parameters during the period of observations as mentioned in Table {\ref{tab1}}.

\vspace{-0.45cm}
\subsection{\textbf{Broadband Features of `Canonical' Class Variabilities}}
\label{4.2}
\subsubsection{\textbf{Class $\delta$ \textbf{[Epochs - 1, 17, 18, 19 and 20]}}}
\label{4.2.1}

\begin{figure}
    \centering
         \hspace*{-1.3cm}
         \vspace*{-2.5cm}
        \includegraphics[height=10cm, width=6.8cm, angle=-90]{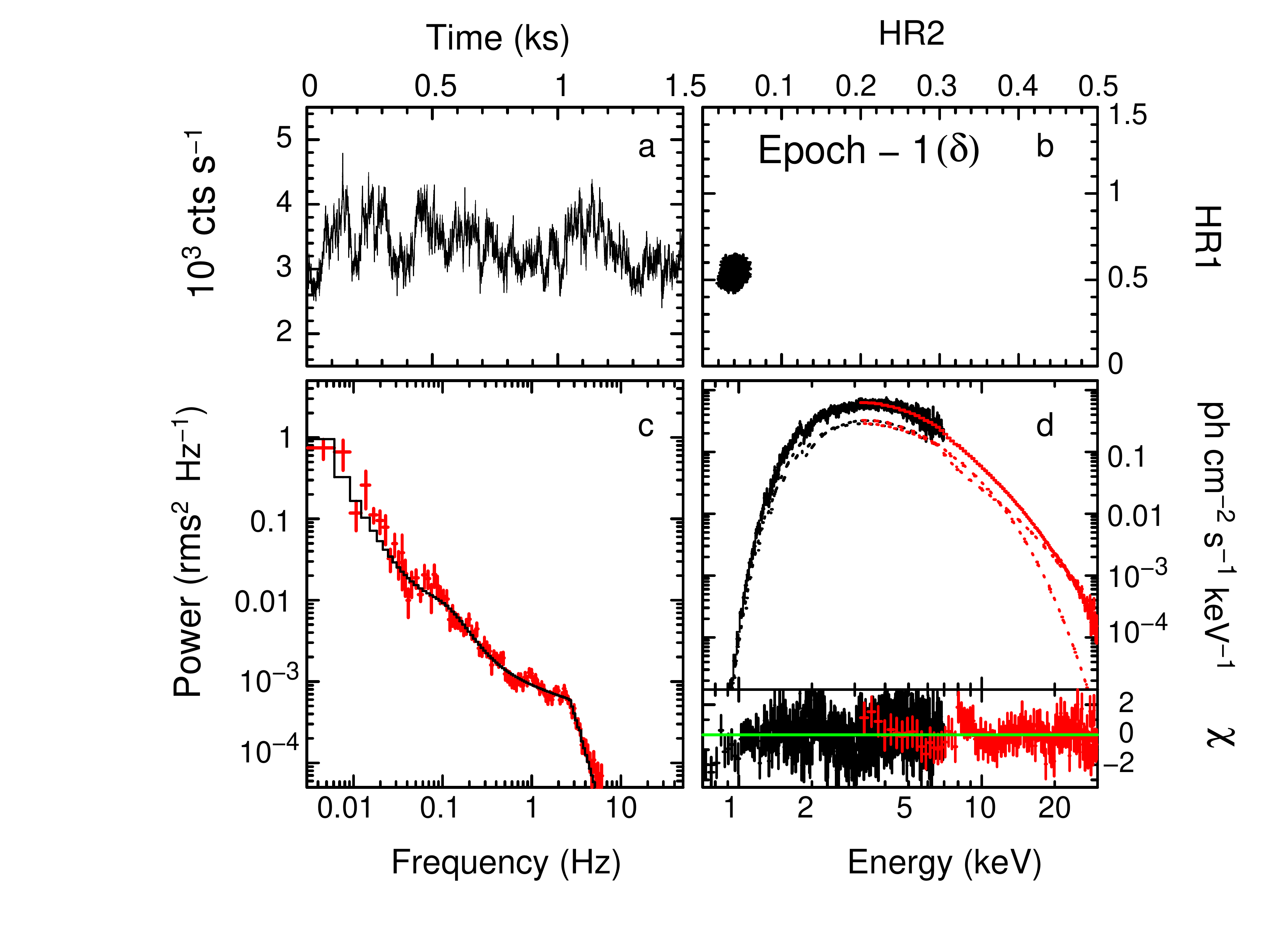}
    \vspace*{+2cm} \caption{Background subtracted 1~s binned light curve and CCD of the source exhibiting $\delta$ variability class during Segment 3 of Epoch 1, are shown in panels~{\it a} and {\it b}. Panel~\textit{c} shows the PDS generated from the 0.01~s binned light curve in $3 - 60$ keV energy band, while panel~{\it d} depicts the broadband energy spectrum from the \textit{SXT} and \textit{LAXPC} for the energy range $0.7 - 30$ keV. The \textit{SXT} and \textit{LAXPC} data are plotted in black and red colour, respectively. The spectrum is best fitted using Model-1. See text for details.}
    \label{fig5}
     \end{figure}

\vspace{-0.1cm}
{The X-ray light curves obtained from Epochs - 1, 17, 18, 19 and 20 displayed a rather stable count rate superposed with red noise variability. These specific features along with identifiable dips in the light curve, are distinctive of the source during $\delta$ variability class, with reference to \cite{2000A&A...355..271B}.} The source during these Epochs are thus classified to belong to $\delta$ class (see also Figure \ref{fig3}). The flux in the light curve varied between $2.5 - 5$ {kcts~s$^{-1}$}. The data points in the CCD {lie} between 0.3 $\leq$ HR1 $\leq$ 0.6 and HR2 $\leq$ 0.06. The panels~\textit{a} and \textit{b} of Figure~\ref{fig5} show the typical light curve and CCD respectively, corresponding to the {$\delta$ variability class.} The PDS does not show any distinct narrow frequency features (panel~\textit{c} of Figure~\ref{fig5}), and has a \textit{powerlaw} noise for the broad frequency range. The total rms power of the PDS was found to vary within  {{5.3$  _{-1.8}^{+2.0}$ -- 7.7$_{-0.5}^{+0.3}$}}\%. The results obtained from all the Epochs belonging to class $\delta$ are given in Table~\ref{tab2}.

The broadband spectra for the $\delta$ class observations were well fitted with Model-1 along with additional \textit{gaussian} or \textit{smedge} components for certain Epochs. During Epoch 1, the spectrum was described with a high index ($\Gamma$) value of 3.6 with 55\% disc flux contribution, indicating a soft nature (panel~\textit{d} of Figure~\ref{fig5}). The $\Gamma$ value was $\sim${3.0$_{-0.1}^{+0.1}$} during Epochs 17 -- 20, while a gradual decrease of seed photon temperature ({$kT_\textrm{in}$}) value from  1.9$_{-0.1}^{+0.1}$ to 1.3$_{-0.1}^{+0.1}$ keV, accompanied by a constantly increasing electron cloud temperature ({{$kT_\textrm{e}$}}) from  {2.7$_{-0.1}^{+0.1}$ to 3.5$_{-0.2}^{+0.5}$} keV was observed. The best-fit spectral parameters are mentioned in Table~\ref{tab3}.

\vspace{-0.5cm}
\subsubsection{\textbf{Class $\chi$ \textbf{[Epochs - 2, 3, 21$-$31]}}}
\label{4.2.2}
\vspace{-0.1cm}

\begin{figure}
    \centering
        \hspace*{-1.35cm}
        \includegraphics[height=10cm, width=7.4cm, angle=-90]{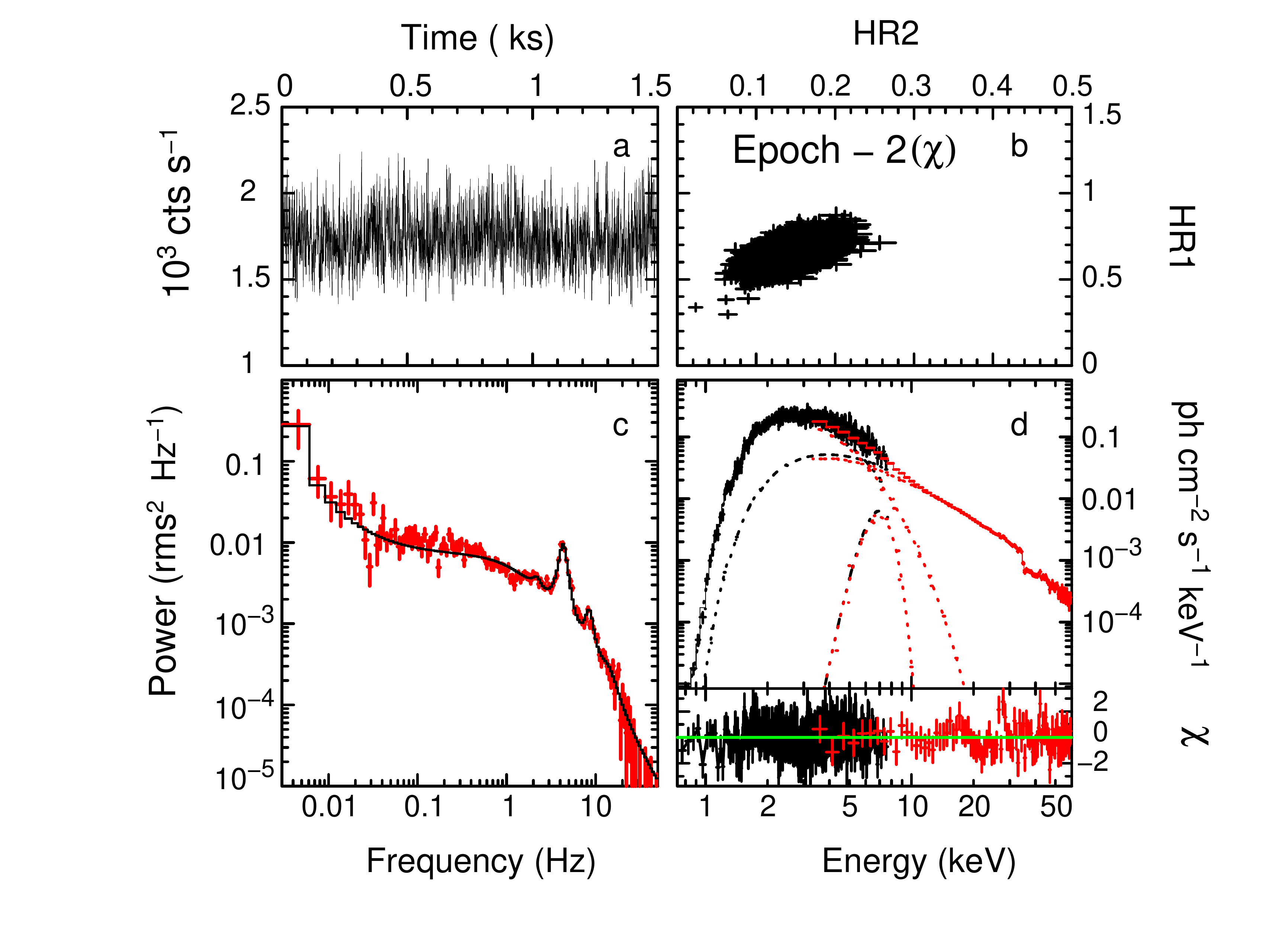}
        \vspace*{-0.65cm}
  \caption{Plots for observations belonging to $\chi$ variability class (Segment 1 of Epoch~2). LFQPO at 4.31 Hz is observed in the PDS. The broadband energy spectrum (0.7 -- 60 keV) is well fitted using Model-1. See text for details.}
    \label{fig6}
    \end{figure}

Flux exhibited by the source during the initial segments of Epochs~2 and 3 were relatively low over long time scales ($\sim$3 ks). The light curve was devoid of any strong variability. The structure of the CCD was diagonally elongated for most of the observations with HR2 $\geq$ 0.06. This position in CCD is associated with state~C of \cite{2000A&A...355..271B}.{ Source exhibiting these characteristics is categorised to belong to $\chi$ variability class \citep{2000A&A...355..271B}.} A few segments in {  Epochs} 2 and 3 and Epochs 21 - 31 exhibit $\chi$ variability class. As seen in Figure \ref{fig1}, Epochs 21 to 31 shows the steady decrease in X-ray flux. The average count rate from the \textit{LAXPC} light curve during Epochs~21 to 31 is seen to have decreased from  {$\sim$ 1700 to 90  cts~s$^{-1}$} with a corresponding gradual increase in the HR values. A drastic increase in the HR values is seen during Epoch~31. In Figure~\ref{fig6}, the data corresponding to Epoch~2 is plotted. The PDS obtained from the observations belonging to class $\chi$ displayed a flat-top noise and QPOs with harmonics (panel~\textit{c} of Figure~\ref{fig6}). Fundamental QPOs for all the observations belonging to this class were varying between 1.3 -- 4.3 Hz. Power spectra for all the Epochs also showed narrow 1$^\textrm{st}$ harmonic while Epoch~22 displayed 1$^{{\textrm{st}}}$ harmonic with less power. The PDS of Epoch~25 showed a sub-harmonic (Q-factor = 1.6) at 1.69 Hz, while Epochs~22 and 24 showed faint sub-harmonics. The total rms varied within the range of  {{18$_{-1}^{+1}$ -- 23$_{-1}^{+1}$}}\%. The QPO rms was found to be  $>12.6$\% during this class (see Table~\ref{tab2} for details).

The broadband energy spectra were well fitted using Model-1 with   {$kT_\textrm{in}$} varying within  {1.1$_{-0.1}^{+0.1}$ -- 1.5$_{-0.1}^{+0.1}$} keV range and  {$kT_\textrm{e}$} varying between  {$11 -55$} keV. $\Gamma$ was always found to be  {$\sim$1.8$_{-0.1}^{+0.1}$} during Epochs 2, 3, and 21 -- 25. $\Gamma$ value dropped to $\sim$ {1.6$_{-0.1}^{+0.1}$} during Epochs~26, 27, and 28 and further decreased to  {1.2$_{-0.1}^{+0.1}$} during Epoch~31. A Gaussian peak was always found { with line energy varying around  {6.3$_{-0.1}^{+0.1}$ -- 7.2$_{-0.2}^{+0.2}$} keV having a width { in the range of} 0.6$_{-0.1}^{+0.1}$ -- 1.1$_{-0.2}^{+0.2}$ keV respectively.} Details of temporal and  spectral properties of $\chi$ class are presented in Tables {\ref{tab2} and \ref{tab3}}.

\vspace{-0.5cm}
\subsubsection{\textbf{Class $\rho$ \textbf{[Epoch - 7]}}}
\label{4.2.3}
\begin{figure}
    \centering
        \hspace*{-1.4cm}
        \includegraphics[height=10cm, width=8.2cm, angle=-90]{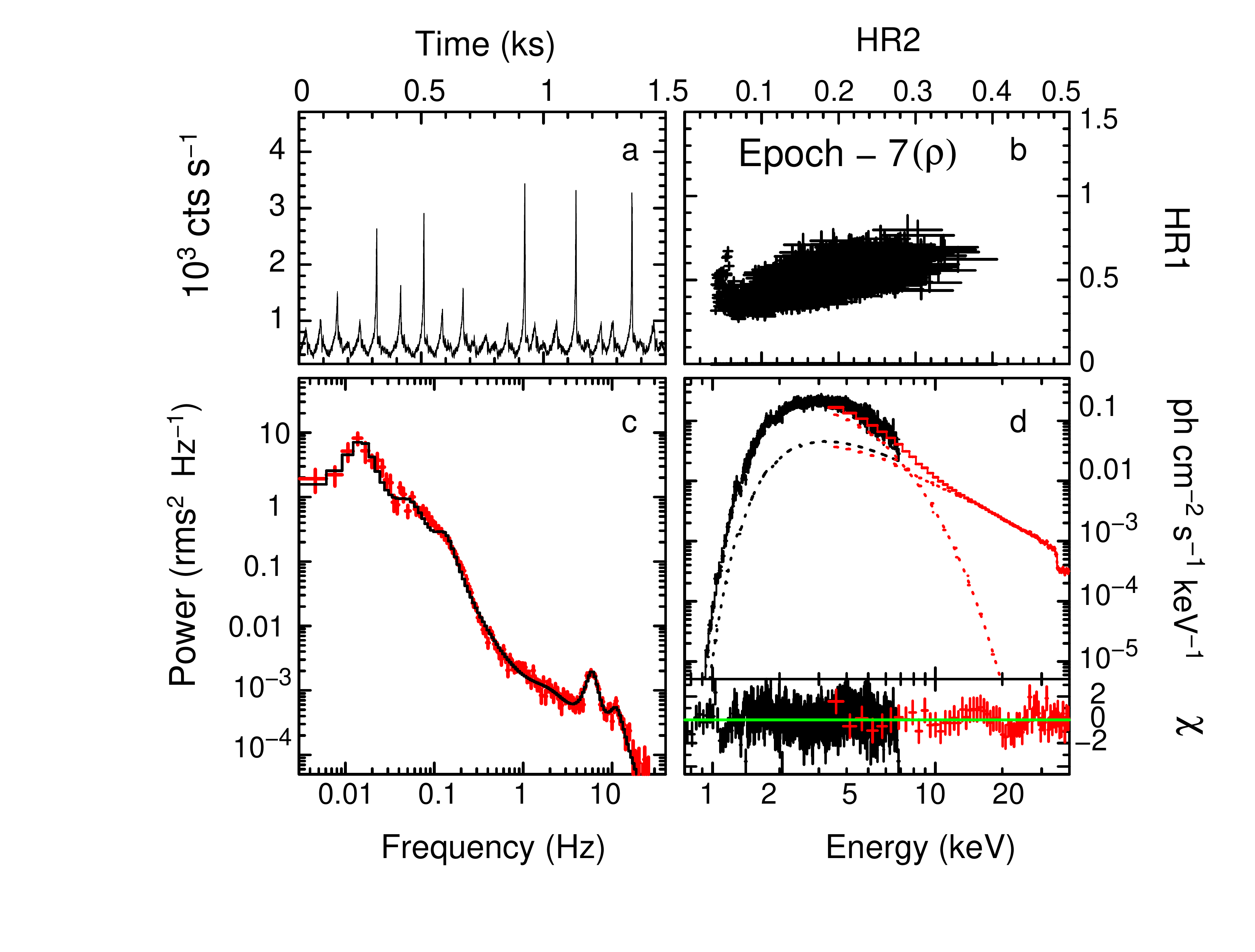}
   \vspace*{-0.87cm} \caption{Plots for observation belonging to $\rho$ variability class (Segment 3 of Epoch~7$^{a}$). A strong QPO was obtained in the PDS at 5.64 Hz. mHz QPOs were also observed for this Epoch at 14.07 mHz (see Section \ref{sec4.3.2} for details). The broadband energy spectrum was fitted in the energy range of $0.7-60$ keV using Model-1. See text for details.}
     \label{fig7}
    \end{figure}

\vspace{-0.1cm}
Epoch~7$^{a}$ and 7$^{b}$ showed distinct regular flares in the light curve (see panel \textit{a} of Figure~\ref{fig7}) repeating on a time scale of $50 - 60$~s with flux varying between 0.5 -- 4   {kcts~s$^{-1}$}. {This regular and characteristic pattern recurring on a timescale of 1 -- 2 min, is categorised as $\rho$ variability class (\citealt{2000A&A...355..271B}).} The peak amplitude of these flare profiles varied non-uniformly unlike that already seen in this source so far - which is typically uniform with constant amplitude \citep{2000A&A...355..271B}.
These burst profiles were single peaked bursts whereas the regular bursts usually {consist} of a two-peak structure \citep{1999ApJ...517..935Y, 2000A&A...355..271B}.
The points in the CCD vary between 0.2 $\leq$ HR1 $\leq$ 0.8 and 0.03 $\leq$ HR2 $\leq$ 0.3 for Epoch~7$^{a}$ (see panel \textit{b} of Figure~\ref{fig7}), whereas the HR2 value was seen to be increasing drastically to 0.6 during Epoch~7$^{b}$. There exists a small branch extending towards higher HR1 values. However, the CCD from our analysis slightly differs in structure from that observed by {\it Rossi X-ray Timing Explorer (RXTE)}, which exhibited a loop like structure \citep{2000A&A...355..271B}. The PDS from our observations belonging to $\rho$ class showed a QPO at $\sim$5 Hz along with a faint 1$^{\rm st}$ harmonic  (panel~\textit{c} of Figure~\ref{fig7}). {The PDS exhibited a powerlaw noise component with peaked noise at $\sim$0.1 Hz, which could be the result of quasi-periodic variations, between the flares, lasting for 25 -- 30 s}. The total rms amplitude varied between  {$34 - 60$}\% (see Table~\ref{tab2}). 

The broadband energy spectra for Epoch~7$^{a}$ and 7$^{b}$ were well fitted using Model-1 along with a \textit{smedge} component (panel \textit{d }of Figure~\ref{fig7}). The disc temperature was found to be $\sim$1.3 keV with the photon index of $\sim$1.9,  {{$kT_\textrm{e}$}} at 14.8
\ keV and a disc flux contribution of $\sim$51\%. All model parameters are summarised in Table~\ref{tab3}.

\vspace{-0.5cm}
\subsubsection{\textbf{Class $\kappa$ \textbf{[Epochs - 10, 11, 12 and 13]}}}
\label{4.2.4}
\begin{figure}
    \centering
       \hspace*{-0.6cm}
        \includegraphics[height=9cm, width=8cm, angle=-90]{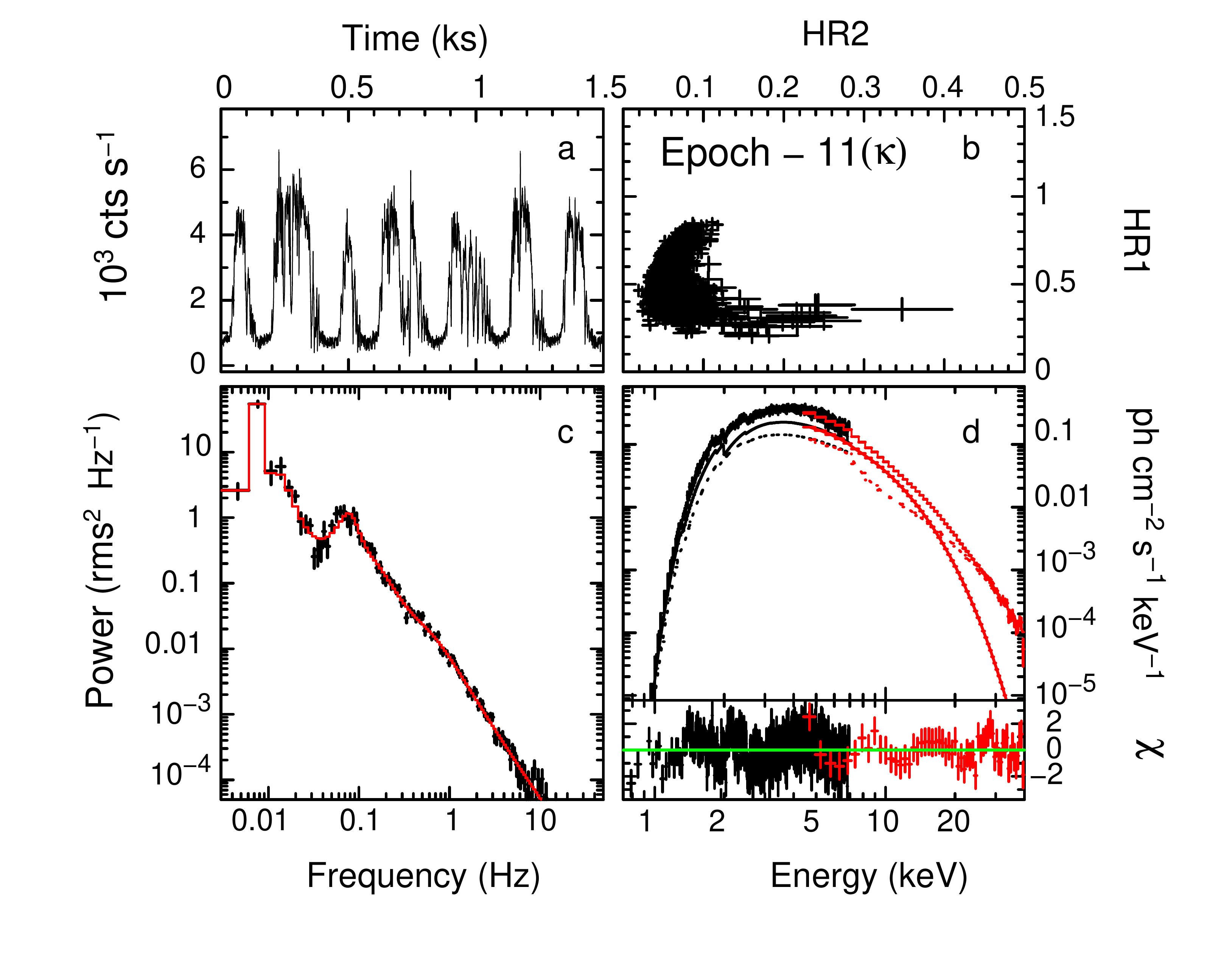}
        \vspace*{-0.77cm}
    \caption{Plots for observations belonging to $\kappa$ variability class (Segment 14 of Epoch 11). PDS shows a broad QPO at $\sim$0.1 Hz, along with a mHz QPO at 6 mHz. The energy spectrum ($0.7-40$ keV) is fitted with Model-1. See text for details.}
\label{fig8}
\end{figure}

\vspace{-0.1cm}
The light curves of Epochs~10 to 13 showed large amplitude variability. A steady rise and fall in luminosity, both lasting for $\sim$100 s was observed. This cycle repeated at a period of $\sim$250~s. The flux varied between $2 - 6$ {kcts~s$^{-1}$} during the rise, while in the dip it was at $\sim$1 kcts~s$^{-1}$ (panel~\textit{a} of Figure~\ref{fig8}). A C-shaped distribution was observed in the CCD (panel~\textit{b} of Figure~\ref{fig8}) with HR values ranging between 0.2 $\leq$ HR1 $\leq$ 1 and 0.03 $\leq$ HR2 $\leq$ 0.3. {This quasi-periodic low-quiet, high-variable and oscillating features in the light curve, with a C-shaped CCD  are distinctive of $\kappa$ variability class, with respect to \citet{2000A&A...355..271B}. Hence, we classify Epochs 10 -- 13 to belong to the same class.} Associated with the periodicity of the flare, a QPO at 6 mHz was observed in the power spectrum. {A broad QPO at $\sim$0.1 Hz was also observed corresponding to the multiple flares in the segment. This could be associated with the quasi-regular oscillations of { shorter} time scale ($\sim$20 s) observed within the cycles}. The PDS showed a \textit{powerlaw} noise with total rms variability decreasing from   {{55.5$_{-0.3}^{+0.2}$ to 51.4$_{-0.2}^{+0.2}$}\%} (see Table~\ref{tab2}).

{Model-1 along with a \textit{smedge} component described the energy spectra for Epochs 10, 11 and 13. The spectrum corresponding to Epoch 12 was fitted using Model – 2, requiring an additional {\it powerlaw} component. In order to identify the origin of this \textit{powerlaw} component, the light curve corresponding to Epoch 12 was further segregated into two categories, based on the count rates: one consisting of only the low counts and the other corresponding to the high counts in rise, peak and the fall of the flux points. Spectral analysis revealed the presence of an additional \textit{powerlaw} component, both during the low-count and the high-count states in the light curve with a powerlaw index of $\sim$1.15$_{-0.01}^{+0.01}$.} From the overall analysis, it is observed that {$kT_\textrm{in}$} varied between  {1.6$_{-0.1}^{+0.1}$ -- 2.5$_{-0.1}^{+0.1}$ } keV, while  {$kT_\textrm{e}$} varied between {11.6$_{-0.2}^{+0.1}$ -- 3.2$_{-0.2}^{+0.2}$ keV} with $\Gamma$ in the range of  {2.3$_{-0.1}^{+0.1}$ -- 3.8$_{-0.1}^{+0.1}$} (see Table \ref{tab3}) as the source evolves from Epoch 10 to Epoch 13. The disc flux contribution varied between {30 -- 55\%} of the total flux.

\vspace{-0.5cm}
\subsubsection{ {Class $\omega$ \textbf{[Epochs - 14, 15]}}}
\begin{figure}
    \centering
       \hspace*{-1.3cm}
        \includegraphics[height=10cm, width=7.4cm, angle=-90]{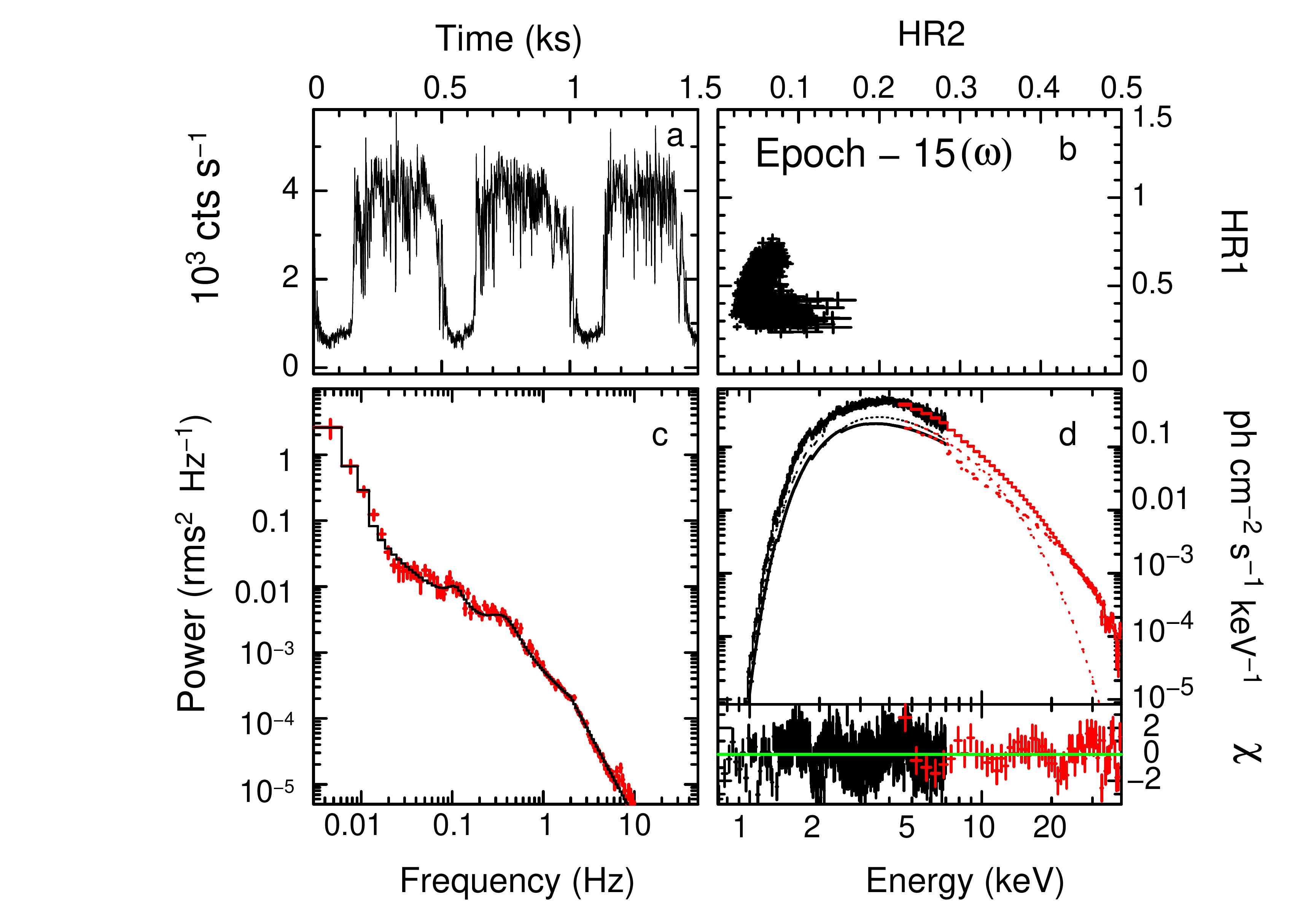}
    \caption{The light curve, CCD, PDS and broadband energy spectrum of the source during $\omega$ class (Segment 2 of Epoch~15) are shown in panel \textit{a},\textit{ b}, \textit{c} and \textit{d}. Model-1 produced a good fit for the broadband energy spectrum for the energy range of $0.7 - 40$ keV. See text for details.}
    \label{fig9}
\end{figure}

\vspace{-0.1cm}
Observations belonging to Epochs~14 and 15 showed periodic fluctuations between high ($2 - 6$ {kcts~s$^{-1}$}) and low intensities ($\sim$0.8 {kcts~s$^{-1}$}) in their light curves with a branched distribution in the CCD (see panels~\textit{a} and \textit{b} of Figure~\ref{fig9}). The points occupying the left branch, extending towards higher HR1 values, are of higher intensity, while those points extending towards higher HR2 values have lower intensity. The HR values range between 0.2 $\leq$ HR1 $\leq$ 0.8 and 0.02 $\leq$ HR2 $\leq$ 0.15. {These transitions between two intensity states at these timescales are distinctive of $\omega$ variability class as mentioned in \citet{2002MNRAS.331..745K, 2002JApA...23..213N, 2005MNRAS.360..825Z}.  Epochs 14 and 15 are thus classified into $\omega$ variability class.} The PDS obtained from these observations showed a \textit{powerlaw} noise component. No definite signature of QPOs were observed in any of the observations and the total rms for these observations varied between {30 -- 45}\%. The temporal results for the observations during Epoch~14 and 15 are given in Table~\ref{tab2}. The energy spectra were well fitted using Model-1 with an additional \textit{smedge} component. {$kT_\textrm{in}$} { was estimated to be $\sim$2.5 keV,} $\Gamma$ was $\sim$2.8 and {$kT_\textrm{e}$} lay between { {6.2$_{-0.1}^{+0.1}$ -- 8.4$_{-0.1}^{+0.1}$}} keV, respectively. A dominant disc flux contribution was observed which increased from {61 to 75\%} during Epochs 14 and 15. Best fit model parameters are listed in Table \ref{tab3}.
 
 \vspace{-0.4cm}
 \subsection{\textbf{`Spectro-temporal' Characteristics during Class Transition}} 
 \label{sec4.3}
 \vspace{-0.1cm}
  In the earlier section, we presented the `spectro-temporal' features corresponding to canonical classes. However, using \textit{AstroSat} observations, we were also able to detect class transitions within a few hours in an Epoch. We discuss the behaviour displayed by the source during each of these transitions. 
  
  \vspace{-0.5cm}
  \subsubsection{\textit{\textbf{$\chi \longrightarrow \rho$ [Epochs - 2, 3]}}}
 \label{sec4.3.1}
 \vspace{-0.1cm}

\begin{figure*}
 \hspace*{-0.5cm}
    \centering
          	\includegraphics[height=7cm, width=17.5cm]{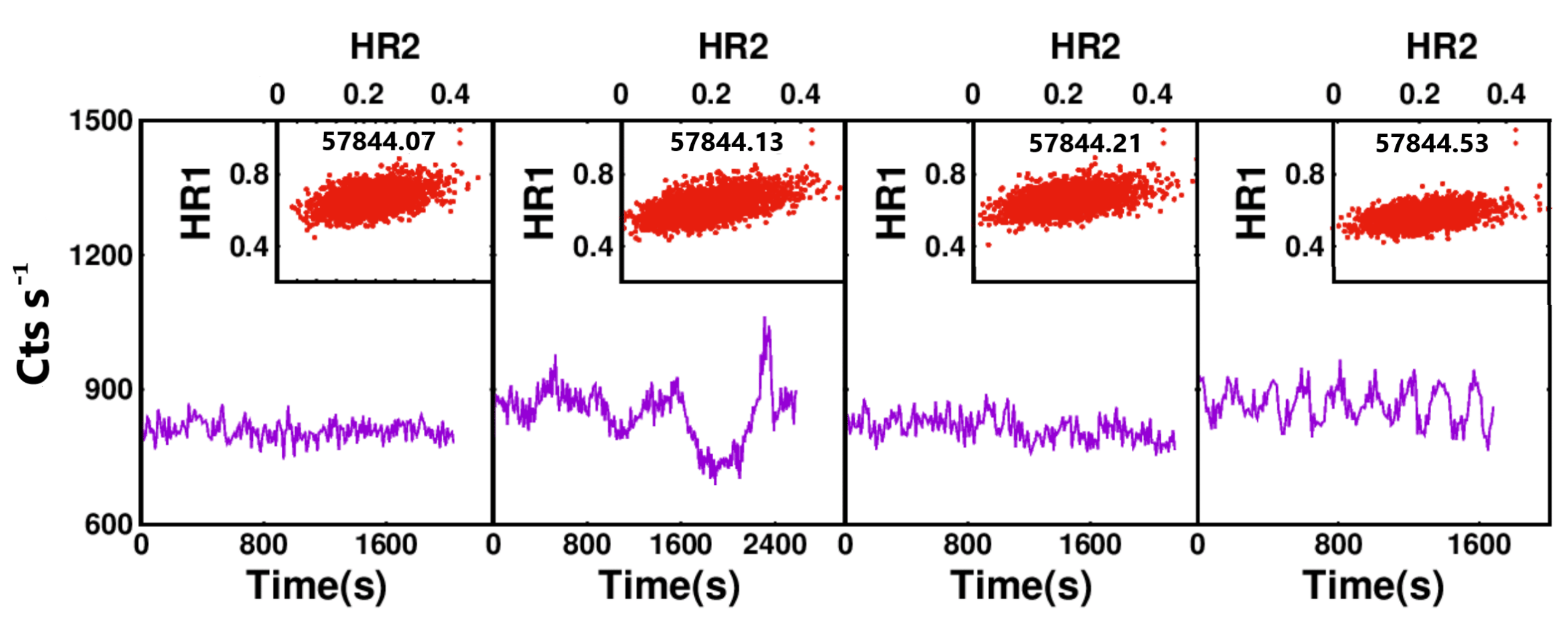}
	
	\caption{{\it LAXPC} light curves and corresponding CCDs of GRS~1915$+$105 indicating transition from the class $\chi \rightarrow$ HB ($\rho^\prime$) {of GRS~1915$+$105} during Epochs~3$^{a}$, 3$^{b}$, 3$^{c}$ and 3$^{d}$ (\textit{left to right panels}). To be noted that for better representation the light curve has been shown with a bin-time of 10 s, while the CCD corresponds to 1s binning as mentioned in section \ref{sec3}. {The start time  (T{$_{\rm start}$}) during each segment is also marked on the top panel.}}
    \label{fig10}
\end{figure*}

As mentioned in Section \ref{4.2.2}, the source initially exhibited $\chi$ variability class during Epoch 3 (segment 1, MJD 57844.07; Epoch 3$^a$ in Figure~\ref{fig10}). The light curve exhibited a drift out of $\chi$ variability class, showing an arbitrary broad profile with an increase in intensity, indicating an intermediate (IM) state during segment 2 (Epoch 3$^b$ in Figure~\ref{fig10}). But for the  {third} and  {fourth} segments (i.e. MJD 57844.21 and 57844.28) the $\chi$ variability class is regained (see Epoch 3$^c$ in Figure~\ref{fig10}), followed by intermediate profiles until the  {seventh} segment (MJD 57844.53). A `heart-beat' state, which is similar to a $\rho$ class (hence denoted here as $\rho^\prime$), is exhibited since the  {eighth} segment i.e. MJD 57844.53 (Epoch 3$^d$, Figure~\ref{fig10}). Interestingly, the CCD for all these segments during the transition remained almost the same (see top panels of Figure~\ref{fig10}), except the `heart-beat' state which exhibited a marginal decrease in the hardness. HR2 decreased from the range $0.14 - 0.36$ to $0.1 - 0.3$. We also observed a  very significant change in the PDS, with segments 1 and 3 ($\chi$ profile) displaying a very narrow QPO along with a first harmonic, whereas, the first harmonic vanished during the  {second} (intermediate state) and the last segment (`heart-beat' state).{ The total rms is seen to decrease from  {19.2$_{-0.1}^{+0.1}$} {\%} to  {17.4$_{-0.1}^{+0.1}$} {\%}, from $\chi$ class (Epoch 3$^a$) to `heart-beat' state (Epoch 3$^d$). This decrease however, cannot be considered significant (see Table \ref{tab2}). Although, the variation in other parameters, like the decrease in the Q-factor from {5.9$_{-0.4}^{+0.4}$} to {3.1$_{-0.1}^{+0.3}$}, increase in $\Gamma$ from 1.83$_{-0.01}^{+0.01}$ to 1.95$_{-0.01}^{+0.01}$, and increase in {$kT_\textrm{e}$} from {12.6$_{-1}^{+2}$} to {33$_{-3}^{+4}$}, from 3$^a$ to 3$^d$ respectively, is significant.}

{A} similar transition pattern is observed during Epoch 2 i.e., the source makes a transition in the sequence : $\chi$ $\longrightarrow$ IM $\longrightarrow$ $\chi$ $\longrightarrow$ $\rho^\prime$ during Epoch 2 also. Following this, the source transited back to $\chi$ class ($\sim$217 ks later) during the initial segment of Epoch 3. It is interesting to note that even after exhibiting one transition of $\chi$ $\longrightarrow$ $\rho^\prime$, the source reverted to $\chi$ class and followed a similar transition to $\rho^\prime$ during Epoch 3 (see Section \ref{sec5} for a comparison with \citealt{2019ApJ...870....4R} and \citealt{2020ApJ...889L..36M}).

\vspace{-0.5cm}
 \subsubsection{\textit{\textbf{\textbf{$\rho^\prime$, the variants of $\rho$ [Epochs - 4, 5, 6$^{a}$ and 6$^{b}$]}}}}
 \label{sec4.3.2}
 \vspace{-0.1cm}
 
\begin{figure*}
	\hspace*{-0.5cm}
				\centering
				  	\includegraphics[height=7cm, width=17.5cm]{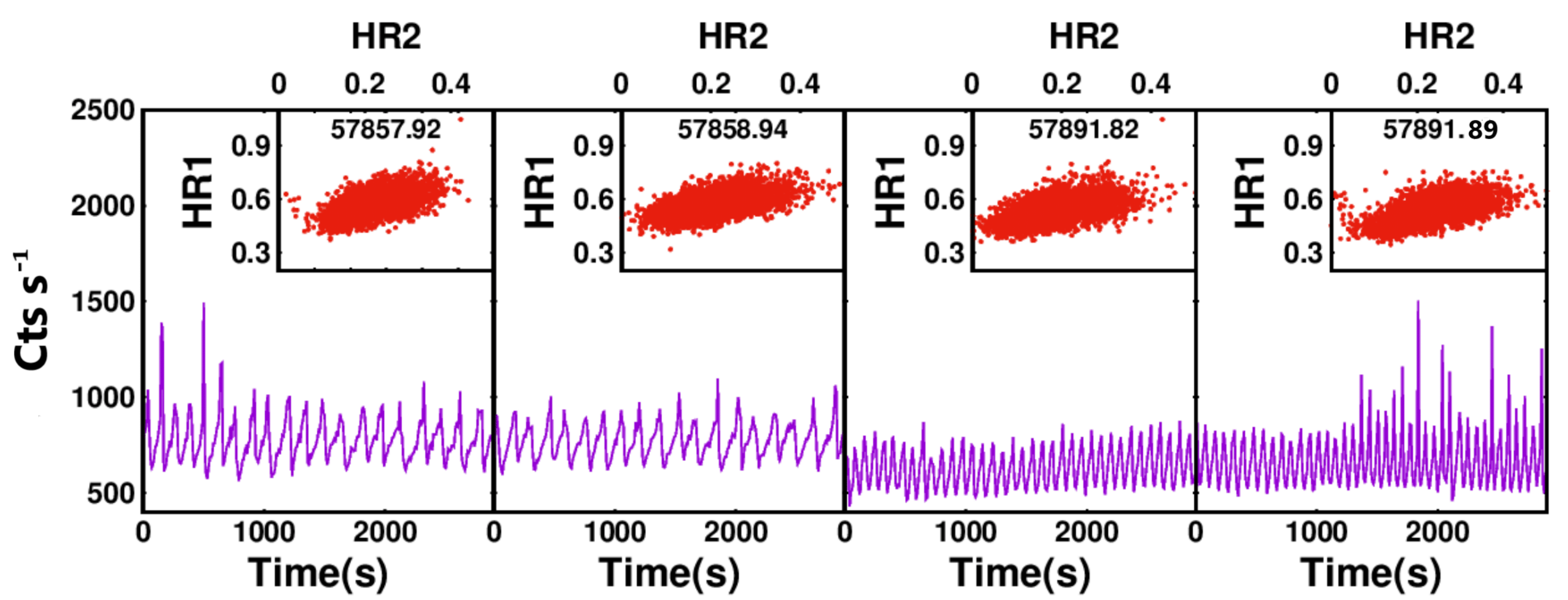}
				\caption{{\it LAXPC} light curves and corresponding CCDs of GRS~1915$+$105 exhibiting `heart-beat'$(\rho^\prime)$ profiles during Epochs~4, 5, 6$^{a}$, and 6$^{b}$ (\textit{left to right panels}). It can be seen that the burst profiles have varying intensity and are skewed during these observations. See text for details.}
				\label{fig11}
			\end{figure*}

\begin{figure}
    \centering
        
  	\hspace*{-0.3cm}\includegraphics[height=8.5cm, width=7cm,angle=-90]{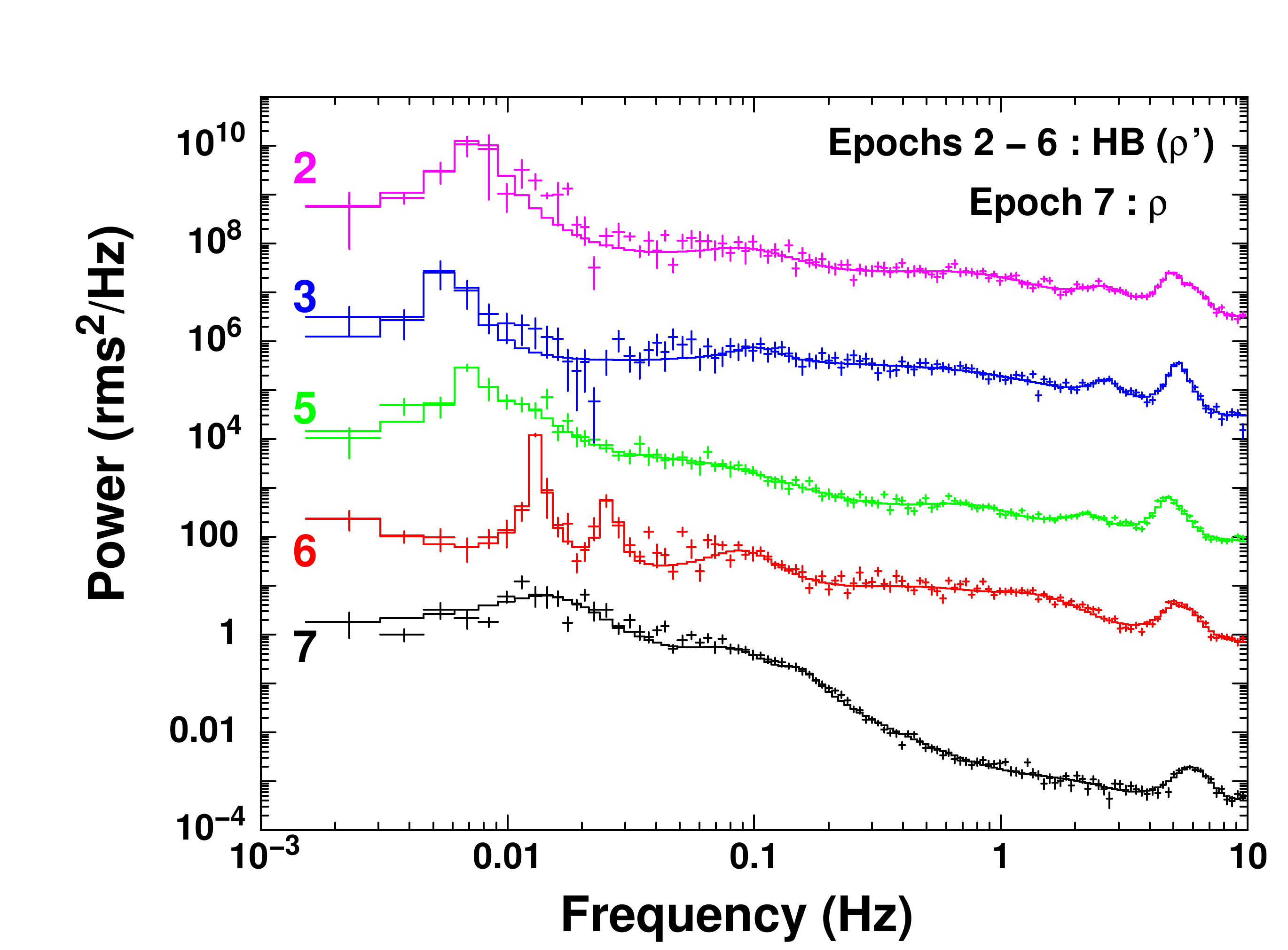}
			\caption{Over-plot of PDS obtained during Epochs 2, 3, 5, 6 and 7, for segments which exhibited a `heart-beat' profile. The PDS during Epochs 2, 3, 5 and 6 are scaled w.r.t that of Epoch 7 by a factor of 10$^8$, 10$^6$, 10$^4$ and 10$^2$ respectively. QPO frequency ($\nu$)  {evolved} from 5.84 mHz to 14.07 mHz { during Epochs} 2 to 7.}

    \label{fig12}
\end{figure}

 Regular burst profiles were observed in the last few segments of Epochs 2 and 3. Epochs~4 to 6$^{b}$ also showed regular burst profiles of varying peak intensity, where the profiles became narrower but skewed with every Epoch. The flares during Epochs 2 and 3 were broader and {less} intense relative to the flares observed during Epochs 4 to 6$^{b}$. Most of the flares during Epochs~4, 5, 6$^{a}$ and 6$^{b}$ showed a steady rise initially, while a few flares showed a sudden increase in the peak intensity, as shown in Figure \ref{fig11}. The periodicity of the bursts was $80 - 90$~s during Epochs 4 and 5, while it decreased to $\sim$60 s during Epochs~6$^{a}$ and 6$^{b}$. Considering the evolving nature of the `heart-beat' profiles, we classify these as $\rho^\prime$. Attributing to the periodicity of the flares, the source, during these Epochs, displayed mHz QPOs. Epochs 2 to 6 displayed narrow mHz features in the PDS while Epoch 7 was seen displaying a broad mHz QPO. The broadness could be due to the varying intensity of the burst (see Figure \ref{fig7}). The QPO frequency increased from $\sim${5 mHz to 14 mHz} from Epochs 2 to 7 (see Figure \ref{fig12}). This evolution of the mHz QPO frequency is consistent with the evolving `heart-beat' profile where the skewness, periodicity and intensity varies with every Epoch.  It appears that, during \textit{AstroSat} observations the source never reached the a canonical $\rho$ class, unlike during {\it RXTE} era where a constant peak intensity is observed for $\rho$ class. A much refined structure and periodicity is exhibited only when the source reached Epoch~7 (see Section \ref{4.2.3})

Low frequency QPOs were also observed during Epochs 2 -- 6$^{b}$, with the frequency increasing from 4.3 Hz to 5.1 Hz and the Q-factor simultaneously decreasing from {5.1$_{-0.1}^{+0.4}$} to {2.8$_{-0.3}^{+0.2}$}. The total rms of the PDS is seen to increase from {18\% to 32\%}. From the spectral fits, a steady photon index at $\sim$1.9 and disc temperature of $\sim$1.3 keV respectively, were observed. Even though $\Gamma$ and {$kT_\textrm{in}$} were consistent, { {$kT_\textrm{e}$}} was initially seen to increase to {33 }keV (Epochs~2 and 3) and then further decrease to $\sim$14 keV during Epochs~4 to 7. These observations  {show a} corresponding relative increase in the disc flux contribution of {36 -- 43}\%.

\vspace{-0.6cm}
 \subsubsection{\textit{\textbf{$\rho \longrightarrow \kappa$ via an `unknown' class [Epochs - 8, 9$^{a}$ and 9$^{b}$]}}}
 \label{sec4.3.3} 
\begin{figure*}
	\hspace*{-0.5cm}
				\centering
				  	\includegraphics[height=7cm, width=17.5cm]{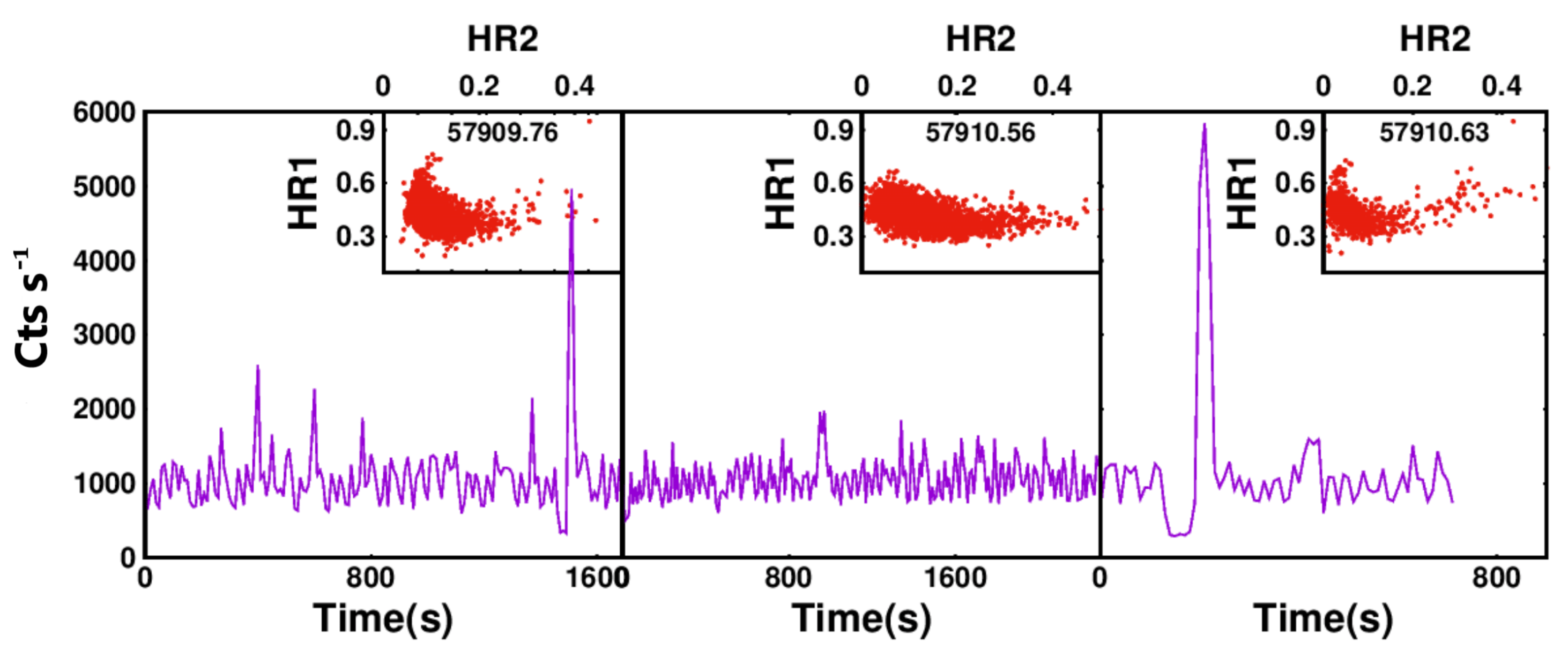}
				\caption{{\it LAXPC} light curves and corresponding CCDs of GRS~1915$+$105 during Epochs~8, 9$^{a}$  and 9$^{b}$ (\textit{left to right panels}). Aperiodically appearing  sudden dip followed by a sharp flare is seen in the light curves of this `unknown' class. See text for details.}
				\label{fig13}
			\end{figure*}

 \vspace{-0.1cm}
The X-ray light curves during Epochs~8 and 9 displayed random noise throughout the observation, interrupted by an aperiodically appearing steady dip lasting for $\sim$70 s, followed by a large amplitude flare. Figure~\ref{fig13} shows the light curves and corresponding CCDs during Epochs 8, 9{$^a$} and 9{$^b$}, where 8 and 9{$^b$} showed an aperiodic flare while Epoch 9{$^a$} did not show any flare. The appearance of the dips {and} flares are observed within a few seconds duration in each segment. The CCDs for Epochs~8 and 9{$^b$} show a cloud of points between 0.18 $\leq$ HR1 $\leq$ 0.8 and 0.01 $\leq$ HR2 $\leq$ 0.38, with the points tending to elongate towards higher HR values in the lower-right and the upper-left part. These change to 0.25 $\leq$ HR1 $\leq$ 0.6 and 0.03 $\leq$ HR2 $\leq$ 0.2 respectively, for Epoch~9{$^a$}. Considering the hybrid nature of the light curve, we classify these Epochs as an `unknown' class. We note that the previous observation during Epoch 7 showed a $\rho$ variability class, and occurred 16 days prior to Epoch 8 (see Section \ref{4.2.3}, Table \ref{tab1}). Even though Epochs 8 and 9 are continuous observations, the next observation took place only 33 days later on Epoch 10 which depicted a $\kappa$ variability class (see Section \ref{4.2.4}, Table \ref{tab1}). Since there is a significant observational gap, we limit to mentioning Epochs 8 and 9 as a transition from $\rho$ to $\kappa$ through an `unknown' class. The temporal analysis during this `unknown' class showed no detection of QPOs in the power spectra. { A total rms value of $\sim$63\% was obtained for both Epochs}, and the PDS nature exhibited a slight flattening at lower frequencies with dominant Poisson noise above 8 Hz. 

The energy spectral fits for this `unknown' class showed a disc dominant emission with the flux contribution decreasing from {66 to 58\%} during Epochs 8 to 9{$^b$}. The disc temperature also {slightly} decreased from  {{1.8$_{-0.1}^{+0.1}$} to {1.6$_{-0.1}^{+0.1}$}} keV while the photon index and the electron temperature varied between  {{2.4$_{-0.1}^{+0.1}$ -- 2.7$_{-0.1}^{+0.1}$}} and  {{4$_{-1}^{+1}$ -- 14$_{-2}^{+1}$ keV}} respectively.

 \vspace{-0.4cm}
 \subsubsection{\textit{\textbf{$\omega \longrightarrow \gamma$ [Epoch - 16]}}}
\label{sec4.3.4}
\begin{figure*}
 	\hspace*{-0.5cm}
				\centering
				  	\includegraphics[height=7cm, width=17.5cm]{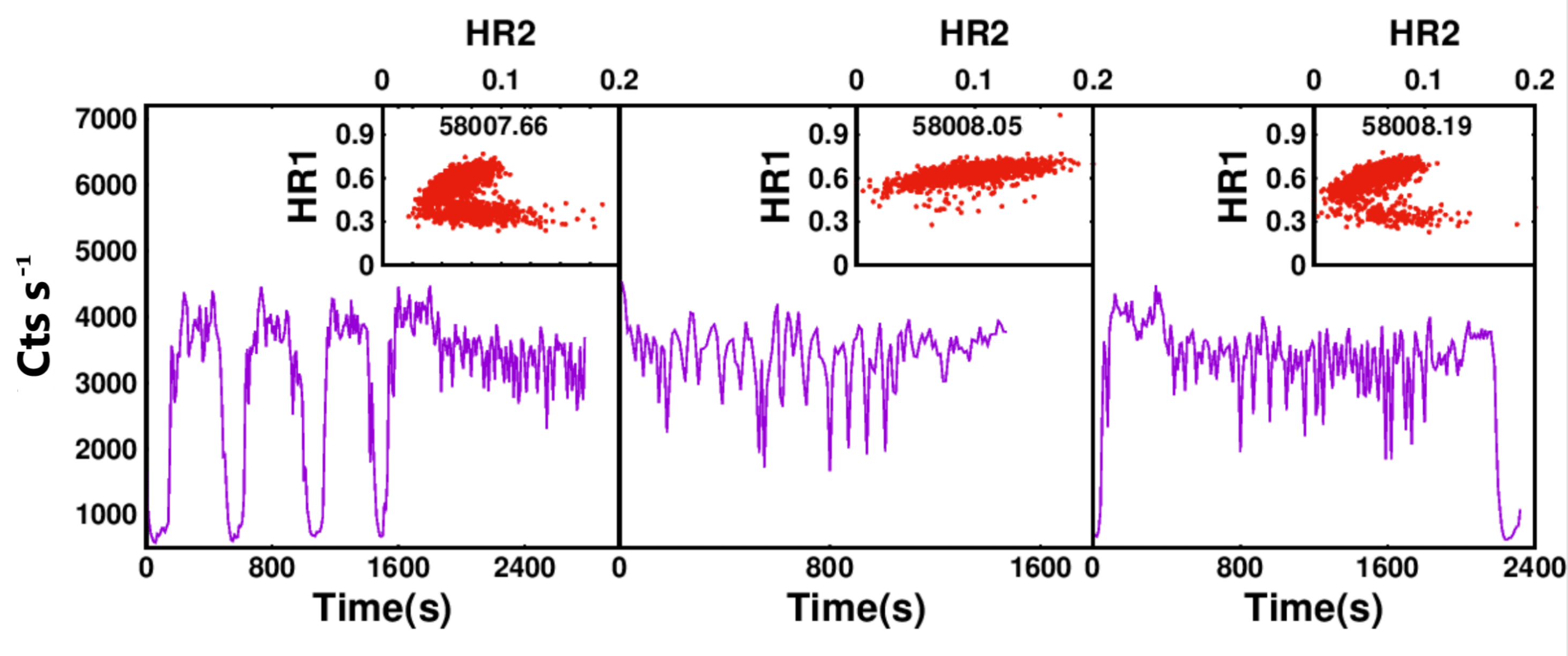}
				\caption{{\it LAXPC} light curves and corresponding CCDs of GRS~1915$+$105 during Epochs 16$^{a}$, 16$^{b}$, and 16$^{c}$ (\textit{left to right panels}). Indications of an $\omega$ class during 16$^{a}$, periodic sharp dips of a $\gamma$ class in 16$^b$ and a combination of both $\omega$ and $\gamma$ in 16$^c$ are evident in this figure. See text for details.} 
				\label{fig14}
			\end{figure*}

 \vspace{-0.1cm}
 A careful inspection of the light curve during Epoch 16 showed distinct pattern variation over a few segments. Figure \ref{fig14} shows the light curve from 2$^\textrm{nd}$, 11$^\textrm{th}$ and 14$^\textrm{th}$ segments during Epoch 16 and the inset displays the corresponding CCDs. The 2$^\textrm{nd}$ segment (i.e. Epoch 16$^a$, MJD 58007.66 - left panel in Figure \ref{fig14}) displays $\omega$ variability class for $\sim$1800 s followed by a steady high count rate in the light curve. The 11$^\textrm{th}$ segment observed after $\sim$2.21 ks (Epoch 16$^b$, MJD 58008.05 - middle panel in Figure \ref{fig14}) displays sharp dips similar to $\gamma$ variability class, as in  \cite{2000A&A...355..271B}. A very interesting profile is seen during the 14$^\textrm{th}$ segment (Epoch 16$^c$, MJD 58008.19 - right panel in Figure \ref{fig14}), $\sim$10.2 ks after 11$^\textrm{th}$ segment, where the light curve has narrow and sharp dips of a $\gamma$ class sandwiched between the dips of $\omega$ class. The CCD for the 2$^\textrm{nd}$ and 14$^\textrm{th}$ segments appear branched, with the upper branch extending towards higher HR1 and higher HR2 values and the lower branch towards higher HR2 values. In comparison, for the 11$^\textrm{th}$ segment the CCD does not appear branched, instead it is elongated towards higher HR1 and HR2 values (see inset panels of Figure \ref{fig14}).  
It is evident that the lower branch in the CCDs, during 2$^\textrm{nd}$ and 14$^\textrm{th}$ segments, originate from the dip corresponding to $\omega$ variability class.  
We presume a possible transition of the source from $\omega$ to $\gamma$ variability class, but ending up with a combined $\omega+\gamma$ structure during the final segment of the observation. 
 
\begin{figure}
    \centering
  	\hspace*{-0.85cm}\includegraphics[height=8.5cm, width=6.1cm,angle=-90]{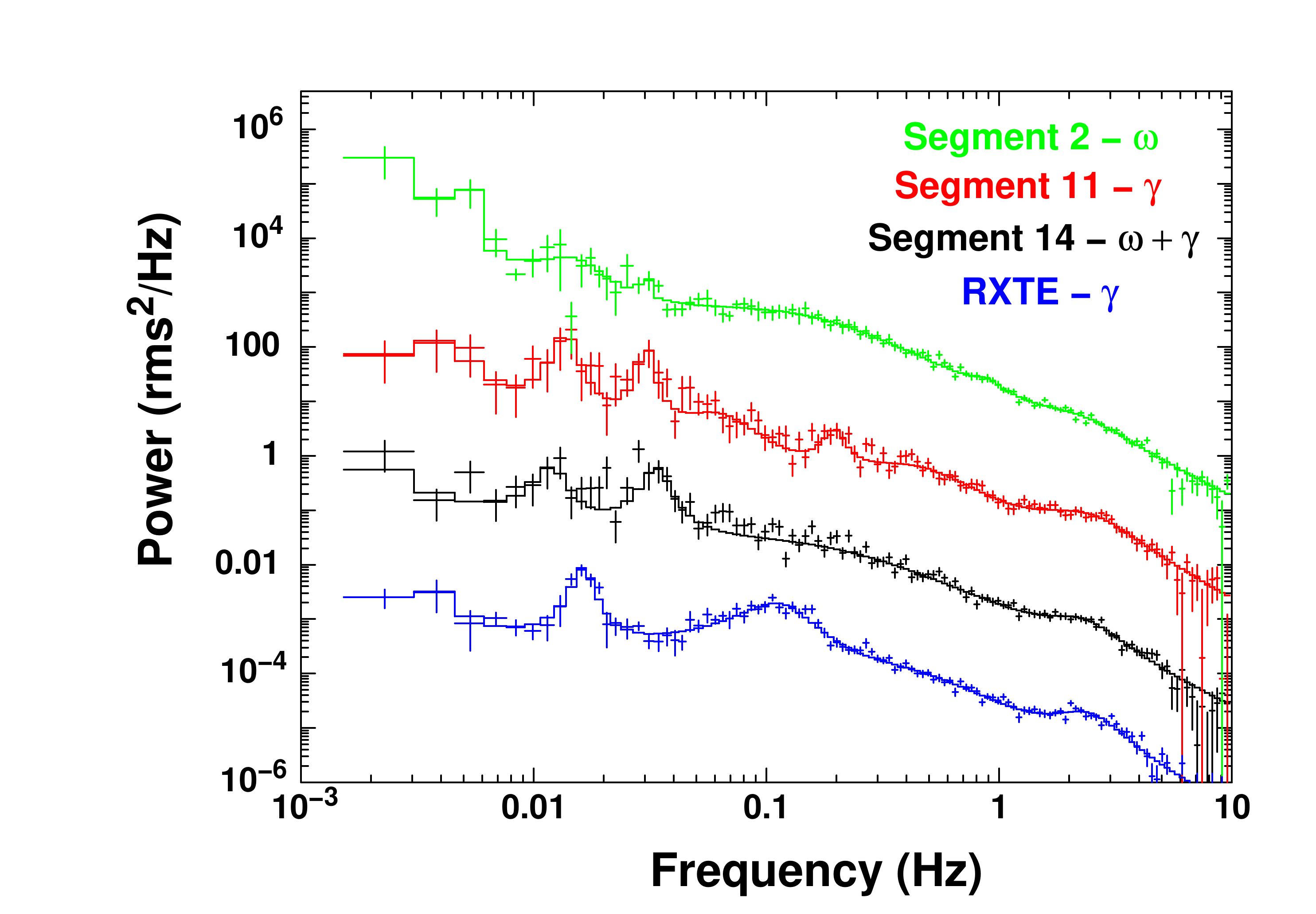}
			\caption{Over-plot of PDS obtained from segments 2, 11 and 14 during Epoch 16 along with that obtained from {\it RXTE-PCA} observation of the source during $\gamma$ variability class (plotted in blue). The PDS of segments 2, 11, and {\it RXTE} are scaled by a factor of 10$^{4}$, 10$^{2}$, 10$^{-2}$ respectively w.r.t that of segment 14. See text for details.}
    \label{fig15}
\end{figure}

 From an independent study of the source during the same Epoch by \cite{2019MNRAS.489.1037B}, the source was classified to belong to $\mu$ variability class, whereas we did not observe any indication of $\mu$ variability class. In order to substantiate our case, we studied the {\it RXTE-Proportional Counter Array (PCA)} observation of the source particular to $\gamma$ variability class during MJD 50654.02 (the same observation considered in \cite{2000A&A...355..271B}; following the standard data reduction procedures\footnote{\url{https://heasarc.gsfc.nasa.gov/docs/xte/recipes/cook_book.html}}(see also \citealt{2016MNRAS.460.4403R})). A variability profile similar to 11$^\textrm{th}$ segment (MJD 58008.05 - middle panel in Figure \ref{fig14}) and an elongated CCD without any branches was observed. In addition, a mHz QPO feature was obtained in the power spectra from the {\it RXTE-PCA} observation during $\gamma$ class. An over-plot of PDS for all the 4 segments during Epoch 16 along with that obtained from {\it RXTE-PCA} during $\gamma$ variability class, is plotted in Figure \ref{fig15}. Segments 11 and 14 show mHz QPOs at $\sim$13.6 and 18.1 mHz respectively with a broad Lorentzian peak at $\sim$2 Hz. The total rms variability of { {16.9$_{-1.0}^{+1.2}$}}\% obtained  {for the} 11$^\textrm{th}$ segment is comparable to the value obtained from {\it RXTE-PCA} observation of the source during $\gamma$ class i.e., 18.8$_{-1.3}^{+0.9}$\%. The total rms for 14$^\textrm{th}$ segment was  {{14.9$_{-1.2}^{+1.2}$\%}}.  Segment 2 showed a total rms variability of  {{34.4$_{-2.0}^{+1.6}$}}\% without any detection of QPO features in the PDS.

The energy spectra for these 3 classes: $\omega$, $\gamma$ and  $\omega+\gamma${, are} well fitted using Model-1. The spectral fit during $\omega$ class shows a $\Gamma$ of  {2.3$_{-0.1}^{+0.1}$} with {$kT_\textrm{in}$} of 2.3$_{-0.1}^{+0.1}$ keV and { {$kT_\textrm{e}$}} of 5$_{-0.1}^{+0.4}$ keV, whereas the spectral fits during $\gamma$ and $\omega+\gamma$ classes exhibited  {a} decreased $\Gamma$ of {3.4$_{-0.1}^{+0.1}$} and 3.1$_{-0.1}^{+0.1}$ respectively, with a steady {$kT_\textrm{in}$} of 2.6$_{-0.1}^{+0.1}$ keV and {$kT_\textrm{e}$} varying between 13.7$_{-0.7}^{+0.7}$  {and} 8.3$_{-0.3}^{+0.6}$ keV. All the three classes showed a dominant disc flux contribution varying between {71 and 75\%}. Best fit model parameters for timing and spectral results are listed in Tables \ref{tab2} and \ref{tab3}.

 \vspace{-0.4cm}  
\subsection{Evolution of `Spectro-Temporal' Features}
\label{sec4.4}
\vspace{-0.08cm}
\begin{figure}
    \centering \hspace*{-1.2cm}
    \includegraphics[height=8.8cm, width=10.5cm]{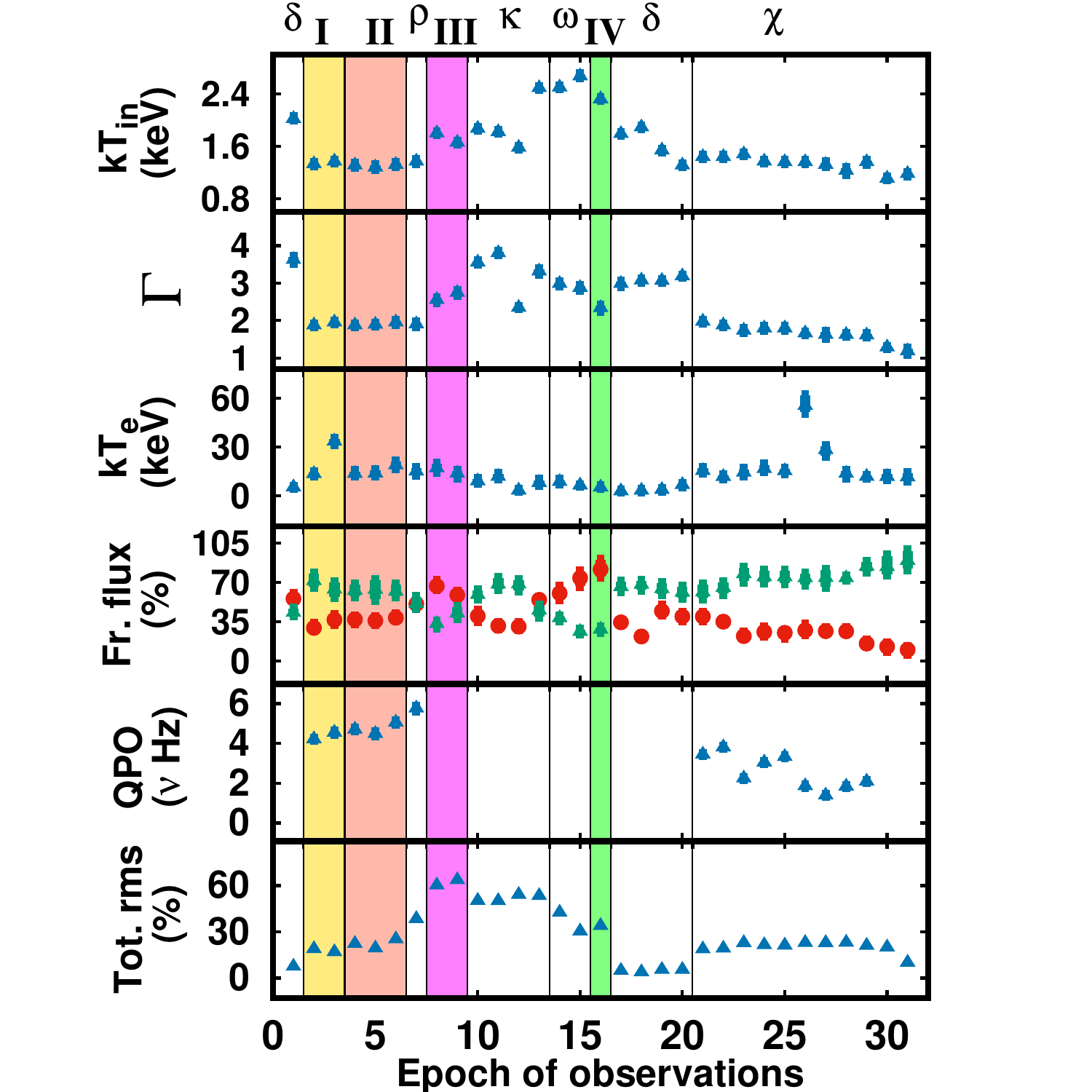}
    \caption{Overall time evolution of the spectral and temporal parameters of the source during the course of \textit{AstroSat} {observations} are shown. Different class variabilities detected are denoted. The shaded regions in yellow (I), coral (II), magenta (III), and green (IV) represent the class transition from $\chi \rightarrow $ HB ($\rho^\prime$) (see Section~{\ref{sec4.3.1}}), variants of $\rho$ - $\rho^\prime$ (see Section~{\ref{sec4.3.2}}), $\rho \rightarrow \kappa$ via an `unknown' class (see Section~{\ref{sec4.3.3}}), and $\omega \rightarrow \gamma$ (see Section~{\ref{sec4.3.4}}) respectively. The evolution of spectral parameters i.e. the electron temperature ({${kT_\textrm{e}}$}), disc temperature ({$kT_\textrm{in}$}) in keV, Photon Index ($\Gamma$), fractional contribution of flux (fr. flux) i.e. disc flux (red points) and Comptonised flux (green points) are shown in the top four panels, respectively. {The last two panels at the bottom} show the evolution of temporal parameters; QPO frequency ($\nu$) in Hz and the total rms variability (Tot. rms) of the PDS (in \%) respectively.}
\label{fig16}
\end{figure}

Figure~\ref{fig16} shows the overall time evolution of the spectral and temporal parameters during different Epochs of \textit{AstroSat} {observations}. The variation of disc temperature ({${kT_\textrm{in}}$}), photon index ($\Gamma$), electron temperature ({${kT_\textrm{e}}$}) and fractional flux from the model components -- Comptonised flux fraction (green points), and the disc flux fraction (red points) in \%, are shown in the first 4 panels \textit{(top to bottom)}. Evolution of QPO frequency ($\nu$) and the total rms (\%) are shown in the bottom two panels in the figure.

\begin{figure}
    \centering
          \hspace*{-0.9cm} 
  	\includegraphics[height=6.8cm, width=9.7cm]{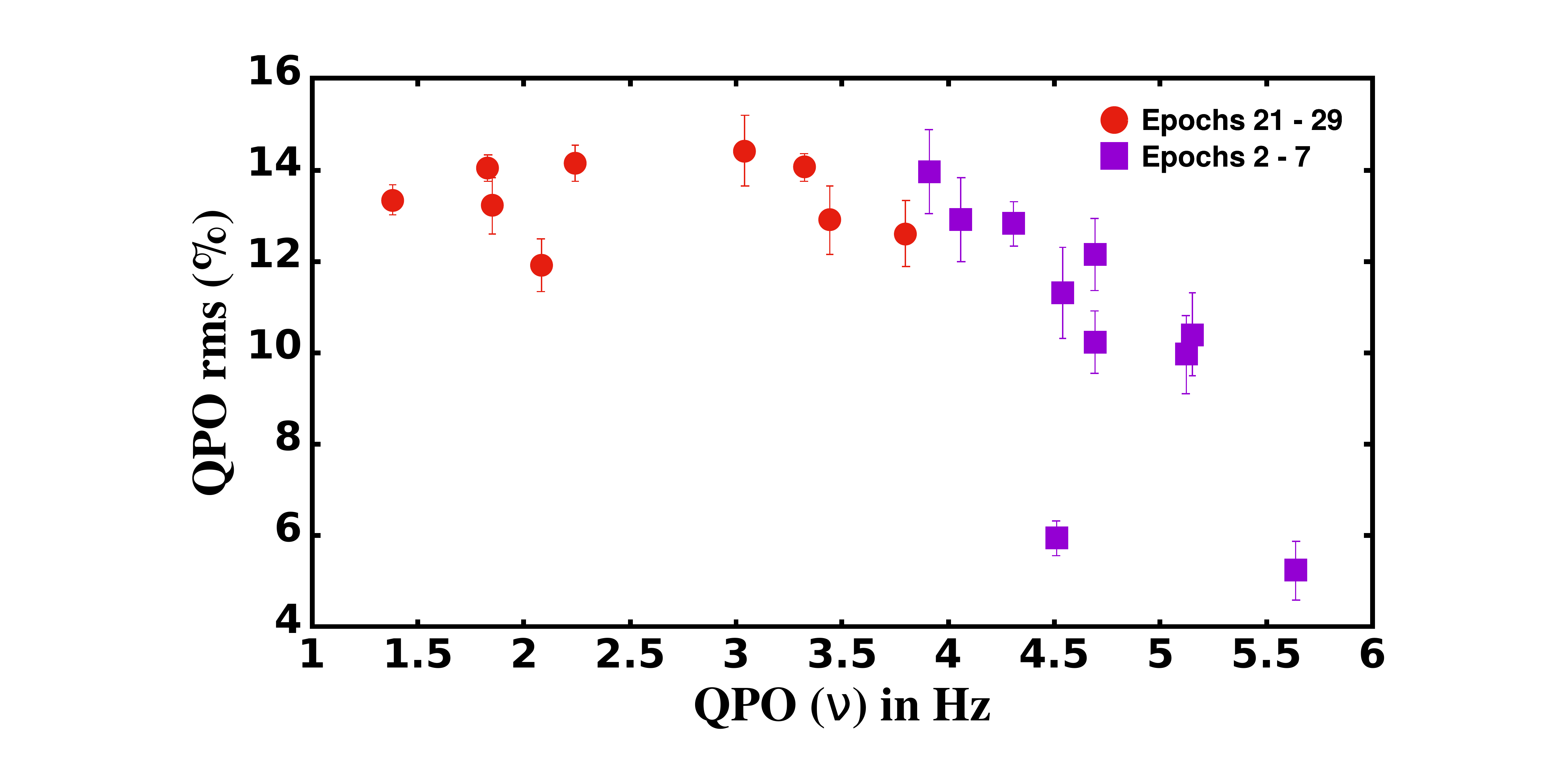}
  	 \vspace*{-0.8cm}
			\caption{ {The plot shows the relationship between QPO rms and the QPO frequency. The red points represent the Type-C QPOs corresponding to Epochs 21 to 29. The purple points represent the parameters corresponding to Epochs 2 to 7. See text for details.}}
    \label{fig17}
\end{figure}

The initial low intensity phase (Epochs 2 \& 3 in Figure \ref{fig1}), is characterised by an energy spectra with photon index of $\sim$1.8, {${kT_\textrm{e}}$} and {${kT_\textrm{in}}$} varying between  {11.5$_{-1.1}^{+1.1}$ and 13.3$_{-0.9}^{+0.8}$} keV, and  {1.87$_{-0.01}^{+0.01}$ and 1.95$_{-0.01}^{+0.01}$} keV respectively, with the disc flux contributing {21 -- 29\%}. The PDS  {shows} prominent detection { of} low frequency Type-C QPOs of 4.3 -- 3.9 Hz and has a flat-top noise of {total} rms $\sim$18\% (see Table \ref{tab2}). These variations of the `spectro-temporal' parameters are indicative of the source occupying a hard state in the initial phase. During the rising phase of the source intensity (Epochs 4 to 13), a further increase is observed in $\Gamma$ from {1.9} to {3},  {${kT_\textrm{in}}$} from {1.3} to {2.4} keV with the disc flux contribution increasing from {37 to 54\%}.  {${kT_\textrm{e}}$} is observed to decrease from {33 to 7} keV (see Figure~\ref{fig16} and Table \ref{tab3}).{ Meanwhile, between Epochs 2 -- 7, the QPO frequency shows an anti-correlation with the QPO rms (see Figure \ref{fig17}). As the QPO frequency increases from {3.9} Hz to {5.}6 Hz, the QPO rms amplitude is seen decreasing from 13\% to 5\%.} The PDS nature changes from a flat-top noise to powerlaw, and total rms increases from $\sim$ 22 to 51\%. The LFQPOs are seen transiting from a strong narrow Type-C to a broad feature. Based on the evolution of these ‘spectro-temporal’ characteristics of the source, it is evident that the source undergoes a spectral state transition from hard to intermediate state.

As the source forges to the peak phase of light curve (from Epoch 14 to 20; Figure \ref{fig1}), $\Gamma$ reaches {3.4}, and ${kT_\textrm{e}}$ decreases to 2.7 keV. An increase in disc flux contribution, from {60\% to 74\%} is seen initially followed by an unusual decrease to {22\%}. The source attains a softer state during this period. {The source also exhibits a  maximum bolometric luminosity of {$\sim$36\%} L$_\textrm{EDD}$ (calculated using the formula as stated in Section \ref{sec 3.2}) during Epoch 17 ($\delta$ class)}. The PDS has a powerlaw nature with rare occurrence of a broad 0.1 Hz QPO (see panel \textit{d} of Figure \ref{fig8}). Finally, during the decline of source intensity in the light curve (Epochs~21 to 31 in Figure \ref{fig1}), a steady decrease is observed in $\Gamma$ from  {{1.9 to 1.1}} with an associated increase in the Comptonised flux contribution from {61\% to 89\%} (see Table \ref{tab3}). The {minimum} bolometric luminosity of {$\sim$2.4\%} L$_\textrm{EDD}$ was registered during Epoch 29 {($\chi$ class; see also section \ref{sec5}).} The PDS exhibited a flat-top noise with Type-C QPOs of frequency decreasing from 3.4 to 1.3 Hz and total rms remained in the range of 19$_{-1}^{+1}$ -- 23$_{-1}^{+1}$\% (see Figure~\ref{fig16}; Table \ref{tab2}). {In review of the correlation between QPO frequency and QPO rms during Epochs 21 -- 29, it is observed that as QPO frequency increases from {1.3 to 2.2} Hz, QPO rms is seen to increase from {  {13.2$_{-0.6}^{+0.5}$\% to 14.1$_{-0.6}^{+0.7}$\%}} (see Figure \ref{fig17}). However, the QPO rms shows a decreasing trend on further increase in the QPO frequency. This correlation pattern for Type-C QPOs, is in accordance with the results reported by \cite{2021MNRAS.503.5522K, 2020MNRAS.494.1375Z, 2013MNRAS.434...59Y},  {who} showed an initial increase in the QPO rms, reaching its maximum at $\sim$2 Hz and then decrease beyond 2 Hz.} The source  {exhibited} a sharp decline in the total rms amplitude, {22\% to 7\%}, during the last observations. These characteristics trace the change in the state of the source, to hard state via an intermediate state. Following this, multiple flarings were seen in the light curve (Epoch 31 - Figure \ref{fig1}) which probably held back the source from attaining quiescence.

\begin{figure}
    \centering
    \hspace*{-0.35cm}
      	\vspace*{-0.2cm}
		\includegraphics[height=9.1cm, width=6.8cm,angle=-90]{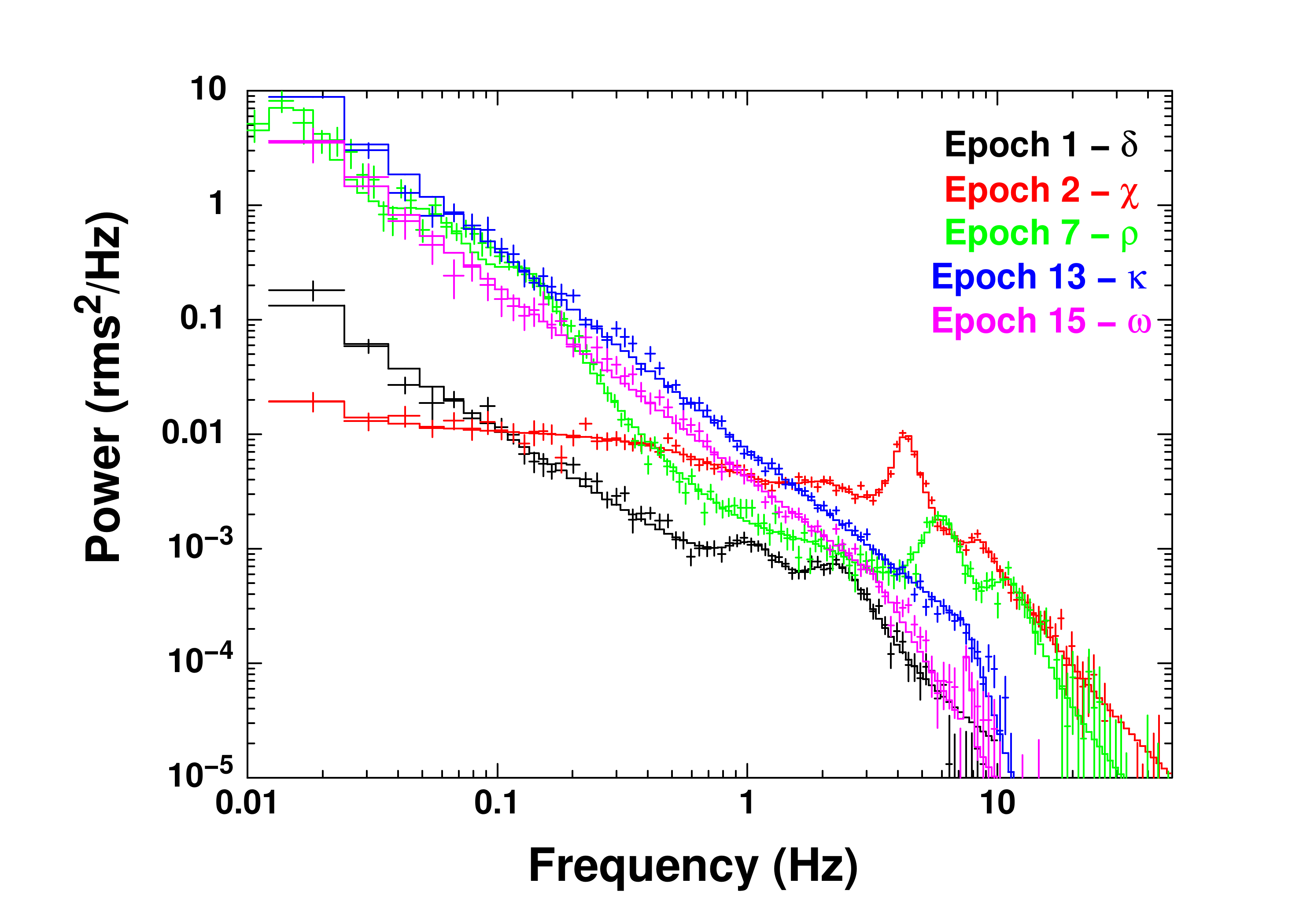}
		\caption{Over-plot of PDS for the `canonical' classes exhibited by the source indicating variation of total rms variability within the frequency range, 0.01 -- 50 Hz. Lower values of rms power is observed for harder $\chi$ class, while the value increases for intermediate and softer classes ($\rho, \kappa, \omega, \delta$) in accordance with the rms-flux relation. See text for details.}
		 \label{fig18}
\end{figure}

The evolutionary trend seen in the `spectro-temporal' features of the source, both during the rise and the decay, is analogous to that observed for any other outbursting black holes (see Figure \ref{fig16}). Yet, the source fails in  displaying a canonical HID (see Figure \ref{fig4}). Another widely known attribute for an outbursting black hole is the total rms variability which is known to decrease as the source makes a transition from hard to soft state and increase as it makes a transition back to hard state. However, as seen in Figure \ref{fig16}, for GRS~1915$+$105 the total rms variability has a different evolution pattern. Here, the total rms is seen to increase as the source evolves from hard to soft states. Significant change in the rms power is evident from Figure \ref{fig18}, where we have shown the PDS for various Epochs corresponding to different classes. While the harder class ($\chi$) have less power, the softer observations ($\rho,\  \kappa \  \& \  \omega$) have high power in low frequencies. The increasing total rms variability during the rising phase (intermediate states), stands in accordance with the linear `total rms-flux' relationship \citep{2001MNRAS.323L..26U} that has been, so far, observed in a few BHs like Cyg X-1 (see also \citealt{2012MNRAS.422.2620H}).

\begin{figure}
    \centering
    \hspace*{-0.8cm}
		\includegraphics[height=9cm, width=7cm,angle=-90]{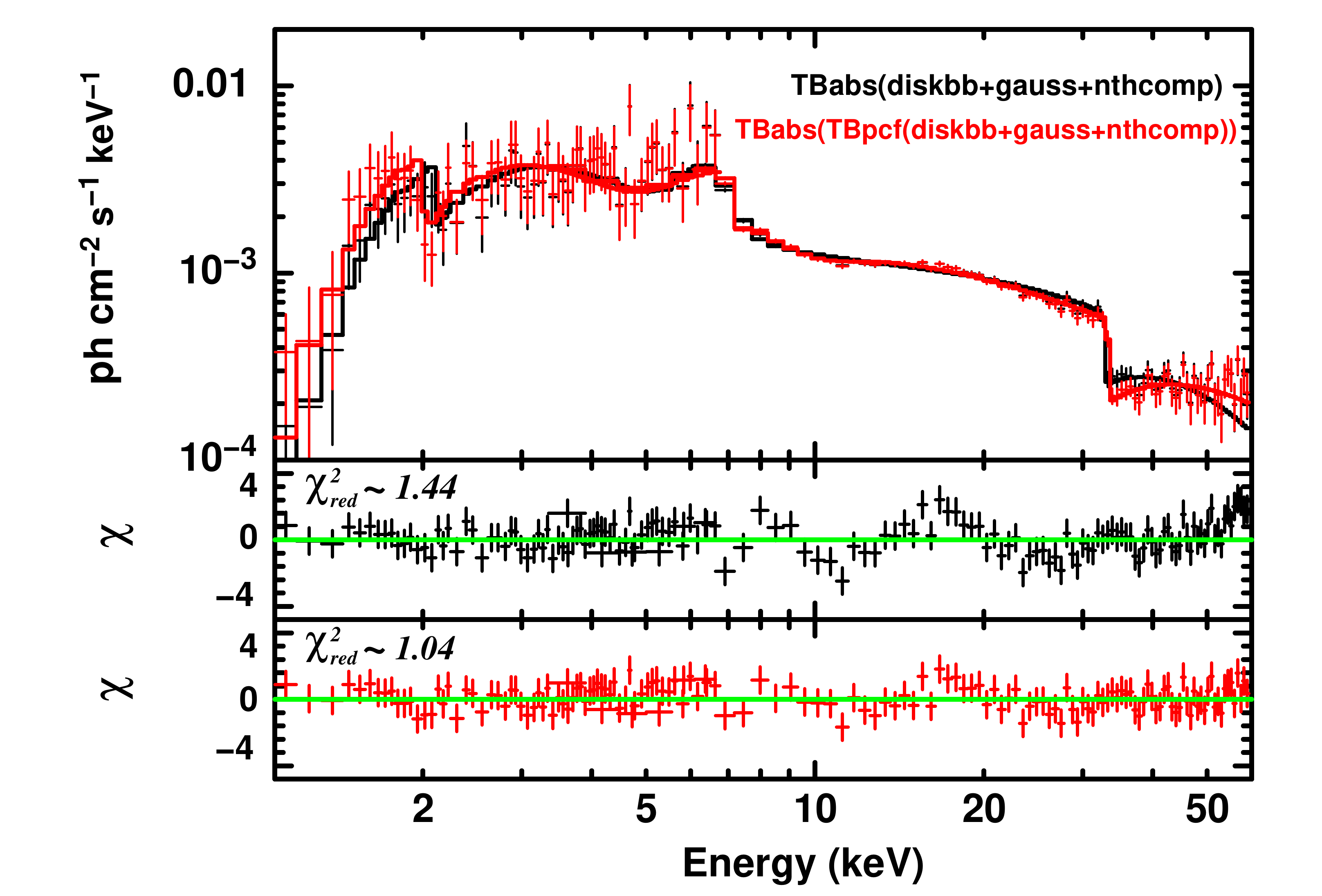}
		\caption{{Over-plot of the broadband energy spectra of the source corresponding to Epoch 31 obtained with and without the partial covering absorption model, \textit{TBpcf}. The corresponding residuals obtained from both model combinations are shown in the middle (without \textit{TBpcf}) and the bottom (with \textit{TBpcf}) panels. The $\chi^{2}_{\rm red}$  values for both cases are also mentioned in the respective panels. See text for details.}}
		 \label{fig19}
\end{figure}

     \begin{table*}
     \vspace*{-0.3cm}  
      \centering
       \caption{Details of model fitted parameters of power spectra obtained from {\it LAXPC} 20 observations of GRS~1915$+$105 in $3 - 60$ keV energy band for each Epoch. Multiple \textit{Lorentzians} were used to fit the PDS. { Here, $\nu$ represents the QPO frequency in \textit{Hz}, $\sigma$ denotes the QPO significance and the} total rms for the entire PDS is given in percentage. }
             \label{tab2}

	\begin{tabular}{lccccccc} 
	 
	 \hline
		      {Epoch} & \multicolumn{4}{c}{Fundamental QPO}  &  Total rms &  {$\chi^{2}$/dof }  \\
		     \cline{2-7}    
	&\textit{v} (Hz) &   Q-factor& rms (\%)& $\sigma$ & (\%)& \\
	\hline
	\hline
1& -&-&-&-&{  {{7.7}$  _{-0.5}^{+0.3}$}}&{ {228.84/214}}\\
2&4.31$_{-0.02}^{+0.01}$& {{  5.1}$_{-0.1}^{+0.4}$}&  {{12.8}$  _{-0.4}^{+0.5}$}& {13.3}& { { {18.9}}$  _{-0.8}^{+0.9}$}&{ {235.61/199}}\\
3$^a$&3.91$_{-0.02}^{+0.02}$& {{  5.9}$_{-0.1}^{+0.2}$}&  {{13.9}$  _{-0.9}^{+0.9}$}& {12.5}&  {{ 19.2}$_{-1.0}^{+1.2}$} &{ {209.51/192}}\\
3$^b$&4.69$_{-0.05}^{+0.05}$& {{  4.3}$_{-0.1}^{+0.3}$}&  {{12.1}$  _{-0.8}^{+0.8}$}& {12.1}&  {{18.4}$  _{-1.2}^{+1.4}$}& {194.75/169}\\
3$^c$&4.06$_{-0.02}^{+0.03}$& {{  6.4}$_{-0.2}^{+0.3}$}&  {{12.9}$  _{-0.9}^{+0.8}$}& {11.2}&  {{18.7}$  _{-1.1}^{+1.0}$}& {441.84/374} \\
3$^d$&4.54$_{-0.05}^{+0.05}$& {{  3.2}$_{-0.1}^{+0.3}$}&  {{11.3}$  _{-1.0}^{+0.9}$}& {16.1}&  {{17.4}$  _{-1.6}^{+1.0}$}& {363.75/293 }\\
4&4.69$_{-0.03}^{+0.03}$& {{  4.0}$_{-0.1}^{+0.4}$}&  {{10.2}$  _{-0.7}^{+0.7}$}& {16.1}&  {{22.0}$  _{-1.7}^{+1.8}$}& {320.70/294} \\
5&4.51$_{-0.08}^{+0.01}$& {{  3.1}$_{-0.1}^{+0.3}$}&  {{5.9}$  _{-0.4}^{+0.4}$}& {10.0}&  {{20.0}$  _{-2.4}^{+1.7}$}& {283.34/216}\\
6$^a$&5.15$_{-0.06}^{+0.03}$& {{  2.8}$_{-0.2}^{+0.3}$}&  {{10.4}$  _{-0.9}^{+0.9}$}& {9.3}&  {{25.1}$  _{-2.4}^{+2.4}$}& {349.08/329}\\
6$^b$&5.12$_{-0.01}^{+0.02}$& {{  2.9}$_{-0.1}^{+0.5}$}&  {{9.9}$  _{-0.8}^{+0.8}$}& {10.6}&  {{31.8}$  _{-2.6}^{+2.8}$}& {509.61/380}\\
7&5.64$_{-0.08}^{+0.09}$& {{  4.3}$_{-0.2}^{+0.6}$}&  {{5.2}$  _{-0.6}^{+0.6}$}& {7.9}&  {{34.6}$  _{-3.1}^{+2.6}$}& {529.59/445} \\
8&-&-&-&-&  {{62.8}$  _{-6.5}^{+4.3}$}& {371.76/307}\\
9$^a$&-&-&-&-&  {{61.9}$  _{-5.6}^{+3.6}$}& {284.86/229}\\
9$^b$&-&-&-&-&  {{63.0}$  _{-4.2}^{+3.4}$}& {267.81/241}\\
10&-&-&-&-&  {{55.5}$  _{-2.9}^{+2.1}$}& {199.17/146}\\
11&-&-&-&-&  {{53.1}$  _{-2.3}^{+1.7}$}& {241.05/200}\\
12&-&-&-&-&  {{53.7}$  _{-2.6}^{+2.2}$}& {169.57/140}\\
13&-&-&-&-&  {{51.4}$  _{-2.4}^{+2.5}$}& {317.97/286}\\
14&-&-&-&-&  {{44.3}$  _{-2.0}^{+1.9}$}& {153.10/140}\\
15&-&-&-&-&  {{30.5}$  _{-1.8}^{+2.0}$}& {131.23/81}\\
16$^a$&-&-&-&-&  {{34.5}$  _{-2.0}^{+1.6}$}& {167.40/130}\\
16$^b$&-&-&-&-& { {16.9}$  _{-1.0}^{+1.2}$}& {121.18/90}\\
16$^c$&-&-&-&-&  {{14.9}$  _{-1.2}^{+1.2}$}& {149.96/111}\\

17&-&-&-&-&  {{6.8}$  _{-0.9}^{+0.9}$}& {144.80/125}\\
18&-&-&-&-&  {{5.3}$  _{-1.8}^{+2.0}$}& {241.21/219}\\
19&-&-&-&-&  {{6.2}$  _{-1.0}^{+0.9}$}& {260.43/198}\\
20&-&-&-&-&  {{5.9}$  _{-0.5}^{+0.7}$}& {189.36/185}\\
21& 3.44$_{-0.01}^{+0.01}$& {{  3.2}$_{-0.1}^{+0.4}$}&  {{12.9}$  _{-0.7}^{+0.8}$}& {14.6}& { {19.1}$  _{-1.7}^{+1.0}$ }& {262.42/228}\\
22&3.80$_{-0.01}^{+0.02}$& {{  4.5}$_{-0.1}^{+0.4}$}&  {{12.6}$  _{-0.7}^{+0.7}$}& {16.9}&  {{21.8}$  _{-1.5}^{+1.3}$ }& {258.30/203}\\
23&2.24$_{-0.01}^{+0.02}$& {{  6.9}$_{-0.1}^{+0.7}$}&  {{14.1}$  _{-0.4}^{+0.4}$}& {10.4}&  {{21.7}$  _{-1.4}^{+1.6}$} & {310.16/276}\\
24&3.04$_{-0.01}^{+0.01}$& {{  5.8}$_{-0.1}^{+0.6}$}&  {{14.4}$  _{-0.8}^{+0.8}$}& {13.9}&  {{20.9}$  _{-1.5}^{+1.5}$}& {194.35/218}\\
25&3.32$_{-0.03}^{+0.02}$& {{  4.2}$_{-0.1}^{+0.4}$}& { {14.1}$  _{-0.6}^{+0.7}$}& {33.8}& { {21.7}$  _{-1.4}^{+1.6}$}& {163.28/168}\\
26&1.85$_{-0.02}^{+0.02}$& {{  10.2}$_{-0.1}^{+0.5}$}&  {{13.2}$  _{-0.6}^{+0.5}$}& {22.0}&  {{22.5}$  _{-2.1}^{+1.8}$} & {236.04/217}\\
27&1.38$_{-0.01}^{+0.01}$& {{  8.8}$_{-0.1}^{+0.1}$}&  {{13.3}$  _{-0.3}^{+0.4}$}& {18.8}&  {{19.3}$  _{-1.2}^{+0.9}$ }& {243.51/217}\\
28&1.83$_{-0.01}^{+0.01}$& {{  8.9}$_{-0.4}^{+0.1}$}&  {{14.0}$  _{-0.3}^{+0.3}$}& {23.8}&  {{22.7}$  _{-0.9}^{+1.0}$} & {262.13/208}\\
29&2.08$_{-0.01}^{+0.01}$& {{  10.7}$_{-1.9}^{+2.0}$}&  {{11.9}$  _{-0.6}^{+0.6}$}& {4.8}&  {{21.4}$  _{-0.9}^{+1.1}$} & {281.06/260}\\
30&-&-&-&-&  {{22.9}$  _{-0.6}^{+0.8}$}& {310.58/250}\\
31&-&-&-&-& { {7.5}$  _{-0.1}^{+0.2}$}& {196.21/183}\\
\hline
\hline
	\end{tabular}
	\end{table*}

\begin{table*}
\vspace{-0.3cm}
    \centering
    
         \caption{Broadband spectral parameter values for GRS~1915$+$105 obtained from the best-fit model for each Epoch.  {$kT_\textrm{in}$} (in units of keV) corresponds to disc temperature, N$_{\textrm{diskbb}}$ is disc normalisation,  {E$_{\rm line}$ corresponds to the Fe line energy (keV)}  {and the line width denotes the width of the gaussian line (keV)},  {$kT_\textrm{e}$} (keV) corresponds to electron temperature, and N$_{\textrm{nthcomp}}$ corresponds to \textit{nthcomp} normalisation. Both disc and Comptonised flux fractions, in \%, for every Epoch are also quoted. {{L$_{\textrm{bol}}$} corresponds to the bolometric luminosity (0.3 -- 100 keV) in the units of Eddington luminosity (\% L$_{\textrm{EDD}}$). The distance to the source is considered as 8.6 kpc.} Errors for all the parameters are calculated with 90\% confidence.}
         \label{tab3}

       \hspace*{-0.8cm}
		\begin{tabular}{lcccccccccccc} 
		\hline
		      {Epoch} & $n_\textrm{H} \times 10^{22}$  &   $\Gamma$ &  $kT_{\textrm{in}}$ & N$_{\textrm{diskbb}}$ & { E$_{\rm line}$} &   Line width &  {$kT_\textrm{e}$} &  N$_{\textrm{nthcomp}}$&  Disc fl. & Comp. fl. & {L$_{\textrm{bol}}$} & {$\chi^{2}$/dof }   \\
		      &{  atoms cm$^{-2}$}&&(keV)&& {(keV)}&(keV)&(keV)&&fr. (\%)&fr. (\%)& {(\% L$_{\textrm{EDD}}$})&\\
		    
	\hline
	\hline
	1& 5.82$_{-0.24}^{+0.37}$&3.62$_{-0.08}^{+0.08}$&2.02$_{-0.01}^{+0.02}$& {{61$_{-7}^{+8}$}}&   {- }& {-}& {5.1$_{-0.2}^{+0.2}$}& 1.44$_{-0.02}^{+0.03}$&   {{55.6$_{-4.2}^{+3.9}$}}&   {{43.1$_{-2.1}^{+3.6}$}}& {16.0}& {{719.5/630}}\\
	
	2& 5.86$_{-0.14}^{+0.18}$&1.87$_{-0.02}^{+0.02}$&1.33$_{-0.01}^{+0.01}$&  {{87$_{-7}^{+7}$}}&   {6.97$_{-0.16}^{+0.12}$} & {0.99$_{-0.23}^{+0.26}$}&  {13.3$_{-0.9}^{+0.8}$}& 0.36$_{-0.01}^{+0.01}$&   {{29.7$_{-1.2}^{+3.5}$}}&   {{71.4$_{-5.1}^{+6.8}$}}& {6.4}& {{559.9/513}}\\
	
	3$^a$& 5.41$_{-0.16}^{+0.18}$&1.83$_{-0.01}^{+0.01}$&1.35$_{-0.01}^{+0.01}$&  {{77$_{-5}^{+6}$}}&  { 7.30$_{-0.17}^{+0.22}$} & {0.7*}& { 12.6$_{-1.3}^{+1.6}$}& 0.32$_{-0.01}^{+0.02}$&   {{21.3$_{-2.6}^{+4.1}$}}&    {{77.5$_{-6.5}^{+6.3}$}}& {6.5}& {{565.4/472}}\\
	
	3$^b$& 5.38$_{-0.10}^{+0.16}$&1.83$_{-0.02}^{+0.02}$&1.37$_{-0.01}^{+0.01}$&  {{81$_{-6}^{+7}$}}&   {6.84$_{-0.21}^{+0.26}$} & {0.7*}&  {11.6$_{-1.1}^{+1.1}$}& 0.30$_{-0.01}^{+0.01}$&   {{23.1$_{-2.3}^{+3.2}$}}&   {{75.2$_{-5.4}^{+7.2}$}}& {6.1} & {{582.5/476}}\\
	
	3$^c$& 5.46$_{-0.24}^{+0.11}$&1.94$_{-0.02}^{+0.01}$&1.25$_{-0.02}^{+0.01}$&  {{80$_{-4}^{+3}$}}&   {6.18$_{-0.28}^{+0.27}$ }& {0.7*}&  {13.1$_{-2.1}^{+1.9}$}& 0.50$_{-0.09}^{+0.06}$&   {{28.2$_{-2.1}^{+4.8}$}}&   {{71.1$_{-6.1}^{+5.0}$}}& {6.4 }& {{548.7/538}}\\
	
		3$^d$& 5.35$_{-0.19}^{+0.31}$&1.95$_{-0.01}^{+0.01}$&1.37$_{-0.02}^{+0.01}$&  {{95$_{-5}^{+7}$}}&  { 6.84$_{-0.34}^{+0.28}$} & {0.91$_{-0.20}^{+0.21}$}&  {33.4$_{-3.5}^{+4.7}$}& 0.35$_{-0.01}^{+0.01}$&   {{36.9$_{-3.2}^{+4.1}$}}&   {{63.8$_{-4.2}^{+5.7}$}}& {6.1} & {{562.2/552}}\\
	
	4&5.46$_{-0.13}^{+0.13}$&1.87$_{-0.03}^{+0.03}$&1.31$_{-0.03}^{+0.02}$&  {{107$_{-13}^{+15}$}}&   {7.16$_{-0.29}^{+0.25}$} & {0.7*}& {13.4$_{-1.1}^{+1.5}$}& 0.29$_{-0.02}^{+0.03}$&   {{37.2$_{-3.1}^{+2.6}$}}&  { {62.9$_{-4.9}^{+5.3}$}}&{5.8}& {{612.1/536}}\\
	
	5& 5.41$_{-0.21}^{+0.24}$& 1.89$_{-0.02}^{+0.02}$& 1.28$_{-0.03}^{+0.03}$&  {{118$_{-11}^{+13}$}}&   {7.16$_{-0.35}^{+0.34}$}& {1.02$_{-0.35}^{+0.38}$} & {13.7$_{-0.9}^{+1.8}$}& 0.31$_{-0.01}^{+0.01}$&   {{36.1$_{-3.0}^{+3.1}$}}&   {{63.6$_{-8.7}^{+8.2}$}}& {5.8}&{{525.1/525}}\\
	
6$^a$&5.55$_{-0.46}^{+0.31}$&1.95$_{-0.04}^{+0.04}$&1.32$_{-0.02}^{+0.02}$&  {{87$_{-4}^{+3}$}}&   {-} & {-}&  {18.7$_{-1.4}^{+2.4}$}& 0.26$_{-0.01}^{+0.01}$&   {{38.6$_{-1.6}^{+2.8}$}}&   {{61.9$_{-4.7}^{+6.1}$}}& {4.7}& {{635.6/528}}\\

6$^b$&5.42$_{-0.37}^{+0.28}$&1.87$_{-0.04}^{+0.04}$&1.38$_{-0.03}^{+0.03}$&  {{83$_{-6}^{+5}$}}&  { -} & {-}& {15.6$_{-2.4}^{+3.5}$}& 0.21$_{-0.01}^{+0.01}$&   {{42.5$_{-2.8}^{+3.8}$}}&   {{57.2$_{-3.2}^{+6.1}$}}&{4.8} & {392.8/384}\\

7&5.76$_{-0.33}^{+0.10}$&1.90$_{-0.05}^{+0.04}$&1.37$_{-0.01}^{+0.02}$&  {{98$_{-6}^{+5}$}}&   {-} & {-}&  {14.9$_{-1.9}^{+2.1}$} &0.19$\dagger$&  {{51.0$_{-2.6}^{+4.5}$}}&  {{51.0$_{-2.1}^{+3.8}$}}& {4.7}&{{730.5/542}}\\

  8& 5.79$_{-0.12}^{+0.11}$&2.55$_{-0.09}^{+0.08}$&1.80$_{-0.01}^{+0.02}$&   {{52$_{-3}^{+4}$}}&   {-}& {-} & {16.8$_{-2.0}^{+2.8}$}& 0.24$_{-0.01}^{+0.01}$&   {{66.9$_{-4.8}^{+3.1}$}}&   {{33.1$_{-3.1}^{+2.0}$}}& {6.5}& {{555.3/502}}\\
   
9$^a$& 5.82$_{-0.56}^{+0.79}$&2.36$_{-0.05}^{+0.04}$&1.72$_{-0.01}^{+0.01}$&   {{57$_{-5}^{+9}$}}&  {-} & {-}&  {4.3$_{-0.1}^{+0.1}$}& 0.32$_{-0.01}^{+0.01}$&   {{64.7$_{-3.1}^{+6.4}$}}&   {{34.6$_{-2.3}^{+3.6}$}}& {7.2}& {{624.3/508}}\\
    	
9$^b$& 6.04$_{-0.12}^{+0.10}$&2.74$_{-0.05}^{+0.04}$&1.66$_{-0.01}^{+0.01}$&   {{66$_{-4}^{+6}$}}&  { -} & {-} & {13.6$_{-2.2}^{+1.6}$}& 0.47$_{-0.01}^{+0.01}$&   {{58.8$_{-4.2}^{+5.3}$}}&   {{42.6$_{-4.3}^{+5.6}$}}&{ 7.1} & {{818.6/573}}\\

10& 5.69$_{-0.11}^{+0.18}$& 3.55$_{-0.02}^{+0.02}$&1.87$_{-0.01}^{+0.01}$&  {{29$_{-3}^{+4}$}}&   {- }& {-}&  {8.8$_{-0.9}^{+1.4}$} &0.80$_{-0.02}^{+0.02}$&   {{40.2$_{-4.1}^{+4.2}$}}&   {{59$_{-5.5}^{+3.7}$}}& {7.6}&{{765.1/568}}\\

11&6.36$_{-0.12}^{+0.18}$&3.80$_{-0.03}^{+0.03}$ &1.82$_{-0.01}^{+0.01}$&  {{24$_{-2}^{+4}$}}&   {-} & {-}&  {11.6$_{-0.9}^{+1.8}$} &1.32$_{-0.04}^{+0.03}$ & { {31.4$_{-2.1}^{+2.2}$}}&   {{68.2$_{-3.9}^{+6.0}$}}& {8.2 }& {{662.6/578}}\\

12&6.26$_{-0.14}^{+0.17}$&2.34$_{-0.01}^{+0.01}$&1.58$_{-0.01}^{+0.01}$& { {48$_{-3}^{+4}$}}&   {-} & {-}&  {3.2$_{-0.1}^{+0.1}$}& 0.26$_{-0.01}^{+0.01}$&  { {30.8$_{-2.1}^{+2.8}$}}&  { {68.4$_{-3.4}^{+5.2}$}}& {9.9}&{{629.4/527}}\\

13&6.44$_{-0.40}^{+0.43}$&3.31$_{-0.05}^{+0.05}$&2.49$_{-0.01}^{+0.01}$& { {21$_{-3}^{+4}$}}&  { -} & {-}&  {7.8$_{-1.0}^{+1.7}$}& 0.64$_{-0.01}^{+0.01}$&  { {54.6$_{-0.9}^{+1.9}$}}&  { {45.1$_{-5.5}^{+3.8}$}}& {12.0}& {{665.4/574}}\\

 14&6.17$_{-0.25}^{+0.20}$&2.98$_{-0.04}^{+0.04}$&2.5$_{-0.01}^{+0.01}$& { {28$_{-4}^{+6}$}}&  { - }& {-}&  {8.4$_{-1.0}^{+1.0}$}& 0.60$_{-0.03}^{+0.02}$&  { {60.5$_{-4.9}^{+5.9}$}}&  { {39.3$_{-1.4}^{+1.5}$}}& {15.5}&{{767.6/620}}\\
 
    15&6.09$_{-0.37}^{+0.21}$&2.87$_{-0.04}^{+0.04}$&2.67$_{-0.01}^{+0.01}$& { {24$_{-3}^{+3}$}}&   {-} & {-}&  {6.3$_{-0.1}^{+0.2}$}& 0.57$_{-0.02}^{+0.02}$&  { {74.1$_{-6.8}^{+6.0}$}}&  { {26.2$_{-1.4}^{+1.7}$}}& {16.2}&{{910.6/660}}\\
    
    16$^a$&6.60$_{-0.12}^{+0.11}$&2.33$_{-0.06}^{+0.06}$&2.32$_{-0.01}^{+0.03}$& { {51$_{-4}^{+6}$}}&   {-}& {-} &  {5.0$_{-0.1}^{+0.4}$}& 0.43$_{-0.02}^{+0.01}$&  { {71.8$_{-6.4}^{+8.2}$}}&  { {28.0$_{-1.8}^{+1.9}$}}& {16.7}&{{591.9/502}}\\
    
    16$^b$&6.56$_{-0.51}^{+0.69}$&3.39$_{-0.04}^{+0.05}$&2.61$_{-0.02}^{+0.02}$& { {31$_{-3}^{+4}$}}&  { -} & {-}&  {13.7$_{-0.7}^{+0.7}$}& 0.78$_{-0.04}^{+0.04}$&   {{71.0$_{-8.1}^{+5.3}$}}&  { {29.3$_{-3.1}^{+3.9}$}}&{17.1} & {{881.1/659}}\\
    
    16$^c$&6.56$_{-0.40}^{+0.68}$&3.13$_{-0.07}^{+0.06}$&2.61$_{-0.02}^{+0.02}$& { {30$_{-4}^{+4}$}}&  { -} & {-}&  {8.3$_{-0.3}^{+0.6}$}& 0.66$_{-0.01}^{+0.01}$&  { {74.6$_{-5.3}^{+4.2}$}}&  { {27.7$_{-6.2}^{+3.1}$}}& {16.9} &{{752.2/590}}\\
    
 17& 6.57$_{-0.15}^{+0.14}$&2.98$_{-0.06}^{+0.06}$&1.79$_{-0.01}^{+0.01}$& { {73$_{-5}^{+7}$}}&   {-} & {-}&  {2.7$_{-0.1}^{+0.1}$}& 3.99$_{-0.09}^{+0.09}$&  { {34.5$_{-2.0}^{+2.1}$}}&  { {66.0$_{-3.7}^{+4.8}$}}& {35.7} & {853.6/625}\\
 
 18&6.38$_{-0.12}^{+0.12}$&3.07$_{-0.03}^{+0.03}$&1.89$_{-0.01}^{+0.01}$& { {61$_{-5}^{+5}$}}&  { -} & {-}&  {2.7$_{-0.1}^{+0.1}$}& 4.12$_{-0.01}^{+0.04}$&  { {22.0$_{-1.0}^{+1.9}$}}&  { {68.1$_{-4.4}^{+3.4}$}}& {32.4}&{{699.0/551}}\\
 
 19&6.50$_{-0.13}^{+0.11}$&3.06$_{-0.02}^{+0.02}$&1.54$_{-0.01}^{+0.01}$& { {44$_{-3}^{+6}$}}&   {-} & {-}&  {3.1$_{-0.2}^{+0.2}$}& 4.28$_{-0.16}^{+0.09}$&  { {44.6$_{-2.6}^{+4.6}$}}&  { {64.2$_{-4.2}^{+5.2}$}}& {23.8}& {{860.5/649}}\\
 
 20&6.08$_{-0.21}^{+0.19}$&3.18$_{-0.01}^{+0.01}$&1.31$_{-0.01}^{+0.01}$& { {67$_{-5}^{+6}$}}&   {-} & {-}&  {3.5$_{-0.2}^{+0.5}$}& 2.65$_{-0.07}^{+0.05}$&  { {39.7$_{-1.3}^{+3.6}$}}&  { {61.1$_{-4.5}^{+5.3}$}}& {21.9}&{{714.8/577}}\\
 
 21&5.20$_{-0.13}^{+0.13}$&1.97$_{-0.01}^{+0.02}$&1.44$_{-0.01}^{+0.01}$& { {593$_{-24}^{+28}$}} &   {-} & {-}& { 15.2$_{-1.0}^{+1.8}$} &1.03$_{-0.05}^{+0.05}$ & { {39.7$_{-3.3}^{+2.8}$}} & { {61.2$_{-4.9}^{+6.7}$}} & {12.7} & {{626.9/573}}\\
 
 22&5.10$_{-0.18}^{+0.20}$&1.87$_{-0.01}^{+0.01}$&1.44$_{-0.01}^{+0.01}$ & { {117$_{-20}^{+25}$}}&   {8.04$_{-0.42}^{+0.37}$} & {0.7*}&  {11.5$_{-1.0}^{+2.0}$} &0.55$_{-0.02}^{+0.02}$&  { {35.2$_{-1.2}^{+1.6}$}}&  { {65.4$_{-5.3}^{+4.5}$}}& {10.9}&  {{1029.7/880}}\\
 
 23&4.26$_{-0.09}^{+0.11}$&1.73$_{-0.03}^{+0.03}$&1.48$_{-0.01}^{+0.01}$ & { {33$_{-2}^{+4}$}}&   {7.02$_{-0.36}^{+0.33}$} & {0.82$_{-0.24}^{+0.28}$}&  {14.2$_{-1.5}^{+2.1}$} &0.25$_{-0.01}^{+0.01}$& { {22.3$_{-1.3}^{+2.8}$}}& { {77.6$_{-6.4}^{+5.4}$}}& {6.4}& {{537.0/556}}\\

 24&4.55$_{-0.11}^{+0.11}$&1.79$_{-0.04}^{+0.04}$&1.37$_{-0.01}^{+0.01}$ & { {46$_{-3}^{+5}$}}&   {6.89$_{-0.21}^{+0.19}$} & {0.74$_{-0.29}^{+0.32}$}&  {16.4$_{-1.1}^{+2.5}$ }&0.24$_{-0.01}^{+0.01}$& { {26.1$_{-2.8}^{+3.6}$}}& { {74.7$_{-4.5}^{+5.5}$}}& {4.7}&{{536.5/539}}\\

 25&4.56$_{-0.20}^{+0.16}$&1.79$_{-0.03}^{+0.04}$ & 1.36$_{-0.01}^{+0.02}$ & { {52$_{-3}^{+4}$}} &   {6.84$_{-0.18}^{+0.20}$} & {0.78$_{-0.28}^{+0.35}$}&  {15.1$_{-1.2}^{+1.5}$}& 0.22$_{-0.01}^{+0.01}$&  { {25.2$_{-3.9}^{+1.1}$}}&  { {74.7$_{-5.1}^{+6.0}$}}& {4.6}&  {{516.9/493}}\\

 26&4.63$_{-0.14}^{+0.14}$&1.66$_{-0.02}^{+0.02}$ &1.36$_{-0.02}^{+0.02}$&  { {21$_{-1}^{+2}$}}&   {7.02$_{-0.26}^{+0.24}$} & {0.82$_{-0.303}^{+0.27}$}&  {55.4$_{-4.1}^{+6.7}$}&0.12$\dagger$& { {27.5$_{-3.2}^{+4.9}$}}& { {73.9$_{-3.8}^{+5.6}$}}& {3.1}& {{442.4/438}}\\

 27&4.61$_{-0.13}^{+0.14}$&1.62$_{-0.08}^{+0.08}$&1.32$_{-0.02}^{+0.03}$& { {18$_{-1}^{+2}$}}&   {7.05$_{-0.14}^{+0.16}$} & {0.96$_{-0.25}^{+0.27}$}&  {28.1$_{-3.5}^{+2.4}$}&0.10$\dagger$& { {26.7$_{-2.0}^{+4.1}$}}& { {74.1$_{-7.7}^{+7.1}$}}& {2.8}&{{400.2/369}}\\

 28&5.01$_{-0.65}^{+0.51}$&1.62$_{-0.02}^{+0.01}$&1.23$_{-0.05}^{+0.04}$& { {27$_{-1}^{+3}$}}&   {7.13$_{-0.20}^{+0.22}$} & {1.12$_{-0.22}^{+0.24}$}&  {13.5$_{-2.0}^{+1.8}$}&0.10$\dagger$& { {26.5$_{-1.2}^{+2.6}$}}& { {73.9$_{-1.1}^{+1.6}$}}& {2.7}& {{94.1/89}}\\
 
 29&5.13$_{-0.21}^{+0.23}$&1.60$_{-0.03}^{+0.03}$&1.36$_{-0.02}^{+0.02}$& { {21$_{-2}^{+3}$}}&   {6.88$_{-0.15}^{+0.18}$} & {0.94$_{-0.33}^{+0.37}$}&  {11.6$_{-0.4}^{+0.4}$}& 0.08$\dagger$ & { {15.7$_{-2.0}^{+2.4}$}}&  { {84.2$_{-4.2}^{+3.9}$}}& {2.4} & {{445.6/473}}\\
 
 30&3.59$_{-0.11}^{+0.13}$&1.28$_{-0.01}^{+0.01}$&1.11$_{-0.01}^{+0.01}$& { {10$_{-1}^{+2}$}}&   {6.46$_{-0.06}^{+0.06}$} & {0.6*}&  {11.6$_{-1.2}^{+1.8}$}& 0.04$\dagger$& { {13.0$_{-3.4}^{+2.6}$}}& { {83.8$_{-6.3}^{+10.0}$}}& {0.7} & {{99.8/90}}\\
 
 31&6.15$_{-0.24}^{+0.21}$&1.18$_{-0.08}^{+0.07}$&1.18$_{-0.01}^{+0.02}$& { {5$_{-1}^{+1}$}}&   {6.34$_{-0.06}^{+0.07}$} & {0.61$_{-0.09}^{+0.09}$}&  {11.4$_{-1.7}^{+2.5}$}& 0.01$\dagger$& { {10.1$_{-3.3}^{+2.0}$}}& {{89.4$_{-7.2}^{+8.8}$}}& {1.1} &{197.6/137}\\
\hline
\hline
	\end{tabular}%

\begin{flushleft}
      * Frozen ;  $\dagger$ Error values are insignificant 
\end{flushleft}

      \end{table*}

\begin{table*}
\centering
\caption{{Best-fit broadband spectral parameters corresponding to Epochs 30 and 31 obtained by using the model combination \textit{TBabs(TBpcf(diskbb+gauss+nthcomp)}, where  \textit{TBpcf} is the partial covering absorption model along the continuum. The $n_\textrm{H}$ ($\times 10^{22}$  atoms cm$^{-2}$) indicates the local absorption value; the PCF denotes the local partial covering fraction. The other parameters in this table are defined in the caption of Table \ref{tab3}. Errors for all the parameters are calculated with 90\% confidence. } }
         \label{tab4}
\begin{tabular}{lccccccccccccc}

 \hline        
{Epoch} & {$n_\textrm{H} \times 10^{22}$} & {PCF}  &   {$\Gamma$} &  {$kT_{\textrm{in}}$} &{ N$_{\textrm{diskbb}}$} &{{E$_{\rm line}$}} &   {Line width} &  {{$kT_\textrm{e}$}} &  {N$_{\textrm{nthcomp}}$} & {L$_{\textrm{bol}}$} &{$\chi^{2}$/dof} &  \\
&  {atoms cm$^{-2}$} & &&{(keV)}&& {{(keV)}}&{(keV)}&{(keV)}&& {(\% L$_{\textrm{EDD}}$)}& \\
\hline\hline
{30}     &  { 45.2$_{-4.6}^{+8.1}$}   &{ 0.13$_{-0.02}^{+0.02}$} & {1.13$_{-0.06}^{+0.06}$} &  {1.32$_{-0.04}^{+0.05}$} &  {3.2$_{-0.7}^{+0.7}$} &{6.47$_{-0.14}^{+0.21}$} &{0.6*}&   {9.5$_{-0.8}^{+1.2}$} & {0.01} {$\dagger$ }& {0.6} &{79.6/87}  \\
{31 }    &   {108.0$_{-18.3}^{+11.7}$}   & {0.41$_{-0.01}^{+0.01}$}    & {1.21$_{-0.05}^{+0.06}$} &  {1.09$_{-0.07}^{+0.09}$} &  {7.6$_{-1.2}^{+0.9}$} &{6.63$_{-0.31}^{+0.29}$} &{0.90$_{-0.07}^{+0.05}$}&   {16.6$_{-1.4}^{+1.7}$} &{ 0.01 $\dagger$} &{ 0.8} &{139.7/134 }\\
 
\hline\hline
\end{tabular}

\begin{flushleft}
      * Frozen ;  $\dagger$ Error values are insignificant 
\end{flushleft}

\end{table*}

 \vspace{-0.6cm}  
 \section{Discussions and Conclusions}
\label{sec5}
\vspace{-0.1cm}

 In this paper, we perform an in-depth analysis of 31 \textit{AstroSat} observations of GRS~1915$+$105 during the period of November 2016 to June 2019 by studying the broadband `spectro-temporal' features of the source. Based on the long term variability of the source using \textit{MAXI} light curve (see Figure~\ref{fig1}), we delineate the evolution track of the source. The source is found making large amplitude {variations} between MJD 57700 to 57760, where the X-ray intensity of the source is relatively high. The \textit{AstroSat} observation of the source during MJD 57704 (Epoch~1) asserts the same. The flux of the source shows a slow increase during MJD 57800 to 58050, which corresponds to Epochs~2 to 17, followed by a {highly varying} phase (Epochs 17 -- 20), and finally a gradual decrease beyond MJD 58200 (Epochs 21 -- 31). The X-ray intensity variations since MJD 57760 resembled the light curve of a `canonical' outbursting BH, exhibiting a slow-rise and slow-decay profile \citep{2010A&A...520A..98D}. It has to be noted that the rise time ($\sim$300 days) and decay time ($\sim$200 days) of the light curve for this phase of GRS~1915$+$105, are longer in comparison to  {the} canonical sources (except GX~339{--}4; \citealt{2005A&A...440..207B, 2012A&A...542A..56N}). 

Although the HR values exhibit a decrease during high intensity state and an increase in the low intensity state, there appears {a  {patternless} variation} in the overall HR track and the typical HR variation seen during an outburst is not observed. This resulted in the HID not achieving a complete `q'-shape or hysteresis (see also \citealt{2016MNRAS.462.1834R, 2020MNRAS.497.1197B, 2021MNRAS.tmp.2464B}), unlike canonical sources (see Figure~\ref{fig4}). Previous studies of the HID of GRS~1915$+$105, performed by \cite{1997ApJ...477L..41C} using {\it RXTE} observations only during the peak (May - October 1996), also did not detect hysteresis. A mechanism for hysteresis is postulated by \cite{2014ApJ...782L..18B}, where Magnetic Rotational Instability (MRI) drives the viscosity parameter ($\alpha$) to 1,  {resulting in} L $\sim$ L{$_{\textrm{EDD}}$} during the peak. This MRI strength is augmented during  {the} quiescence by the net fields. This field diffuses during the outburst and hence leads to the drop in the luminosity, which is seen as a hysteresis. In the present scenario, GRS~1915$+$105 displays a maximum { bolometric luminosity of $\sim$36\% L{$_{\textrm{EDD}}$}} during the peak. We refer to the aforementioned phenomena and presume a possibility as to why the hysteresis is not observed during the {\it AstroSat} observation period. Only in the event of quiescence, MRI is built up which further leads to L{$_{\textrm{EDD}}$} during the outburst.  {Since} GRS~1915$+$105 has never attained quiescence,  {it has not been driven to} L{$_{\textrm{EDD}}$} in this cycle. Thus, we do not observe any evidence of hysteresis in the HID during this period.

Throughout the 31 Epochs, the source is seen to exhibit various structured and non-structured variability classes. Referring to the light curve and CCD profiles as shown in Section \ref{4.2} and with reference to \cite{2000A&A...355..271B}, we deduce that the source was initially in a $\delta$ class during the peak, followed by a sequence of class transitions: $\chi \rightarrow \rho \rightarrow \kappa \rightarrow \omega {  \rightarrow \gamma} \rightarrow \delta$ in the rising phase, and $\delta \rightarrow \chi$ in { declining phase of the three year observation period considered}. During the $\delta$ class, the source has a softer energy spectrum and  {less }electron temperature, along with  {the }absence of QPOs (see Figure \ref{fig5}). Similar characteristics, based on {\it AstroSat} observations, have been reported by \cite{2020MNRAS.499.5891S}. These are indicative of a disc dominated emission, where the Keplerian disc is closer to the central object. During the low flux phase of the light curve, the source exhibits $\chi$ variability class for a few segments during Epochs~2 and 3, where the energy spectra are harder with contribution of hard photons till 60 keV and a dominating Comptonised flux { (see Figure \ref{fig6} \& Table \ref{tab3}).} These Epochs also exhibit relatively lower photon index and higher total rms variability with strong Type-C LFQPOs in the PDS (see also \cite{2019ApJ...870....4R, 2020ApJ...889L..36M}). During the `heart-beat' $\rho$ class, the source exhibited periodic single burst peaks with multiplicity 1 as referred by \cite{1999ApJ...517..935Y,2010A&A...513A..21M}. The burst profiles have varying peak intensity (see Figure \ref{fig7}), in contrast to previous observations  {with} {\it RXTE} \citep{2000A&A...355..271B}. The  burst sequence is interpreted as an emptying/refilling cycle of the inner portion of the accretion disc. In our observations, we notice the gradual decrease in the recurrence time of the pulse accompanied by an increase in the burst strength over time (see Figure \ref{fig11}). This observation also stands compatible with the anti-correlation between recurrence time and amplitude of the burst as proposed by \cite{2018ApJ...865...19W}. 

During transition to $\kappa$ class, a gradual increase is observed in $\Gamma$ from {2.5} to {3.8} and {$kT_\textrm{in}$} from {1.8} keV to {2.4} keV (see Table \ref{tab3}). The disc flux, however, does not show any increase, whereas an increase in the Comptonised flux is noticed. The additional {\it powerlaw} coupled with the Comptonising corona model, indicates the presence of an extended corona above the disc (see also \cite{2010MNRAS.409..903P,2020MNRAS.499.5891S} for a similar reasoning). It is assumed that the soft photons from the disc are Comptonised from this extended corona. These are suggestive of an intermediate state where strong wind outflows are generated with a possible re-entry of matter into the Corona \citep{2013AdSpR..52..740P}. During Epochs~14 { and} 15, where the source displayed $\omega$ variability class \citep{2002MNRAS.331..745K}, the light curve showed aperiodic dip (low) and non-dip (high) regions (see Figure \ref{fig9}). A similar pattern was observed during the long-term observation study by \cite{2002JApA...23..213N,2010MNRAS.409..903P}, where the source is seen to be switching between high intensity and low intensity states and finally the variability disappears and settles down at a high intensity state. In this manuscript, we do not proceed with the detailed behaviour of the source during the low intensity and high intensity states. But an overall analysis of the source during this period shows an increasing disc flux contribution (see also \citealt{2005MNRAS.360..825Z}), from {60 -- 74\%}, along with a relative decrease in {{{$kT_\textrm{e}$}}} from  {{8.4$_{-1.0}^{+1.0}$ to 6.3$_{-0.1}^{+0.2}$}} keV. No narrow QPOs were observed during this period{,} similar to \cite{2002JApA...23..213N}. The $\chi$ class appears once again during the declining phase of the source flux from Epoch~21 to 31. This is supported by a constant decrease in the $\Gamma$ and {$kT_\textrm{in}$} values and a corresponding increase in the Comptonised flux (see Figure \ref{fig16} and Table \ref{tab3}). This refers to a very low Keplerian rate with a larger Comptonised corona resulting in a harder spectra \citep{2013AdSpR..52..740P}. Very strong QPOs with harmonics have also been observed, and the fundamental frequency of the QPO decreases during the decline phase (see Table \ref{tab2} { and} Figure \ref{fig16}). {A positive correlation is seen between the QPO frequency and rms amplitude of the QPO till 2.24 Hz, and beyond that an anti-correlation is seen between the two (see Figure \ref{fig17}). \citet{2021MNRAS.503.5522K}  {give} a schematic picture of the evolution of the corona and the inner radius of the disc of the source, corresponding to the change in the behaviour of QPO rms-frequency around $\sim$1.8 Hz. Their analysis reveals that, as the QPO frequency decreases from 5 to 2 Hz, the size of the corona decreases and at a critical frequency $\sim$1.8 Hz and below that, the size of corona begins to increase again. They however do not explain the mechanism that triggers this switch in behaviour.}

We also witness class transitions in the source (Section~\ref{sec4.3}) within an Epoch itself. These transitions were accompanied by evident change in the light curve, CCD, spectral and temporal parameters. Interestingly, during the transition of the source from $\chi \longrightarrow \rho$ through an intermediate state (Figure \ref{fig10}), the source returns to a $\chi$ class before proceeding to the intermediate state once again and settling at the `heart-beat' state. In addition, the `heart-beat' profile is broad and less intense during these Epochs, and not similar to what has been observed previously for this source \citep{2000A&A...355..271B}. It has to be noted that \cite{2019ApJ...870....4R, 2020ApJ...889L..36M} had also reported on the transition from $\chi \longrightarrow \rho$. For comparison, the segment which they have quoted as intermediate state (segment-2 of Epoch~3 for an example) has similar `spectro-temporal' characteristics as that of a `heart-beat' class (see Section~\ref{sec4.3.1}; last segment of Epoch~3 in Figure~\ref{fig10}). From our analysis based on the {\it AstroSat} observations, the burst profile of the `heart-beat' state ($\rho^\prime$) is seen pursuing evolution over a few days time (see Section \ref{sec4.3.2}). The profiles have significant variation in the source intensity, skewness of the flares (Figure \ref{fig11}) and varying periodicity indicated by an increase in frequency of mHz QPOs from 5.8 to 14.1 mHz (Figure \ref{fig12}; see also \citealt{2021MNRAS.501.6123K} on the report of mHz QPO during Epoch 2). Detection of these characteristics are unique and novel, and not been discussed till date for any observations of the source. 

An unusual characteristic is exhibited in the light curve and CCD as the source transits from $\rho$ to $\kappa$ class, where a random appearance of dip {  and} flares are observed, resulting in an `unknown' class (see Section \ref{sec4.3.3}; Figure \ref{fig13}). Since {\it AstroSat} did not observe the source continuously, it is difficult to comment on the exact Epoch which tracks the beginning of this feature and how it evolved until the source reached a complete $\kappa$ class.  In review of the transition from $\omega \longrightarrow \gamma$, an earlier study by \cite{2019MNRAS.489.1037B} had reported about transition of the source from $\omega$ to $\mu$ class. Our in-depth analysis of the light curve, CCD and the PDS showing the presence of mHz QPOs (see Figure \ref{fig15}), prove the transition to be from $\omega$ to $\gamma$ and a combination of both classes during the final segment of the observation (see Figure \ref{fig14}). A comparative study with {\it RXTE-PCA} observations corroborate the result (see Section \ref{sec4.3.4} {  and} Figure \ref{fig15}). All such transitions occurring within a few hours duration are very rare in this source.

Previous studies by \cite{2005A&A...431..825C} using {the Indian X-ray Astronomy Experiment ({\it IXAE})} and {\it RXTE} observations have shown {transitions} from $\kappa \rightarrow \rho$, $\chi \rightarrow \rho$, $\chi \rightarrow \theta$ and $\rho \rightarrow \alpha$. \cite{2002MNRAS.330..487N} further deduced that a transition from $\chi$ to $\rho$ state happens through an intermediate class $\alpha$. However, we note that even though we observe multiple intermediate states during the transition $\chi \rightarrow \rho$, they do not resemble that of an $\alpha$ class. {From our studies, we observe a transition from $\chi \longrightarrow \rho^\prime$ within a timescale of $\sim$2.1 ks. The transition from $\omega$ to $\gamma$ variability classes in  $\sim$2.2 ks, and the emergence of a combination of both $\omega$ and $\gamma$ variability classes after a {  time-span} of $\sim$10.2 ks was also observed. Initial studies  {account} these timescales to be associated with the time to fill up the inner disc, on a viscous timescale \citep{1997ApJ...488L.109B}. However, \citet{2005A&A...431..825C}  {state }that these timescales are smaller in comparison to the viscous time for the keplerian disc of this source and therefore conclude that the accretion flow is sub-keplerian in nature and is originating close to the compact object. Another study by \citet{2001NewAR..45..449L} states that although the thermal timescale is much shorter than viscous timescale in an equilibrium disc, during a local disc instability, both timescales are comparable. In conjunction with  \citet{2001NewAR..45..449L} and \citet{2006MNRAS.368..379M}, we speculate that the timescales for the transitions observed in this source  {correspond} to the local disc instabilities induced by the radiation pressure domination within the inner disc, at thermal-viscous timescales.}

Although there have been many efforts in explaining the features and evolution of the source associated to a certain class, there are only a few attempts to explain the physical processes which trigger a specific class transition. \cite{2004ApJ...609..313M, 2020MNRAS.492.4033A, 2020MNRAS.496.1697M} use a Non-Linear Time Series (NLTS) analysis to explain the various types of long-term variability in terms of deterministic non-linear system with inherent stochastic noise. They attempt to derive the global equations that govern the temporal behaviour of the system. Alternatively,  \cite{2000astro.ph.12526C, 2005A&A...431..825C} consider the advective disc paradigm of black hole accretion to study the variation of light curve of the source. It is postulated that the propagation of the  {source characteristics} through three fundamental states, which primarily differ by the varied Keplerian and sub-Keplerian accretion rates, accounts for all the variabilities observed in the light curves (see also \citealt{2001MNRAS.324..267N}).

Owing to \textit{AstroSat}'s broadband spectral and temporal coverage, the evolution of `spectro-temporal' features of the source during the different Epochs could be thoroughly studied. {  In the} PDS during different Epochs in the rising phase, as the source evolves from hard state to intermediate states, {Type-C QPOs are seen evolving into a broader feature with a monotonic increase in the QPO frequency (see Figure \ref{fig16})}.  The noise associated with the PDS slowly advances from {  a flat-top to} power law noise. This rising phase is also associated with an increase in photon index and the disc flux contribution along with the decrease in {${kT_\textrm{e}}$}. During the decline, the contribution of Comptonised flux and {${kT_\textrm{e}}$} is seen to increase with a gradual reduction in photon index. Type-C QPOs reappear as the source descends into hard state (see Section \ref{sec4.4}). Such correlated `spectro-temporal' features and systematic evolution of  {QPO frequencies} of BH sources including GRS~1915$+$105, have been identified earlier by \cite{2007ApJ...663..445S,2009ApJ...699..453S} and is also in accordance with canonical sources like H~1743$-$322 \citep{2013AdSpR..52.2143D}, GX~339{--}4 \citep{2005A&A...440..207B,2012A&A...542A..56N,2019AdSpR..63.1374S}, XTE~J1859$+$226 \citep{2005ApJ...629..403C,2014AdSpR..54.1678R, 2018Ap&SS.363...90N}, MAXI~J1535$-$571 \citep{2019MNRAS.488..720B}. Our results are in agreement with the hypothesis postulated by  \cite{1996ApJ...457..805M,1997ApJ...474..378R,2000A&A...360L..25R, 2014MNRAS.442..251D}, where it is speculated that the QPOs are generated by the oscillation of post shock regions. As the post shock region stretches closer to the BH, the QPO frequency increases along with the disc flux. The increased number of soft photons from the close-by disc cools down the Compton cloud, thereby decreasing the electron temperature as the source softens. A similar coronal oscillatory model proposed by \cite{2004ApJ...612..988T} also favours the scenario.

{We also examined \textit{AstroSat} data for evidences referring to internal obscuration beyond MJD 58600.  {The presence of intrinsic obscuration has been reported very recently using different partial covering absorption models along the continuum, in addition to the interstellar absorption model. Various researchers (\citealt{2020ApJ...904...30M, 2021ApJ...909...41B, 2021MNRAS.503..152M}) used different combination of models to describe the observed obscuration. Our attempts with the model \textit{(phabs$_{los}$(phabs$_1$*cutoffpl+const*phabs$_2$*cutoffpl+pexmon))} suggested by \citet{2021ApJ...909...41B} do not imply a statistically significant fit which gave a local absorption value of 0.07$\times10^{22}$ atoms cm$^{-2}$ and a reduced chi-square of 1.72 during Epoch 31, whereas the local absorption column density value could not be constrained during Epoch 30. The model combination in \citet{2020ApJ...904...30M} uses  {photoionised} absorption models on three highly resolved \textit{Chandra-High Energy Transmission Grating (HETG)} spectra, one as the source enters the internal obscuration and the latter two when the source is deeply obscured. This model combination reveals strong absorption lines. Considering the moderate spectral resolution of {\it SXT} in comparison to {\it Chandra}, we did not implement the methodology used by \citet{2020ApJ...904...30M}. We were able to adopt the simplest model that copes with the already existing model i.e., a partially covering absorber \textit{TBpcf} along the continuum (\citealt{2021MNRAS.503..152M}). Thus, we used the model form
\textit{TBabs*constant(TBpcf(diskbb+gauss+nthcomp))} to fit the broadband spectrum from \textit{AstroSat}.} {Figure \ref{fig19} shows an over-plot of two energy spectra corresponding to Epoch 31 obtained  with (plotted in red) and without (plotted in black) considering the \textit{TBpcf} model. The energy spectrum modeled without \textit{TBpcf} yielded a  $\chi^{2}$/dof of 197.6/137. Inclusion of \textit{TBpcf} along the continuum significantly improved the fit with  $\chi^{2}$/dof of 139.7/134. The absorption column density along the line of sight for both Epochs remained $\sim$5 $\times10^{22}$ atoms cm$^{-2}$, although our analysis showed an additional absorption of $45 \times10^{22}$ atoms cm$^{-2}$ with a local partial covering fraction (PCF) of 0.13 during Epoch 30. A local absorption of $108  \times10^{22}$ atoms cm$^{-2}$ with a corresponding PCF of 0.41 was seen during Epoch 31. The best-fit spectral parameters obtained for Epochs 30 and 31 are mentioned in Table \ref{tab4}.} These absorption values are comparable to that obtained by \citet{2021MNRAS.503..152M}, even though the observations considered by them falls 200 days behind our observations. The $n_{\textrm{H}}$ estimation from \citet{2021ApJ...909...41B} during these days also affirms our results.  { {The intrinsic bolometric luminosity values  were also estimated for Epochs 30 and 31 by using the \textit{cflux} model, i.e.,  \textit{TBabs(cflux(TBpcf(diskbb+gauss+nthcomp)))}. Epoch 30 exhibited an intrinsic bolometric luminosity of 0.6\% L$_\textrm{EDD}$, while Epoch 31 showed a value of 0.8\% L$_\textrm{EDD}$. We also estimated the intrinsic bolometric luminosity for these two Epochs without including the \textit{TBpcf} model along the continuum, i.e., \textit{TBabs(TBpcf(cflux(diskbb+gauss+nthcomp)))}. The bolometric luminosities were found to be 0.7\% L$_\textrm{EDD}$ during Epoch 30 and 1.5\% L$_\textrm{EDD}$ during Epoch 31.  The decrease in the intrinsic bolometric luminosity, upon including the partial covering absorber along the continuum, is the result of obscuration around the source. Further detailed correlation study between the obscuration and the intrinsic bolometric luminosity of the source is beyond the scope of the present paper.}} Considering the low flux along with the high absorption values, we conclude that internal obscuration is a dominant effect at lower energies during the later stages in the light curve (after MJD 58600), in agreement with \citet{2020ApJ...904...30M, 2021ApJ...909...41B, 2021MNRAS.503..152M}.  However, we did not observe any absorption lines as reported by \citet{2020ApJ...904...30M} (due to reduced spectral resolution of {\it SXT/AstroSat} in comparison to {\it Chandra}). \citet{2020ApJ...904...30M} also computed the magnetic field strength required to launch a disc wind from small radii. The magnetic field during this period was found to be too low, to expel the wind to infinity, causing a failed wind, which further envelopes the central engine leading to obscuration. It is intriguing to note that even after returning to a low-flux  {phase}, GRS~1915$+$105 did not attain a quiescence.} Repeated flaring in X-rays and radio have been detected (\citealt{2019ATel12773....1M,2019ATel12787....1I,2019ATel13304....1T,2019ATel13308....1H,2020ATel13974....1R}); {with a recent report of re-brightening in the source intensity {  by {\it MAXI}, {\it Neutron Star Interior Composition Explorer - NICER} and {\it AstroSat} observations indicating an evolving energy spectrum and PDS with LFQPOs \citep{2021N,2021RBT}}}. A detailed study of the disc-jet coupling during {\it the multiple flaring events} based on near-simultaneous multi-mission observations will be performed and presented elsewhere (Athulya et al. in preparation).

\vspace{-0.5cm}
\section*{Acknowledgements}
\vspace{-0.1cm}
\hspace{1cm}
{We thank the anonymous reviewer for his/her suggestions and comments that helped to improve the quality of this manuscript.} AMP, RD, SN, SM, AN acknowledge the financial support of Indian Space Research Organisation (ISRO) under RESPOND program Sanction order No. DS-2B-13012(2)/19/2019-Sec.II. This publication uses data from the \textit{AstroSat} mission of the ISRO archived at the Indian Space Science Data Centre (ISSDC). This work has been performed utilising the calibration databases and auxiliary analysis tools developed, maintained and distributed by {\it AstroSat-SXT} team with members from various institutions in India and abroad. This research has made use of {\it MAXI} data provided by RIKEN, JAXA and the {\it MAXI} team. Also this research  {has made} use of software provided by the High Energy Astrophysics Science Archive Research Center (HEASARC) and NASA’s Astrophysics Data System Bibliographic Services. VKA, RBT, AN also thank GH, SAG; DD, PDMSA and Director-URSC for encouragement and continuous support to carry out this research.
\par
\noindent{Facilities: {\it AstroSat}, {\it MAXI},  {{\it RXTE}}.}
\vspace{-0.7cm}
\section*{Data Availability}
\vspace{-0.1cm}
The data used for analysis in this article are available in {\it AstroSat}-ISSDC website (\url{https://astrobrowse.issdc.gov.in/astro_archive/archive/Home.jsp}), {\it MAXI} website (\url{http://maxi.riken.jp/top/index.html}) and {\it RXTE} observation from HEASARC database (\url{https://heasarc.gsfc.nasa.gov/docs/cgro/db-perl/W3Browse/w3browse.pl}).




\bibliographystyle{mnras}
\vspace{-0.5cm}
\bibliography{references} 

\begin{thebibliography}{}
\makeatletter
\relax
\def\mn@urlcharsother{\let\do\@makeother \do\$\do\&\do\#\do\^\do\_\do\%\do\~}
\def\mn@doi{\begingroup\mn@urlcharsother \@ifnextchar [ {\mn@doi@}
  {\mn@doi@[]}}
\def\mn@doi@[#1]#2{\def\@tempa{#1}\ifx\@tempa\@empty \href
  {http://dx.doi.org/#2} {doi:#2}\else \href {http://dx.doi.org/#2} {#1}\fi
  \endgroup}
\def\mn@eprint#1#2{\mn@eprint@#1:#2::\@nil}
\def\mn@eprint@arXiv#1{\href {http://arxiv.org/abs/#1} {{\tt arXiv:#1}}}
\def\mn@eprint@dblp#1{\href {http://dblp.uni-trier.de/rec/bibtex/#1.xml}
  {dblp:#1}}
\def\mn@eprint@#1:#2:#3:#4\@nil{\def\@tempa {#1}\def\@tempb {#2}\def\@tempc
  {#3}\ifx \@tempc \@empty \let \@tempc \@tempb \let \@tempb \@tempa \fi \ifx
  \@tempb \@empty \def\@tempb {arXiv}\fi \@ifundefined
  {mn@eprint@\@tempb}{\@tempb:\@tempc}{\expandafter \expandafter \csname
  mn@eprint@\@tempb\endcsname \expandafter{\@tempc}}}

\bibitem[\protect\citeauthoryear{{Adegoke}, {Mukhopadhyay}  \&
  {Misra}}{{Adegoke} et~al.}{2020}]{2020MNRAS.492.4033A}
{Adegoke} O.,  {Mukhopadhyay} B.,   {Misra} R.,  2020, \mn@doi [\mnras]
  {10.1093/mnras/staa071}, \href
  {https://ui.adsabs.harvard.edu/abs/2020MNRAS.492.4033A} {492, 4033}

\bibitem[\protect\citeauthoryear{{Agrawal}}{{Agrawal}}{2006}]{2006AdSpR..38.2989A}
{Agrawal} P.~C.,  2006, \mn@doi [Advances in Space Research]
  {10.1016/j.asr.2006.03.038}, \href
  {https://ui.adsabs.harvard.edu/abs/2006AdSpR..38.2989A} {38, 2989}

\bibitem[\protect\citeauthoryear{{Agrawal}, {Nandi}, {Girish}  \&
  {Ramadevi}}{{Agrawal} et~al.}{2018}]{2018MNRAS.477.5437A}
{Agrawal} V.~K.,  {Nandi} A.,  {Girish} V.,   {Ramadevi} M.~C.,  2018, \mn@doi
  [\mnras] {10.1093/mnras/sty1005}, \href
  {https://ui.adsabs.harvard.edu/abs/2018MNRAS.477.5437A} {477, 5437}

\bibitem[\protect\citeauthoryear{{Alam}, {Dewangan}, {Belloni}, {Mukherjee}  \&
  {Jhingan}}{{Alam} et~al.}{2014}]{2014MNRAS.445.4259A}
{Alam} M.~S.,  {Dewangan} G.~C.,  {Belloni} T.,  {Mukherjee} D.,   {Jhingan}
  S.,  2014, \mn@doi [\mnras] {10.1093/mnras/stu2048}, \href
  {https://ui.adsabs.harvard.edu/abs/2014MNRAS.445.4259A} {445, 4259}

\bibitem[\protect\citeauthoryear{{Antia} et~al.,}{{Antia}
  et~al.}{2017}]{2017ApJS..231...10A}
{Antia} H.~M.,  et~al., 2017, \mn@doi [\apjs] {10.3847/1538-4365/aa7a0e}, \href
  {https://ui.adsabs.harvard.edu/abs/2017ApJS..231...10A} {231, 10}

\bibitem[\protect\citeauthoryear{{Arnaud}}{{Arnaud}}{1996}]{1996ASPC..101...17A}
{Arnaud} K.~A.,  1996, in {Jacoby} G.~H.,  {Barnes} J.,  eds,  Astronomical
  Society of the Pacific Conference Series Vol. 101, Astronomical Data Analysis
  Software and Systems V. p.~17

\bibitem[\protect\citeauthoryear{{Baby}, {Agrawal}, {Ramadevi}, {Katoch},
  {Antia}, {Mandal}  \& {Nandi}}{{Baby} et~al.}{2020}]{2020MNRAS.497.1197B}
{Baby} B.~E.,  {Agrawal} V.~K.,  {Ramadevi} M.~C.,  {Katoch} T.,  {Antia}
  H.~M.,  {Mandal} S.,   {Nandi} A.,  2020, \mn@doi [\mnras]
  {10.1093/mnras/staa1965}, \href
  {https://ui.adsabs.harvard.edu/abs/2020MNRAS.497.1197B} {497, 1197}

\bibitem[\protect\citeauthoryear{{Baby}, {Bhuvana}, {Radhika}, {Katoch},
  {Mandal}  \& {Nandi}}{{Baby} et~al.}{2021}]{2021MNRAS.tmp.2464B}
{Baby} B.~E.,  {Bhuvana} G.~R.,  {Radhika} D.,  {Katoch} T.,  {Mandal} S.,
  {Nandi} A.,  2021, \mn@doi [\mnras] {10.1093/mnras/stab2719}, \href
  {https://ui.adsabs.harvard.edu/abs/2021MNRAS.tmp.2464B} {}

\bibitem[\protect\citeauthoryear{{Baglio} et~al.,}{{Baglio}
  et~al.}{2018}]{2018ApJ...867..114B}
{Baglio} M.~C.,  et~al., 2018, \mn@doi [\apj] {10.3847/1538-4357/aae532}, \href
  {https://ui.adsabs.harvard.edu/abs/2018ApJ...867..114B} {867, 114}

\bibitem[\protect\citeauthoryear{{Balakrishnan}, {Miller}, {Reynolds},
  {Kammoun}, {Zoghbi}  \& {Tetarenko}}{{Balakrishnan}
  et~al.}{2021}]{2021ApJ...909...41B}
{Balakrishnan} M.,  {Miller} J.~M.,  {Reynolds} M.~T.,  {Kammoun} E.,  {Zoghbi}
  A.,   {Tetarenko} B.~E.,  2021, \mn@doi [\apj] {10.3847/1538-4357/abd6cb},
  \href {https://ui.adsabs.harvard.edu/abs/2021ApJ...909...41B} {909, 41}

\bibitem[\protect\citeauthoryear{{Begelman} \& {Armitage}}{{Begelman} \&
  {Armitage}}{2014}]{2014ApJ...782L..18B}
{Begelman} M.~C.,  {Armitage} P.~J.,  2014, \mn@doi [\apjl]
  {10.1088/2041-8205/782/2/L18}, \href
  {https://ui.adsabs.harvard.edu/abs/2014ApJ...782L..18B} {782, L18}

\bibitem[\protect\citeauthoryear{{Belloni} \& {Altamirano}}{{Belloni} \&
  {Altamirano}}{2013}]{2013MNRAS.432...19B}
{Belloni} T.~M.,  {Altamirano} D.,  2013, \mn@doi [\mnras]
  {10.1093/mnras/stt285}, \href
  {https://ui.adsabs.harvard.edu/abs/2013MNRAS.432...19B} {432, 19}

\bibitem[\protect\citeauthoryear{{Belloni} \& {Hasinger}}{{Belloni} \&
  {Hasinger}}{1990}]{1990A&A...230..103B}
{Belloni} T.,  {Hasinger} G.,  1990, \aap, \href
  {https://ui.adsabs.harvard.edu/abs/1990A&A...230..103B} {230, 103}

\bibitem[\protect\citeauthoryear{{Belloni}, {M{\'e}ndez}, {King}, {van der
  Klis}  \& {van Paradijs}}{{Belloni} et~al.}{1997a}]{1997ApJ...479L.145B}
{Belloni} T.,  {M{\'e}ndez} M.,  {King} A.~R.,  {van der Klis} M.,   {van
  Paradijs} J.,  1997a, \mn@doi [\apjl] {10.1086/310595}, \href
  {https://ui.adsabs.harvard.edu/abs/1997ApJ...479L.145B} {479, L145}

\bibitem[\protect\citeauthoryear{{Belloni}, {M{\'e}ndez}, {King}, {van der
  Klis}  \& {van Paradijs}}{{Belloni} et~al.}{1997b}]{1997ApJ...488L.109B}
{Belloni} T.,  {M{\'e}ndez} M.,  {King} A.~R.,  {van der Klis} M.,   {van
  Paradijs} J.,  1997b, \mn@doi [\apjl] {10.1086/310944}, \href
  {https://ui.adsabs.harvard.edu/abs/1997ApJ...488L.109B} {488, L109}

\bibitem[\protect\citeauthoryear{{Belloni}, {Klein-Wolt}, {M{\'e}ndez}, {van
  der Klis}  \& {van Paradijs}}{{Belloni} et~al.}{2000}]{2000A&A...355..271B}
{Belloni} T.,  {Klein-Wolt} M.,  {M{\'e}ndez} M.,  {van der Klis} M.,   {van
  Paradijs} J.,  2000, \aap, \href
  {https://ui.adsabs.harvard.edu/abs/2000A&A...355..271B} {355, 271}

\bibitem[\protect\citeauthoryear{{Belloni}, {Psaltis}  \& {van der
  Klis}}{{Belloni} et~al.}{2002}]{2002ApJ...572..392B}
{Belloni} T.,  {Psaltis} D.,   {van der Klis} M.,  2002, \mn@doi [\apj]
  {10.1086/340290}, \href
  {https://ui.adsabs.harvard.edu/abs/2002ApJ...572..392B} {572, 392}

\bibitem[\protect\citeauthoryear{{Belloni}, {Homan}, {Casella}, {van der Klis},
  {Nespoli}, {Lewin}, {Miller}  \& {M{\'e}ndez}}{{Belloni}
  et~al.}{2005}]{2005A&A...440..207B}
{Belloni} T.,  {Homan} J.,  {Casella} P.,  {van der Klis} M.,  {Nespoli} E.,
  {Lewin} W.~H.~G.,  {Miller} J.~M.,   {M{\'e}ndez} M.,  2005, \mn@doi [\aap]
  {10.1051/0004-6361:20042457}, \href
  {https://ui.adsabs.harvard.edu/abs/2005A&A...440..207B} {440, 207}

\bibitem[\protect\citeauthoryear{{Belloni}, {Sanna}  \& {M{\'e}ndez}}{{Belloni}
  et~al.}{2012}]{2012MNRAS.426.1701B}
{Belloni} T.~M.,  {Sanna} A.,   {M{\'e}ndez} M.,  2012, \mn@doi [\mnras]
  {10.1111/j.1365-2966.2012.21634.x}, \href
  {https://ui.adsabs.harvard.edu/abs/2012MNRAS.426.1701B} {426, 1701}

\bibitem[\protect\citeauthoryear{{Belloni}, {Bhattacharya}, {Caccese},
  {Bhalerao}, {Vadawale}  \& {Yadav}}{{Belloni}
  et~al.}{2019}]{2019MNRAS.489.1037B}
{Belloni} T.~M.,  {Bhattacharya} D.,  {Caccese} P.,  {Bhalerao} V.,  {Vadawale}
  S.,   {Yadav} J.~S.,  2019, \mn@doi [\mnras] {10.1093/mnras/stz2143}, \href
  {https://ui.adsabs.harvard.edu/abs/2019MNRAS.489.1037B} {489, 1037}

\bibitem[\protect\citeauthoryear{{Bhalerao} et~al.,}{{Bhalerao}
  et~al.}{2015}]{2015MNRAS.447.2274B}
{Bhalerao} V.,  et~al., 2015, \mn@doi [\mnras] {10.1093/mnras/stu2495}, \href
  {https://ui.adsabs.harvard.edu/abs/2015MNRAS.447.2274B} {447, 2274}

\bibitem[\protect\citeauthoryear{{Bhalerao} et~al.,}{{Bhalerao}
  et~al.}{2017}]{2017JApA...38...31B}
{Bhalerao} V.,  et~al., 2017, \mn@doi [Journal of Astrophysics and Astronomy]
  {10.1007/s12036-017-9447-8}, \href
  {https://ui.adsabs.harvard.edu/abs/2017JApA...38...31B} {38, 31}

\bibitem[\protect\citeauthoryear{{Bhargava}, {Belloni}, {Bhattacharya}  \&
  {Misra}}{{Bhargava} et~al.}{2019}]{2019MNRAS.488..720B}
{Bhargava} Y.,  {Belloni} T.,  {Bhattacharya} D.,   {Misra} R.,  2019, \mn@doi
  [\mnras] {10.1093/mnras/stz1774}, \href
  {https://ui.adsabs.harvard.edu/abs/2019MNRAS.488..720B} {488, 720}

\bibitem[\protect\citeauthoryear{{Blondin}, {Stevens}  \& {Kallman}}{{Blondin}
  et~al.}{1991}]{1991ApJ...371..684B}
{Blondin} J.~M.,  {Stevens} I.~R.,   {Kallman} T.~R.,  1991, \mn@doi [\apj]
  {10.1086/169934}, \href
  {https://ui.adsabs.harvard.edu/abs/1991ApJ...371..684B} {371, 684}

\bibitem[\protect\citeauthoryear{{Cadolle Bel} et~al.,}{{Cadolle Bel}
  et~al.}{2011}]{2011A&A...534A.119C}
{Cadolle Bel} M.,  et~al., 2011, \mn@doi [\aap] {10.1051/0004-6361/201117684},
  \href {https://ui.adsabs.harvard.edu/abs/2011A&A...534A.119C} {534, A119}

\bibitem[\protect\citeauthoryear{{Casella}, {Belloni}  \& {Stella}}{{Casella}
  et~al.}{2005}]{2005ApJ...629..403C}
{Casella} P.,  {Belloni} T.,   {Stella} L.,  2005, \mn@doi [\apj]
  {10.1086/431174}, \href
  {https://ui.adsabs.harvard.edu/abs/2005ApJ...629..403C} {629, 403}

\bibitem[\protect\citeauthoryear{{Castro-Tirado}, {Brandt}  \&
  {Lund}}{{Castro-Tirado} et~al.}{1992}]{1992IAUC.5590....2C}
{Castro-Tirado} A.~J.,  {Brandt} S.,   {Lund} N.,  1992, \iaucirc, \href
  {https://ui.adsabs.harvard.edu/abs/1992IAUC.5590....2C} {5590, 2}

\bibitem[\protect\citeauthoryear{{Chakrabarti} \& {Manickam}}{{Chakrabarti} \&
  {Manickam}}{2000}]{2000ApJ...531L..41C}
{Chakrabarti} S.~K.,  {Manickam} S.~G.,  2000, \mn@doi [\apjl]
  {10.1086/312512}, \href
  {https://ui.adsabs.harvard.edu/abs/2000ApJ...531L..41C} {531, L41}

\bibitem[\protect\citeauthoryear{{Chakrabarti} \& {Nandi}}{{Chakrabarti} \&
  {Nandi}}{2000}]{2000astro.ph.12526C}
{Chakrabarti} S.~K.,  {Nandi} A.,  2000, arXiv e-prints, \href
  {https://ui.adsabs.harvard.edu/abs/2000astro.ph.12526C} {pp
  astro--ph/0012526}

\bibitem[\protect\citeauthoryear{{Chakrabarti} \& {Titarchuk}}{{Chakrabarti} \&
  {Titarchuk}}{1995}]{1995ApJ...455..623C}
{Chakrabarti} S.,  {Titarchuk} L.~G.,  1995, \mn@doi [\apj] {10.1086/176610},
  \href {https://ui.adsabs.harvard.edu/abs/1995ApJ...455..623C} {455, 623}

\bibitem[\protect\citeauthoryear{{Chakrabarti}, {Nandi}, {Chatterjee},
  {Choudhury}  \& {Chatterjee}}{{Chakrabarti}
  et~al.}{2005}]{2005A&A...431..825C}
{Chakrabarti} S.~K.,  {Nandi} A.,  {Chatterjee} A.~K.,  {Choudhury} A.~K.,
  {Chatterjee} U.,  2005, \mn@doi [\aap] {10.1051/0004-6361:20041662}, \href
  {https://ui.adsabs.harvard.edu/abs/2005A&A...431..825C} {431, 825}

\bibitem[\protect\citeauthoryear{{Chatterjee}, {Debnath}, {Chatterjee}, {Jana}
  \& {Chakrabarti}}{{Chatterjee} et~al.}{2020}]{2020MNRAS.493.2452C}
{Chatterjee} K.,  {Debnath} D.,  {Chatterjee} D.,  {Jana} A.,   {Chakrabarti}
  S.~K.,  2020, \mn@doi [\mnras] {10.1093/mnras/staa391}, \href
  {https://ui.adsabs.harvard.edu/abs/2020MNRAS.493.2452C} {493, 2452}

\bibitem[\protect\citeauthoryear{{Chen}, {Swank}  \& {Taam}}{{Chen}
  et~al.}{1997}]{1997ApJ...477L..41C}
{Chen} X.,  {Swank} J.~H.,   {Taam} R.~E.,  1997, \mn@doi [\apjl]
  {10.1086/310515}, \href
  {https://ui.adsabs.harvard.edu/abs/1997ApJ...477L..41C} {477, L41}

\bibitem[\protect\citeauthoryear{{Das}, {Chattopadhyay}, {Nandi}  \&
  {Molteni}}{{Das} et~al.}{2014}]{2014MNRAS.442..251D}
{Das} S.,  {Chattopadhyay} I.,  {Nandi} A.,   {Molteni} D.,  2014, \mn@doi
  [\mnras] {10.1093/mnras/stu864}, \href
  {https://ui.adsabs.harvard.edu/abs/2014MNRAS.442..251D} {442, 251}

\bibitem[\protect\citeauthoryear{{Debnath}, {Chakrabarti}  \&
  {Nandi}}{{Debnath} et~al.}{2010}]{2010A&A...520A..98D}
{Debnath} D.,  {Chakrabarti} S.~K.,   {Nandi} A.,  2010, \mn@doi [\aap]
  {10.1051/0004-6361/201014990}, \href
  {https://ui.adsabs.harvard.edu/abs/2010A&A...520A..98D} {520, A98}

\bibitem[\protect\citeauthoryear{{Debnath}, {Chakrabarti}  \&
  {Nandi}}{{Debnath} et~al.}{2013}]{2013AdSpR..52.2143D}
{Debnath} D.,  {Chakrabarti} S.~K.,   {Nandi} A.,  2013, \mn@doi [Advances in
  Space Research] {10.1016/j.asr.2013.09.002}, \href
  {https://ui.adsabs.harvard.edu/abs/2013AdSpR..52.2143D} {52, 2143}

\bibitem[\protect\citeauthoryear{{Dunn}, {Fender}, {K{\"o}rding}, {Belloni}  \&
  {Cabanac}}{{Dunn} et~al.}{2010}]{2010MNRAS.403...61D}
{Dunn} R.~J.~H.,  {Fender} R.~P.,  {K{\"o}rding} E.~G.,  {Belloni} T.,
  {Cabanac} C.,  2010, \mn@doi [\mnras] {10.1111/j.1365-2966.2010.16114.x},
  \href {https://ui.adsabs.harvard.edu/abs/2010MNRAS.403...61D} {403, 61}

\bibitem[\protect\citeauthoryear{{Frank}, {King}  \& {Raine}}{{Frank}
  et~al.}{1985}]{1985apa..book.....F}
{Frank} J.,  {King} A.~R.,   {Raine} D.~J.,  1985, {Accretion power in
  astrophysics}

\bibitem[\protect\citeauthoryear{{Gilfanov}}{{Gilfanov}}{2010}]{2010LNP...794...17G}
{Gilfanov} M.,  2010, {X-Ray Emission from Black-Hole Binaries}.
p.~17, \mn@doi{10.1007/978-3-540-76937-8_2}

\bibitem[\protect\citeauthoryear{{Greiner}, {Cuby}, {McCaughrean},
  {Castro-Tirado}  \& {Mennickent}}{{Greiner}
  et~al.}{2001}]{2001A&A...373L..37G}
{Greiner} J.,  {Cuby} J.~G.,  {McCaughrean} M.~J.,  {Castro-Tirado} A.~J.,
  {Mennickent} R.~E.,  2001, \mn@doi [\aap] {10.1051/0004-6361:20010771}, \href
  {https://ui.adsabs.harvard.edu/abs/2001A&A...373L..37G} {373, L37}

\bibitem[\protect\citeauthoryear{{Hameury}, {King}  \& {Lasota}}{{Hameury}
  et~al.}{1990}]{1990ApJ...353..585H}
{Hameury} J.-M.,  {King} A.~R.,   {Lasota} J.-P.,  1990, \mn@doi [\apj]
  {10.1086/168647}, \href
  {https://ui.adsabs.harvard.edu/abs/1990ApJ...353..585H} {353, 585}

\bibitem[\protect\citeauthoryear{{Hannikainen} et~al.,}{{Hannikainen}
  et~al.}{2005}]{2005A&A...435..995H}
{Hannikainen} D.~C.,  et~al., 2005, \mn@doi [\aap]
  {10.1051/0004-6361:20042250}, \href
  {https://ui.adsabs.harvard.edu/abs/2005A&A...435..995H} {435, 995}

\bibitem[\protect\citeauthoryear{{Heil}, {Vaughan}  \& {Uttley}}{{Heil}
  et~al.}{2012}]{2012MNRAS.422.2620H}
{Heil} L.~M.,  {Vaughan} S.,   {Uttley} P.,  2012, \mn@doi [\mnras]
  {10.1111/j.1365-2966.2012.20824.x}, \href
  {https://ui.adsabs.harvard.edu/abs/2012MNRAS.422.2620H} {422, 2620}

\bibitem[\protect\citeauthoryear{{Homan}, {Wijnands}, {van der Klis},
  {Belloni}, {van Paradijs}, {Klein-Wolt}, {Fender}  \& {M{\'e}ndez}}{{Homan}
  et~al.}{2001}]{2001ApJS..132..377H}
{Homan} J.,  {Wijnands} R.,  {van der Klis} M.,  {Belloni} T.,  {van Paradijs}
  J.,  {Klein-Wolt} M.,  {Fender} R.,   {M{\'e}ndez} M.,  2001, \mn@doi [\apjs]
  {10.1086/318954}, \href
  {https://ui.adsabs.harvard.edu/abs/2001ApJS..132..377H} {132, 377}

\bibitem[\protect\citeauthoryear{{Homan}, {Neilsen}, {Steiner}, {Remillard},
  {Altamirano}, {Gendreau}  \& {Arzoumanian}}{{Homan}
  et~al.}{2019a}]{2019ATel12742....1H}
{Homan} J.,  {Neilsen} J.,  {Steiner} J.,  {Remillard} R.,  {Altamirano} D.,
  {Gendreau} K.,   {Arzoumanian} Z.,  2019a, The Astronomer's Telegram, \href
  {https://ui.adsabs.harvard.edu/abs/2019ATel12742....1H} {12742, 1}

\bibitem[\protect\citeauthoryear{{Homan}, {Neilsen}, {Gendreau}, {Arzoumanian},
  {Remillard}, {Steiner}, {Altamirano}  \& {Wolff}}{{Homan}
  et~al.}{2019b}]{2019ATel13308....1H}
{Homan} J.,  {Neilsen} J.,  {Gendreau} K.,  {Arzoumanian} Z.,  {Remillard} R.,
  {Steiner} J.,  {Altamirano} D.,   {Wolff} M.~T.,  2019b, The Astronomer's
  Telegram, \href {https://ui.adsabs.harvard.edu/abs/2019ATel13308....1H}
  {13308, 1}

\bibitem[\protect\citeauthoryear{{Iwakiri} et~al.,}{{Iwakiri}
  et~al.}{2019}]{2019ATel12787....1I}
{Iwakiri} W.,  et~al., 2019, The Astronomer's Telegram, \href
  {https://ui.adsabs.harvard.edu/abs/2019ATel12787....1I} {12787, 1}

\bibitem[\protect\citeauthoryear{{Iyer}, {Nandi}  \& {Mandal}}{{Iyer}
  et~al.}{2015}]{2015ApJ...807..108I}
{Iyer} N.,  {Nandi} A.,   {Mandal} S.,  2015, \mn@doi [\apj]
  {10.1088/0004-637X/807/1/108}, \href
  {https://ui.adsabs.harvard.edu/abs/2015ApJ...807..108I} {807, 108}

\bibitem[\protect\citeauthoryear{{Karpouzas}, {M{\'e}ndez}, {Garc{\'\i}a},
  {Zhang}, {Altamirano}, {Belloni}  \& {Zhang}}{{Karpouzas}
  et~al.}{2021}]{2021MNRAS.503.5522K}
{Karpouzas} K.,  {M{\'e}ndez} M.,  {Garc{\'\i}a} F.,  {Zhang} L.,  {Altamirano}
  D.,  {Belloni} T.,   {Zhang} Y.,  2021, \mn@doi [\mnras]
  {10.1093/mnras/stab827}, \href
  {https://ui.adsabs.harvard.edu/abs/2021MNRAS.503.5522K} {503, 5522}

\bibitem[\protect\citeauthoryear{{Katoch}, {Baby}, {Nandi}, {Agrawal}, {Antia}
  \& {Mukerjee}}{{Katoch} et~al.}{2021}]{2021MNRAS.501.6123K}
{Katoch} T.,  {Baby} B.~E.,  {Nandi} A.,  {Agrawal} V.~K.,  {Antia} H.~M.,
  {Mukerjee} K.,  2021, \mn@doi [\mnras] {10.1093/mnras/staa3756}, \href
  {https://ui.adsabs.harvard.edu/abs/2021MNRAS.501.6123K} {501, 6123}

\bibitem[\protect\citeauthoryear{{King}, {Kolb}  \& {Burderi}}{{King}
  et~al.}{1996}]{1996ApJ...464L.127K}
{King} A.~R.,  {Kolb} U.,   {Burderi} L.,  1996, \mn@doi [\apjl]
  {10.1086/310105}, \href
  {https://ui.adsabs.harvard.edu/abs/1996ApJ...464L.127K} {464, L127}

\bibitem[\protect\citeauthoryear{{Klein-Wolt}, {Fender}, {Pooley}, {Belloni},
  {Migliari}, {Morgan}  \& {van der Klis}}{{Klein-Wolt}
  et~al.}{2002}]{2002MNRAS.331..745K}
{Klein-Wolt} M.,  {Fender} R.~P.,  {Pooley} G.~G.,  {Belloni} T.,  {Migliari}
  S.,  {Morgan} E.~H.,   {van der Klis} M.,  2002, \mn@doi [\mnras]
  {10.1046/j.1365-8711.2002.05223.x}, \href
  {https://ui.adsabs.harvard.edu/abs/2002MNRAS.331..745K} {331, 745}

\bibitem[\protect\citeauthoryear{{Koljonen} \& {Hovatta}}{{Koljonen} \&
  {Hovatta}}{2021}]{2021A&A...647A.173K}
{Koljonen} K.~I.~I.,  {Hovatta} T.,  2021, \mn@doi [\aap]
  {10.1051/0004-6361/202039581}, \href
  {https://ui.adsabs.harvard.edu/abs/2021A&A...647A.173K} {647, A173}

\bibitem[\protect\citeauthoryear{{Kushwaha}, {Agrawal}  \& {Nandi}}{{Kushwaha}
  et~al.}{2021}]{2021MNRAS.507.2602K}
{Kushwaha} A.,  {Agrawal} V.~K.,   {Nandi} A.,  2021, \mn@doi [\mnras]
  {10.1093/mnras/stab2258}, \href
  {https://ui.adsabs.harvard.edu/abs/2021MNRAS.507.2602K} {507, 2602}

\bibitem[\protect\citeauthoryear{{Lasota}}{{Lasota}}{2001}]{2001NewAR..45..449L}
{Lasota} J.-P.,  2001, \mn@doi [\nar] {10.1016/S1387-6473(01)00112-9}, \href
  {https://ui.adsabs.harvard.edu/abs/2001NewAR..45..449L} {45, 449}

\bibitem[\protect\citeauthoryear{{Lightman} \& {Zdziarski}}{{Lightman} \&
  {Zdziarski}}{1987}]{1987ApJ...319..643L}
{Lightman} A.~P.,  {Zdziarski} A.~A.,  1987, \mn@doi [\apj] {10.1086/165485},
  \href {https://ui.adsabs.harvard.edu/abs/1987ApJ...319..643L} {319, 643}

\bibitem[\protect\citeauthoryear{{Lotti} et~al.,}{{Lotti}
  et~al.}{2016}]{2016ApJ...822...57L}
{Lotti} S.,  et~al., 2016, \mn@doi [\apj] {10.3847/0004-637X/822/1/57}, \href
  {https://ui.adsabs.harvard.edu/abs/2016ApJ...822...57L} {822, 57}

\bibitem[\protect\citeauthoryear{{Maccarone} \& {Coppi}}{{Maccarone} \&
  {Coppi}}{2003}]{2003MNRAS.338..189M}
{Maccarone} T.~J.,  {Coppi} P.~S.,  2003, \mn@doi [\mnras]
  {10.1046/j.1365-8711.2003.06040.x}, \href
  {https://ui.adsabs.harvard.edu/abs/2003MNRAS.338..189M} {338, 189}

\bibitem[\protect\citeauthoryear{{Makishima}, {Maejima}, {Mitsuda}, {Bradt},
  {Remillard}, {Tuohy}, {Hoshi}  \& {Nakagawa}}{{Makishima}
  et~al.}{1986}]{1986ApJ...308..635M}
{Makishima} K.,  {Maejima} Y.,  {Mitsuda} K.,  {Bradt} H.~V.,  {Remillard}
  R.~A.,  {Tuohy} I.~R.,  {Hoshi} R.,   {Nakagawa} M.,  1986, \mn@doi [\apj]
  {10.1086/164534}, \href
  {https://ui.adsabs.harvard.edu/abs/1986ApJ...308..635M} {308, 635}

\bibitem[\protect\citeauthoryear{{Massaro}, {Ventura}, {Massa}, {Feroci},
  {Mineo}, {Cusumano}, {Casella}  \& {Belloni}}{{Massaro}
  et~al.}{2010}]{2010A&A...513A..21M}
{Massaro} E.,  {Ventura} G.,  {Massa} F.,  {Feroci} M.,  {Mineo} T.,
  {Cusumano} G.,  {Casella} P.,   {Belloni} T.,  2010, \mn@doi [\aap]
  {10.1051/0004-6361/200912908}, \href
  {https://ui.adsabs.harvard.edu/abs/2010A&A...513A..21M} {513, A21}

\bibitem[\protect\citeauthoryear{{Massaro}, {Capitanio}, {Feroci}, {Mineo},
  {Ardito}  \& {Ricciardi}}{{Massaro} et~al.}{2020}]{2020MNRAS.496.1697M}
{Massaro} E.,  {Capitanio} F.,  {Feroci} M.,  {Mineo} T.,  {Ardito} A.,
  {Ricciardi} P.,  2020, \mn@doi [\mnras] {10.1093/mnras/staa1125}, \href
  {https://ui.adsabs.harvard.edu/abs/2020MNRAS.496.1697M} {496, 1697}

\bibitem[\protect\citeauthoryear{{Matsuoka} et~al.,}{{Matsuoka}
  et~al.}{2009}]{2009PASJ...61..999M}
{Matsuoka} M.,  et~al., 2009, \mn@doi [\pasj] {10.1093/pasj/61.5.999}, \href
  {https://ui.adsabs.harvard.edu/abs/2009PASJ...61..999M} {61, 999}

\bibitem[\protect\citeauthoryear{{Mayer} \& {Pringle}}{{Mayer} \&
  {Pringle}}{2006}]{2006MNRAS.368..379M}
{Mayer} M.,  {Pringle} J.~E.,  2006, \mn@doi [\mnras]
  {10.1111/j.1365-2966.2006.10111.x}, \href
  {https://ui.adsabs.harvard.edu/abs/2006MNRAS.368..379M} {368, 379}

\bibitem[\protect\citeauthoryear{{Meyer-Hofmeister}, {Liu}, {Qiao}  \&
  {Taam}}{{Meyer-Hofmeister} et~al.}{2020}]{2020A&A...637A..66M}
{Meyer-Hofmeister} E.,  {Liu} B.~F.,  {Qiao} E.,   {Taam} R.~E.,  2020, \mn@doi
  [\aap] {10.1051/0004-6361/202037561}, \href
  {https://ui.adsabs.harvard.edu/abs/2020A&A...637A..66M} {637, A66}

\bibitem[\protect\citeauthoryear{{Mihara} et~al.,}{{Mihara}
  et~al.}{2011}]{2011PASJ...63S.623M}
{Mihara} T.,  et~al., 2011, \mn@doi [\pasj] {10.1093/pasj/63.sp3.S623}, \href
  {https://ui.adsabs.harvard.edu/abs/2011PASJ...63S.623M} {63, S623}

\bibitem[\protect\citeauthoryear{{Miller} et~al.,}{{Miller}
  et~al.}{2020}]{2020ApJ...904...30M}
{Miller} J.~M.,  et~al., 2020, \mn@doi [\apj] {10.3847/1538-4357/abbb31}, \href
  {https://ui.adsabs.harvard.edu/abs/2020ApJ...904...30M} {904, 30}

\bibitem[\protect\citeauthoryear{{Mirabel} \& {Rodr{\'\i}guez}}{{Mirabel} \&
  {Rodr{\'\i}guez}}{1994}]{1994Natur.371...46M}
{Mirabel} I.~F.,  {Rodr{\'\i}guez} L.~F.,  1994, \mn@doi [\nat]
  {10.1038/371046a0}, \href
  {https://ui.adsabs.harvard.edu/abs/1994Natur.371...46M} {371, 46}

\bibitem[\protect\citeauthoryear{{Misra}, {Harikrishnan}, {Mukhopadhyay},
  {Ambika}  \& {Kembhavi}}{{Misra} et~al.}{2004}]{2004ApJ...609..313M}
{Misra} R.,  {Harikrishnan} K.~P.,  {Mukhopadhyay} B.,  {Ambika} G.,
  {Kembhavi} A.~K.,  2004, \mn@doi [\apj] {10.1086/421005}, \href
  {https://ui.adsabs.harvard.edu/abs/2004ApJ...609..313M} {609, 313}

\bibitem[\protect\citeauthoryear{{Misra}, {Rawat}, {Yadav}  \& {Jain}}{{Misra}
  et~al.}{2020}]{2020ApJ...889L..36M}
{Misra} R.,  {Rawat} D.,  {Yadav} J.~S.,   {Jain} P.,  2020, \mn@doi [\apjl]
  {10.3847/2041-8213/ab6ddc}, \href
  {https://ui.adsabs.harvard.edu/abs/2020ApJ...889L..36M} {889, L36}

\bibitem[\protect\citeauthoryear{{Mitsuda} et~al.,}{{Mitsuda}
  et~al.}{1984}]{1984PASJ...36..741M}
{Mitsuda} K.,  et~al., 1984, \pasj, \href
  {https://ui.adsabs.harvard.edu/abs/1984PASJ...36..741M} {36, 741}

\bibitem[\protect\citeauthoryear{{Miyamoto}, {Kitamoto}, {Iga}, {Negoro}  \&
  {Terada}}{{Miyamoto} et~al.}{1992}]{1992ApJ...391L..21M}
{Miyamoto} S.,  {Kitamoto} S.,  {Iga} S.,  {Negoro} H.,   {Terada} K.,  1992,
  \mn@doi [\apjl] {10.1086/186389}, \href
  {https://ui.adsabs.harvard.edu/abs/1992ApJ...391L..21M} {391, L21}

\bibitem[\protect\citeauthoryear{{Molteni}, {Sponholz}  \&
  {Chakrabarti}}{{Molteni} et~al.}{1996}]{1996ApJ...457..805M}
{Molteni} D.,  {Sponholz} H.,   {Chakrabarti} S.~K.,  1996, \mn@doi [\apj]
  {10.1086/176775}, \href
  {https://ui.adsabs.harvard.edu/abs/1996ApJ...457..805M} {457, 805}

\bibitem[\protect\citeauthoryear{{Motta}, {Mu{\~n}oz-Darias}, {Casella},
  {Belloni}  \& {Homan}}{{Motta} et~al.}{2011}]{2011MNRAS.418.2292M}
{Motta} S.,  {Mu{\~n}oz-Darias} T.,  {Casella} P.,  {Belloni} T.,   {Homan} J.,
   2011, \mn@doi [\mnras] {10.1111/j.1365-2966.2011.19566.x}, \href
  {https://ui.adsabs.harvard.edu/abs/2011MNRAS.418.2292M} {418, 2292}

\bibitem[\protect\citeauthoryear{{Motta}, {Homan}, {Mu{\~n}oz Darias},
  {Casella}, {Belloni}, {Hiemstra}  \& {M{\'e}ndez}}{{Motta}
  et~al.}{2012}]{2012MNRAS.427..595M}
{Motta} S.,  {Homan} J.,  {Mu{\~n}oz Darias} T.,  {Casella} P.,  {Belloni}
  T.~M.,  {Hiemstra} B.,   {M{\'e}ndez} M.,  2012, \mn@doi [\mnras]
  {10.1111/j.1365-2966.2012.22037.x}, \href
  {https://ui.adsabs.harvard.edu/abs/2012MNRAS.427..595M} {427, 595}

\bibitem[\protect\citeauthoryear{{Motta}, {Williams}, {Fender}, {Titterington},
  {Green}  \& {Perrott}}{{Motta} et~al.}{2019}]{2019ATel12773....1M}
{Motta} S.,  {Williams} D.,  {Fender} R.,  {Titterington} D.,  {Green} D.,
  {Perrott} Y.,  2019, The Astronomer's Telegram, \href
  {https://ui.adsabs.harvard.edu/abs/2019ATel12773....1M} {12773, 1}

\bibitem[\protect\citeauthoryear{{Motta} et~al.,}{{Motta}
  et~al.}{2021}]{2021MNRAS.503..152M}
{Motta} S.~E.,  et~al., 2021, \mn@doi [\mnras] {10.1093/mnras/stab511}, \href
  {https://ui.adsabs.harvard.edu/abs/2021MNRAS.503..152M} {503, 152}

\bibitem[\protect\citeauthoryear{{Naik}, {Rao}  \& {Chakrabarti}}{{Naik}
  et~al.}{2002a}]{2002JApA...23..213N}
{Naik} S.,  {Rao} A.~R.,   {Chakrabarti} S.~K.,  2002a, \mn@doi [Journal of
  Astrophysics and Astronomy] {10.1007/BF02702284}, \href
  {https://ui.adsabs.harvard.edu/abs/2002JApA...23..213N} {23, 213}

\bibitem[\protect\citeauthoryear{{Naik}, {Agrawal}, {Rao}  \& {Paul}}{{Naik}
  et~al.}{2002b}]{2002MNRAS.330..487N}
{Naik} S.,  {Agrawal} P.~C.,  {Rao} A.~R.,   {Paul} B.,  2002b, \mn@doi
  [\mnras] {10.1046/j.1365-8711.2002.05077.x}, \href
  {https://ui.adsabs.harvard.edu/abs/2002MNRAS.330..487N} {330, 487}

\bibitem[\protect\citeauthoryear{{Nandi}, {Manickam}, {Rao}  \&
  {Chakrabarti}}{{Nandi} et~al.}{2001}]{2001MNRAS.324..267N}
{Nandi} A.,  {Manickam} S.~G.,  {Rao} A.~R.,   {Chakrabarti} S.~K.,  2001,
  \mn@doi [\mnras] {10.1046/j.1365-8711.2001.04339.x}, \href
  {https://ui.adsabs.harvard.edu/abs/2001MNRAS.324..267N} {324, 267}

\bibitem[\protect\citeauthoryear{{Nandi}, {Debnath}, {Mandal}  \&
  {Chakrabarti}}{{Nandi} et~al.}{2012}]{2012A&A...542A..56N}
{Nandi} A.,  {Debnath} D.,  {Mandal} S.,   {Chakrabarti} S.~K.,  2012, \mn@doi
  [\aap] {10.1051/0004-6361/201117844}, \href
  {https://ui.adsabs.harvard.edu/abs/2012A&A...542A..56N} {542, A56}

\bibitem[\protect\citeauthoryear{{Nandi} et~al.,}{{Nandi}
  et~al.}{2018}]{2018Ap&SS.363...90N}
{Nandi} A.,  et~al., 2018, \mn@doi [\apss] {10.1007/s10509-018-3314-1}, \href
  {https://ui.adsabs.harvard.edu/abs/2018Ap&SS.363...90N} {363, 90}

\bibitem[\protect\citeauthoryear{{Neilson}, {Altmirano}, {Homan}, {Gendreau},
  {Steiner}  \& {Arzoumanian}}{{Neilson} et~al.}{2021}]{2021N}
{Neilson} J.,  {Altmirano} D.,  {Homan} J.,  {Gendreau} K.,  {Steiner} J.,
  {Arzoumanian} Z.,  2021, The Astronomer’s Telegram, 14792, 1

\bibitem[\protect\citeauthoryear{{Nowak}, {Vaughan}, {Wilms}, {Dove}  \&
  {Begelman}}{{Nowak} et~al.}{1999}]{1999ApJ...510..874N}
{Nowak} M.~A.,  {Vaughan} B.~A.,  {Wilms} J.,  {Dove} J.~B.,   {Begelman}
  M.~C.,  1999, \mn@doi [\apj] {10.1086/306610}, \href
  {https://ui.adsabs.harvard.edu/abs/1999ApJ...510..874N} {510, 874}

\bibitem[\protect\citeauthoryear{{Obst}, {Pottschmidt}, {Lohfink}, {Wilms},
  {B{\"o}ck}, {Smith}, {Tomsick}  \& {Kreykenbohm}}{{Obst}
  et~al.}{2011}]{2011AcPol..51f..49O}
{Obst} M.,  {Pottschmidt} K.,  {Lohfink} A.,  {Wilms} J.,  {B{\"o}ck} M.,
  {Smith} D.~M.,  {Tomsick} J.~A.,   {Kreykenbohm} I.,  2011, Acta
  Polytechnica, \href {https://ui.adsabs.harvard.edu/abs/2011AcPol..51f..49O}
  {51, 49}

\bibitem[\protect\citeauthoryear{{Pahari} \& {Pal}}{{Pahari} \&
  {Pal}}{2010}]{2010MNRAS.409..903P}
{Pahari} M.,  {Pal} S.,  2010, \mn@doi [\mnras]
  {10.1111/j.1365-2966.2010.17378.x}, \href
  {https://ui.adsabs.harvard.edu/abs/2010MNRAS.409..903P} {409, 903}

\bibitem[\protect\citeauthoryear{{Pal}, {Chakrabarti}  \& {Nandi}}{{Pal}
  et~al.}{2013}]{2013AdSpR..52..740P}
{Pal} P.~S.,  {Chakrabarti} S.~K.,   {Nandi} A.,  2013, \mn@doi [Advances in
  Space Research] {10.1016/j.asr.2013.04.016}, \href
  {https://ui.adsabs.harvard.edu/abs/2013AdSpR..52..740P} {52, 740}

\bibitem[\protect\citeauthoryear{{Petterson}}{{Petterson}}{1978}]{1978ApJ...224..625P}
{Petterson} J.~A.,  1978, \mn@doi [\apj] {10.1086/156411}, \href
  {https://ui.adsabs.harvard.edu/abs/1978ApJ...224..625P} {224, 625}

\bibitem[\protect\citeauthoryear{{Protassov}, {van Dyk}, {Connors}, {Kashyap}
  \& {Siemiginowska}}{{Protassov} et~al.}{2002}]{2002ApJ...571..545P}
{Protassov} R.,  {van Dyk} D.~A.,  {Connors} A.,  {Kashyap} V.~L.,
  {Siemiginowska} A.,  2002, \mn@doi [\apj] {10.1086/339856}, \href
  {https://ui.adsabs.harvard.edu/abs/2002ApJ...571..545P} {571, 545}

\bibitem[\protect\citeauthoryear{{Radhika} \& {Nandi}}{{Radhika} \&
  {Nandi}}{2014}]{2014AdSpR..54.1678R}
{Radhika} D.,  {Nandi} A.,  2014, \mn@doi [Advances in Space Research]
  {10.1016/j.asr.2014.06.039}, \href
  {https://ui.adsabs.harvard.edu/abs/2014AdSpR..54.1678R} {54, 1678}

\bibitem[\protect\citeauthoryear{{Radhika}, {Nandi}, {Agrawal}  \&
  {Seetha}}{{Radhika} et~al.}{2016a}]{2016MNRAS.460.4403R}
{Radhika} D.,  {Nandi} A.,  {Agrawal} V.~K.,   {Seetha} S.,  2016a, \mn@doi
  [\mnras] {10.1093/mnras/stw1239}, \href
  {https://ui.adsabs.harvard.edu/abs/2016MNRAS.460.4403R} {460, 4403}

\bibitem[\protect\citeauthoryear{{Radhika}, {Nandi}, {Agrawal}  \&
  {Mandal}}{{Radhika} et~al.}{2016b}]{2016MNRAS.462.1834R}
{Radhika} D.,  {Nandi} A.,  {Agrawal} V.~K.,   {Mandal} S.,  2016b, \mn@doi
  [\mnras] {10.1093/mnras/stw1755}, \href
  {https://ui.adsabs.harvard.edu/abs/2016MNRAS.462.1834R} {462, 1834}

\bibitem[\protect\citeauthoryear{{Radhika}, {Sreehari}, {Nandi}, {Iyer}  \&
  {Mand al}}{{Radhika} et~al.}{2018}]{2018Ap&SS.363..189D}
{Radhika} D.,  {Sreehari} H.,  {Nandi} A.,  {Iyer} N.,   {Mand al} S.,  2018,
  \mn@doi [\apss] {10.1007/s10509-018-3411-1}, \href
  {https://ui.adsabs.harvard.edu/abs/2018Ap&SS.363..189D} {363, 189}

\bibitem[\protect\citeauthoryear{{Rao} \& {Vadawale}}{{Rao} \&
  {Vadawale}}{2001}]{2001ESASP.459..437R}
{Rao} A.~R.,  {Vadawale} S.~V.,  2001, in {Gimenez} A.,  {Reglero} V.,
  {Winkler} C.,  eds,  ESA Special Publication Vol. 459, Exploring the
  Gamma-Ray Universe. pp 437--440

\bibitem[\protect\citeauthoryear{{Rao}, {Naik}, {Vadawale}  \&
  {Chakrabarti}}{{Rao} et~al.}{2000}]{2000A&A...360L..25R}
{Rao} A.~R.,  {Naik} S.,  {Vadawale} S.~V.,   {Chakrabarti} S.~K.,  2000, \aap,
  \href {https://ui.adsabs.harvard.edu/abs/2000A&A...360L..25R} {360, L25}

\bibitem[\protect\citeauthoryear{{Ravishankar}, {Tilak}, {Athulya}, {Radhika},
  {Agrawal}  \& {Nandi}}{{Ravishankar} et~al.}{2021}]{2021RBT}
{Ravishankar} B.~T.,  {Tilak} K.,  {Athulya} M.~P.,  {Radhika} D.,  {Agrawal}
  V.~K.,   {Nandi} A.,  2021, The Astronomer's Telegram, \href
  {https://ui.adsabs.harvard.edu/abs/2021ATel14811....1R} {14811, 1}

\bibitem[\protect\citeauthoryear{{Rawat} et~al.,}{{Rawat}
  et~al.}{2019}]{2019ApJ...870....4R}
{Rawat} D.,  et~al., 2019, \mn@doi [\apj] {10.3847/1538-4357/aaefed}, \href
  {https://ui.adsabs.harvard.edu/abs/2019ApJ...870....4R} {870, 4}

\bibitem[\protect\citeauthoryear{{Reid}, {McClintock}, {Steiner}, {Steeghs},
  {Remillard}, {Dhawan}  \& {Narayan}}{{Reid}
  et~al.}{2014}]{2014ApJ...796....2R}
{Reid} M.~J.,  {McClintock} J.~E.,  {Steiner} J.~F.,  {Steeghs} D.,
  {Remillard} R.~A.,  {Dhawan} V.,   {Narayan} R.,  2014, \mn@doi [\apj]
  {10.1088/0004-637X/796/1/2}, \href
  {https://ui.adsabs.harvard.edu/abs/2014ApJ...796....2R} {796, 2}

\bibitem[\protect\citeauthoryear{{Remillard} \& {McClintock}}{{Remillard} \&
  {McClintock}}{2006}]{2006ARA&A..44...49R}
{Remillard} R.~A.,  {McClintock} J.~E.,  2006, \mn@doi [\araa]
  {10.1146/annurev.astro.44.051905.092532}, \href
  {https://ui.adsabs.harvard.edu/abs/2006ARA&A..44...49R} {44, 49}

\bibitem[\protect\citeauthoryear{{Reynolds}, {Miller}  \&
  {Balakrishnan}}{{Reynolds} et~al.}{2020}]{2020ATel13974....1R}
{Reynolds} M.,  {Miller} J.,   {Balakrishnan} M.,  2020, The Astronomer's
  Telegram, \href {https://ui.adsabs.harvard.edu/abs/2020ATel13974....1R}
  {13974, 1}

\bibitem[\protect\citeauthoryear{Riemann}{Riemann}{1867}]{riemann_1867}
Riemann B.,  1867, Ueber die Darstellbarkeit einer Function durch eine
  trigonometrische Reihe.
G\"{o}ttingen : In der Dieterichschen Buchhandlung

\bibitem[\protect\citeauthoryear{{Rodriguez}, {Durouchoux}, {Mirabel}, {Ueda},
  {Tagger}  \& {Yamaoka}}{{Rodriguez} et~al.}{2002}]{2002A&A...386..271R}
{Rodriguez} J.,  {Durouchoux} P.,  {Mirabel} I.~F.,  {Ueda} Y.,  {Tagger} M.,
  {Yamaoka} K.,  2002, \mn@doi [\aap] {10.1051/0004-6361:20020218}, \href
  {https://ui.adsabs.harvard.edu/abs/2002A&A...386..271R} {386, 271}

\bibitem[\protect\citeauthoryear{{Rybicki} \& {Lightman}}{{Rybicki} \&
  {Lightman}}{1986}]{1986rpa..book.....R}
{Rybicki} G.~B.,  {Lightman} A.~P.,  1986, {Radiative Processes in
  Astrophysics}

\bibitem[\protect\citeauthoryear{{Ryu}, {Chakrabarti}  \& {Molteni}}{{Ryu}
  et~al.}{1997}]{1997ApJ...474..378R}
{Ryu} D.,  {Chakrabarti} S.~K.,   {Molteni} D.,  1997, \mn@doi [\apj]
  {10.1086/303461}, \href
  {https://ui.adsabs.harvard.edu/abs/1997ApJ...474..378R} {474, 378}

\bibitem[\protect\citeauthoryear{{Shakura} \& {Sunyaev}}{{Shakura} \&
  {Sunyaev}}{1973}]{1973A&A....24..337S}
{Shakura} N.~I.,  {Sunyaev} R.~A.,  1973, \aap, \href
  {https://ui.adsabs.harvard.edu/abs/1973A&A....24..337S} {500, 33}

\bibitem[\protect\citeauthoryear{{Shakura} \& {Sunyaev}}{{Shakura} \&
  {Sunyaev}}{1976}]{1976MNRAS.175..613S}
{Shakura} N.~I.,  {Sunyaev} R.~A.,  1976, \mn@doi [\mnras]
  {10.1093/mnras/175.3.613}, \href
  {https://ui.adsabs.harvard.edu/abs/1976MNRAS.175..613S} {175, 613}

\bibitem[\protect\citeauthoryear{{Shang}, {Debnath}, {Chatterjee}, {Jana},
  {Chakrabarti}, {Chang}, {Yap}  \& {Chiu}}{{Shang}
  et~al.}{2019}]{2019ApJ...875....4S}
{Shang} J.~R.,  {Debnath} D.,  {Chatterjee} D.,  {Jana} A.,  {Chakrabarti}
  S.~K.,  {Chang} H.~K.,  {Yap} Y.~X.,   {Chiu} C.~L.,  2019, \mn@doi
  [\textbf{\apj}] {10.3847/1538-4357/ab0c1e}, \href
  {https://ui.adsabs.harvard.edu/abs/2019ApJ...875....4S} {875, 4}

\bibitem[\protect\citeauthoryear{{Shaposhnikov} \& {Titarchuk}}{{Shaposhnikov}
  \& {Titarchuk}}{2007}]{2007ApJ...663..445S}
{Shaposhnikov} N.,  {Titarchuk} L.,  2007, \mn@doi [\apj] {10.1086/518110},
  \href {https://ui.adsabs.harvard.edu/abs/2007ApJ...663..445S} {663, 445}

\bibitem[\protect\citeauthoryear{{Shaposhnikov} \& {Titarchuk}}{{Shaposhnikov}
  \& {Titarchuk}}{2009}]{2009ApJ...699..453S}
{Shaposhnikov} N.,  {Titarchuk} L.,  2009, \mn@doi [\apj]
  {10.1088/0004-637X/699/1/453}, \href
  {https://ui.adsabs.harvard.edu/abs/2009ApJ...699..453S} {699, 453}

\bibitem[\protect\citeauthoryear{{Singh} et~al.,}{{Singh}
  et~al.}{2016}]{2016SPIE.9905E..1ES}
{Singh} K.~P.,  et~al., 2016, in \procspie. p. 99051E,
  \mn@doi{10.1117/12.2235309}

\bibitem[\protect\citeauthoryear{{Singh} et~al.,}{{Singh}
  et~al.}{2017}]{2017JApA...38...29S}
{Singh} K.~P.,  et~al., 2017, \mn@doi [Journal of Astrophysics and Astronomy]
  {10.1007/s12036-017-9448-7}, \href
  {https://ui.adsabs.harvard.edu/abs/2017JApA...38...29S} {38, 29}

\bibitem[\protect\citeauthoryear{{Sreehari} \& {Nandi}}{{Sreehari} \&
  {Nandi}}{2021}]{2021MNRAS.502.1334S}
{Sreehari} H.,  {Nandi} A.,  2021, \mn@doi [\mnras] {10.1093/mnras/stab151},
  \href {https://ui.adsabs.harvard.edu/abs/2021MNRAS.502.1334S} {502, 1334}

\bibitem[\protect\citeauthoryear{{Sreehari}, {Iyer}, {Radhika}, {Nandi}  \&
  {Mandal}}{{Sreehari} et~al.}{2019a}]{2019AdSpR..63.1374S}
{Sreehari} H.,  {Iyer} N.,  {Radhika} D.,  {Nandi} A.,   {Mandal} S.,  2019a,
  \mn@doi [Advances in Space Research] {10.1016/j.asr.2018.10.042}, \href
  {https://ui.adsabs.harvard.edu/abs/2019AdSpR..63.1374S} {63, 1374}

\bibitem[\protect\citeauthoryear{{Sreehari}, {Ravishankar}, {Iyer}, {Agrawal},
  {Katoch}, {Mandal}  \& {Nandi}}{{Sreehari}
  et~al.}{2019b}]{2019MNRAS.487..928S}
{Sreehari} H.,  {Ravishankar} B.~T.,  {Iyer} N.,  {Agrawal} V.~K.,  {Katoch}
  T.~B.,  {Mandal} S.,   {Nandi} A.,  2019b, \mn@doi [\mnras]
  {10.1093/mnras/stz1327}, \href
  {https://ui.adsabs.harvard.edu/abs/2019MNRAS.487..928S} {487, 928}

\bibitem[\protect\citeauthoryear{{Sreehari}, {Nandi}, {Das}, {Agrawal},
  {Mandal}, {Ramadevi}  \& {Katoch}}{{Sreehari}
  et~al.}{2020}]{2020MNRAS.499.5891S}
{Sreehari} H.,  {Nandi} A.,  {Das} S.,  {Agrawal} V.~K.,  {Mandal} S.,
  {Ramadevi} M.~C.,   {Katoch} T.,  2020, \mn@doi [\mnras]
  {10.1093/mnras/staa3135}, \href
  {https://ui.adsabs.harvard.edu/abs/2020MNRAS.499.5891S} {499, 5891}

\bibitem[\protect\citeauthoryear{{Strohmayer}}{{Strohmayer}}{2001}]{2001ApJ...554L.169S}
{Strohmayer} T.~E.,  2001, \mn@doi [\apjl] {10.1086/321720}, \href
  {https://ui.adsabs.harvard.edu/abs/2001ApJ...554L.169S} {554, L169}

\bibitem[\protect\citeauthoryear{{Sunyaev} \& {Titarchuk}}{{Sunyaev} \&
  {Titarchuk}}{1980}]{1980A&A....86..121S}
{Sunyaev} R.~A.,  {Titarchuk} L.~G.,  1980, \aap, \href
  {https://ui.adsabs.harvard.edu/abs/1980A&A....86..121S} {500, 167}

\bibitem[\protect\citeauthoryear{{Tanaka} \& {Lewin}}{{Tanaka} \&
  {Lewin}}{1995}]{1995xrbi.nasa..126T}
{Tanaka} Y.,  {Lewin} W.~H.~G.,  1995, in X-ray Binaries. pp 126--174

\bibitem[\protect\citeauthoryear{{Tandon} et~al.,}{{Tandon}
  et~al.}{2017}]{2017AJ....154..128T}
{Tandon} S.~N.,  et~al., 2017, \mn@doi [\aj] {10.3847/1538-3881/aa8451}, \href
  {https://ui.adsabs.harvard.edu/abs/2017AJ....154..128T} {154, 128}

\bibitem[\protect\citeauthoryear{{Titarchuk} \& {Fiorito}}{{Titarchuk} \&
  {Fiorito}}{2004}]{2004ApJ...612..988T}
{Titarchuk} L.,  {Fiorito} R.,  2004, \mn@doi [\apj] {10.1086/422573}, \href
  {https://ui.adsabs.harvard.edu/abs/2004ApJ...612..988T} {612, 988}

\bibitem[\protect\citeauthoryear{{Tomsick} \& {Kaaret}}{{Tomsick} \&
  {Kaaret}}{2001}]{2001ApJ...548..401T}
{Tomsick} J.~A.,  {Kaaret} P.,  2001, \mn@doi [\apj] {10.1086/318683}, \href
  {https://ui.adsabs.harvard.edu/abs/2001ApJ...548..401T} {548, 401}

\bibitem[\protect\citeauthoryear{{Trushkin}, {Nizhelskij}, {Tsybulev}, {Bursov}
   \& {Shevchenko}}{{Trushkin} et~al.}{2019a}]{2019ATel12855....1T}
{Trushkin} S.~A.,  {Nizhelskij} N.~A.,  {Tsybulev} P.~G.,  {Bursov} N.~N.,
  {Shevchenko} A.~V.,  2019a, The Astronomer's Telegram, \href
  {https://ui.adsabs.harvard.edu/abs/2019ATel12855....1T} {12855, 1}

\bibitem[\protect\citeauthoryear{{Trushkin}, {Nizhelskij}, {Tsybulev}, {Bursov}
   \& {Shevchenko}}{{Trushkin} et~al.}{2019b}]{2019ATel13304....1T}
{Trushkin} S.~A.,  {Nizhelskij} N.~A.,  {Tsybulev} P.~G.,  {Bursov} N.~N.,
  {Shevchenko} A.~V.,  2019b, The Astronomer's Telegram, \href
  {https://ui.adsabs.harvard.edu/abs/2019ATel13304....1T} {13304, 1}

\bibitem[\protect\citeauthoryear{{Ueda}, {Yamaoka}  \& {Remillard}}{{Ueda}
  et~al.}{2009}]{2009ApJ...695..888U}
{Ueda} Y.,  {Yamaoka} K.,   {Remillard} R.,  2009, \mn@doi [\apj]
  {10.1088/0004-637X/695/2/888}, \href
  {https://ui.adsabs.harvard.edu/abs/2009ApJ...695..888U} {695, 888}

\bibitem[\protect\citeauthoryear{{Ueda} et~al.,}{{Ueda}
  et~al.}{2010}]{2010ApJ...713..257U}
{Ueda} Y.,  et~al., 2010, \mn@doi [\apj] {10.1088/0004-637X/713/1/257}, \href
  {https://ui.adsabs.harvard.edu/abs/2010ApJ...713..257U} {713, 257}

\bibitem[\protect\citeauthoryear{{Uttley} \& {McHardy}}{{Uttley} \&
  {McHardy}}{2001}]{2001MNRAS.323L..26U}
{Uttley} P.,  {McHardy} I.~M.,  2001, \mn@doi [\mnras]
  {10.1046/j.1365-8711.2001.04496.x}, \href
  {https://ui.adsabs.harvard.edu/abs/2001MNRAS.323L..26U} {323, L26}

\bibitem[\protect\citeauthoryear{{Verdhan Chauhan} et~al.,}{{Verdhan Chauhan}
  et~al.}{2017}]{2017ApJ...841...41V}
{Verdhan Chauhan} J.,  et~al., 2017, \mn@doi [\apj] {10.3847/1538-4357/aa6d7e},
  \href {https://ui.adsabs.harvard.edu/abs/2017ApJ...841...41V} {841, 41}

\bibitem[\protect\citeauthoryear{{Weng}, {Wang}, {Cai}, {Yuan}  \& {Gu}}{{Weng}
  et~al.}{2018}]{2018ApJ...865...19W}
{Weng} S.-S.,  {Wang} T.-T.,  {Cai} J.-P.,  {Yuan} Q.-R.,   {Gu} W.-M.,  2018,
  \mn@doi [\apj] {10.3847/1538-4357/aad9a6}, \href
  {https://ui.adsabs.harvard.edu/abs/2018ApJ...865...19W} {865, 19}

\bibitem[\protect\citeauthoryear{{Wijnands}, {Homan}  \& {van der
  Klis}}{{Wijnands} et~al.}{1999}]{1999ApJ...526L..33W}
{Wijnands} R.,  {Homan} J.,   {van der Klis} M.,  1999, \mn@doi [\apjl]
  {10.1086/312365}, \href
  {https://ui.adsabs.harvard.edu/abs/1999ApJ...526L..33W} {526, L33}

\bibitem[\protect\citeauthoryear{{Wilms}, {Allen}  \& {McCray}}{{Wilms}
  et~al.}{2000}]{2000ApJ...542..914W}
{Wilms} J.,  {Allen} A.,   {McCray} R.,  2000, \mn@doi [\apj] {10.1086/317016},
  \href {https://ui.adsabs.harvard.edu/abs/2000ApJ...542..914W} {542, 914}

\bibitem[\protect\citeauthoryear{{Yadav}, {Rao}, {Agrawal}, {Paul}, {Seetha}
  \& {Kasturirangan}}{{Yadav} et~al.}{1999}]{1999ApJ...517..935Y}
{Yadav} J.~S.,  {Rao} A.~R.,  {Agrawal} P.~C.,  {Paul} B.,  {Seetha} S.,
  {Kasturirangan} K.,  1999, \mn@doi [\apj] {10.1086/307225}, \href
  {https://ui.adsabs.harvard.edu/abs/1999ApJ...517..935Y} {517, 935}

\bibitem[\protect\citeauthoryear{{Yadav} et~al.,}{{Yadav}
  et~al.}{2016}]{2016ApJ...833...27Y}
{Yadav} J.~S.,  et~al., 2016, \mn@doi [\apj] {10.3847/0004-637X/833/1/27},
  \href {https://ui.adsabs.harvard.edu/abs/2016ApJ...833...27Y} {833, 27}

\bibitem[\protect\citeauthoryear{{Yan}, {Ding}, {Wang}, {Qu}  \& {Song}}{{Yan}
  et~al.}{2013}]{2013MNRAS.434...59Y}
{Yan} S.-P.,  {Ding} G.-Q.,  {Wang} N.,  {Qu} J.-L.,   {Song} L.-M.,  2013,
  \mn@doi [\mnras] {10.1093/mnras/stt968}, \href
  {https://ui.adsabs.harvard.edu/abs/2013MNRAS.434...59Y} {434, 59}

\bibitem[\protect\citeauthoryear{{Zdziarski}, {Johnson}  \&
  {Magdziarz}}{{Zdziarski} et~al.}{1996}]{1996MNRAS.283..193Z}
{Zdziarski} A.~A.,  {Johnson} W.~N.,   {Magdziarz} P.,  1996, \mn@doi [\mnras]
  {10.1093/mnras/283.1.193}, \href
  {https://ui.adsabs.harvard.edu/abs/1996MNRAS.283..193Z} {283, 193}

\bibitem[\protect\citeauthoryear{{Zdziarski}, {Grove}, {Poutanen}, {Rao}  \&
  {Vadawale}}{{Zdziarski} et~al.}{2001}]{2001ApJ...554L..45Z}
{Zdziarski} A.~A.,  {Grove} J.~E.,  {Poutanen} J.,  {Rao} A.~R.,   {Vadawale}
  S.~V.,  2001, \mn@doi [\apjl] {10.1086/320932}, \href
  {https://ui.adsabs.harvard.edu/abs/2001ApJ...554L..45Z} {554, L45}

\bibitem[\protect\citeauthoryear{{Zdziarski}, {Gierli{\'n}ski}, {Rao},
  {Vadawale}  \& {Miko{\l}ajewska}}{{Zdziarski}
  et~al.}{2005}]{2005MNRAS.360..825Z}
{Zdziarski} A.~A.,  {Gierli{\'n}ski} M.,  {Rao} A.~R.,  {Vadawale} S.~V.,
  {Miko{\l}ajewska} J.,  2005, \mn@doi [\mnras]
  {10.1111/j.1365-2966.2005.09112.x}, \href
  {https://ui.adsabs.harvard.edu/abs/2005MNRAS.360..825Z} {360, 825}

\bibitem[\protect\citeauthoryear{{Zdziarski}, {Zi{\'o}{\l}kowski}  \&
  {Miko{\l}ajewska}}{{Zdziarski} et~al.}{2019}]{2019MNRAS.488.1026Z}
{Zdziarski} A.~A.,  {Zi{\'o}{\l}kowski} J.,   {Miko{\l}ajewska} J.,  2019,
  \mn@doi [\mnras] {10.1093/mnras/stz1787}, \href
  {https://ui.adsabs.harvard.edu/abs/2019MNRAS.488.1026Z} {488, 1026}

\bibitem[\protect\citeauthoryear{{Zhang}, {Chen}, {Qu}  \& {Bu}}{{Zhang}
  et~al.}{2015}]{2015AJ....149...82Z}
{Zhang} L.,  {Chen} L.,  {Qu} J.-l.,   {Bu} Q.-c.,  2015, \mn@doi [\aj]
  {10.1088/0004-6256/149/2/82}, \href
  {https://ui.adsabs.harvard.edu/abs/2015AJ....149...82Z} {149, 82}

\bibitem[\protect\citeauthoryear{{Zhang} et~al.,}{{Zhang}
  et~al.}{2020}]{2020MNRAS.494.1375Z}
{Zhang} L.,  et~al., 2020, \mn@doi [\mnras] {10.1093/mnras/staa797}, \href
  {https://ui.adsabs.harvard.edu/abs/2020MNRAS.494.1375Z} {494, 1375}

\bibitem[\protect\citeauthoryear{{{\.Z}ycki}, {Done}  \& {Smith}}{{{\.Z}ycki}
  et~al.}{1999}]{1999MNRAS.309..561Z}
{{\.Z}ycki} P.~T.,  {Done} C.,   {Smith} D.~A.,  1999, \mn@doi [\mnras]
  {10.1046/j.1365-8711.1999.02885.x}, \href
  {https://ui.adsabs.harvard.edu/abs/1999MNRAS.309..561Z} {309, 561}

\bibitem[\protect\citeauthoryear{{van Haaften}, {Nelemans}, {Voss}, {van der
  Sluys}  \& {Toonen}}{{van Haaften} et~al.}{2015}]{2015A&A...579A..33V}
{van Haaften} L.~M.,  {Nelemans} G.,  {Voss} R.,  {van der Sluys} M.~V.,
  {Toonen} S.,  2015, \mn@doi [\aap] {10.1051/0004-6361/201425303}, \href
  {https://ui.adsabs.harvard.edu/abs/2015A&A...579A..33V} {579, A33}

\bibitem[\protect\citeauthoryear{{van der Klis}}{{van der
  Klis}}{1989}]{1989ARA&A..27..517V}
{van der Klis} M.,  1989, \mn@doi [\araa]
  {10.1146/annurev.aa.27.090189.002505}, \href
  {https://ui.adsabs.harvard.edu/abs/1989ARA&A..27..517V} {27, 517}

\bibitem[\protect\citeauthoryear{{van der Klis}}{{van der
  Klis}}{1995}]{1995xrbi.nasa..252V}
{van der Klis} M.,  1995, in X-ray Binaries. pp 252--307

\bibitem[\protect\citeauthoryear{{van der Klis}}{{van der
  Klis}}{2000}]{2000ARA&A..38..717V}
{van der Klis} M.,  2000, \mn@doi [\araa] {10.1146/annurev.astro.38.1.717},
  \href {https://ui.adsabs.harvard.edu/abs/2000ARA&A..38..717V} {38, 717}

\makeatother
\end{thebibliography}




\appendix


\bsp	
\label{lastpage}
\end{document}